\documentclass[longauth]{aa}
\usepackage{graphicx}
\usepackage{nccmath}
\usepackage{txfonts}
\usepackage{longtable}
\usepackage{xcolor}
\usepackage{float}
\usepackage{capt-of}
\usepackage{upgreek}
\usepackage{amsmath}
\usepackage{booktabs}
\usepackage{caption}
\usepackage{subcaption}
\usepackage{hyperref}
\usepackage[capitalise]{cleveref}
\usepackage{etoolbox}
\usepackage{scalerel}
\usepackage[rightcaption]{sidecap}
\newcommand\HII{H\protect\scaleto{$II$~}{1.2ex}}
\newcommand\NeII{Ne\protect\scaleto{$II$~}{1.2ex}}
\usepackage{txfonts}
\hypersetup{
    colorlinks=true,
    linkcolor=blue,
    filecolor=blue,      
    urlcolor=blue,
    citecolor=blue,
}
\begin{document}

   \title{ALMA-IMF XII: Point-process mapping of 15 massive protoclusters\thanks{The luminosity, temperature and column density maps are only available in electronic form at the CDS via anonymous ftp to cdsarc.u-strasbg.fr (130.79.128.5) or via \url{http://cdsweb.u-strasbg.fr}.}}

   \author{P. Dell'Ova
          \inst{1}
          \and
          F. Motte\inst{2}
          \and
          A. Gusdorf\inst{1}
          \and
          Y. Pouteau\inst{2}
          \and
          A. Men’shchikov\inst{3}
          \and
          D., D\'iaz-Gonz\'alez\inst{4}
          \and
          R. Galván-Madrid\inst{4}
          \and
          P.~Lesaffre\inst{1}
          \and
          P.~Didelon\inst{3}
          \and
          A.M. Stutz\inst{5,6}
          \and
          A.P.M. Towner\inst{7}
          \and
          K. Marsh\inst{8}
          \and 
          A. Whitworth\inst{9}
          \and
          M. Armante\inst{1}
          \and
          M. Bonfand\inst{10}
          \and
          T.~Nony\inst{4}
          \and
          M.~Valeille-Manet\inst{11} 
          \and
          S.~Bontemps\inst{11}
          \and
          T.~Csengeri\inst{11}
          \and
          N. Cunningham\inst{2}
          \and
          A. Ginsburg\inst{7}
          \and
          F. Louvet\inst{2,5}
          \and
          R.~H.~\'Alvarez-Guti\'errez\inst{5}
          \and
          N. Brouillet\inst{11}
          \and
          J. Salinas\inst{5,6}
          \and 
          P. Sanhueza\inst{12,13}
          \and 
          F. Nakamura\inst{12,13}
          \and 
          Q. Nguyen Luong\inst{14}
          \and 
          T.~Baug\inst{15}
          \and
          M.~Fern\'andez-L\'opez\inst{16}
          \and 
          H.-L. Liu\inst{5,17}
          \and
          F. Olguin\inst{18}
          }

   \institute{Laboratoire de Physique de l’École Normale Supérieure, ENS, Université PSL, CNRS, Sorbonne Université, Université de Paris, F-75005 Paris, France\\
              \email{pierre.dellova@ens.fr}
        \and
            Université Grenoble Alpes, CNRS, IPAG, 38000 Grenoble, France
        \and
            Université Paris-Saclay, Université Paris Cité, CEA, CNRS, AIM, 91191 Gif-sur-Yvette, France
        \and
            Instituto de Radioastronom\'ia y Atrof\'isica, Universidad Nacional Aut\'onoma de México, Morelia, Michoac\'an 58089, México
        \and
            Departamento de Astronom\'{i}a, Universidad de Concepci\'{o}n, Casilla 160-C, Concepci\'{o}n, Chile
        \and
            Franco-Chilean Laboratory for Astronomy, IRL 3386, CNRS and Universidad de Chile, Santiago, Chile
        \and
            Department of Astronomy, University of Florida, PO Box 112055, USA
        \and 
            Infrared Processing and Analysis Center, CalTech, 1200E California Boulevard Pasadena, CA91125, USA
        \and
            School of Physics and Astronomy, Cardiff University, Queen’s Buildings, The Parade, Cardiff CF24 3AA, UK
        \and
            Departments of Astronomy \& Chemistry, University of Virginia, Charlottesville, VA 22904, USA
        \and 
            Laboratoire d’astrophysique de Bordeaux, Univ. Bordeaux, CNRS, B18N, allée Geoffroy Saint-Hilaire, 33615 Pessac, France
        \and
        National Astronomical Observatory of Japan, National Institutes of
Natural Sciences, 2-21-1 Osawa, Mitaka, Tokyo 181-8588, Japan
        \and 
        Department of Astronomical Science, SOKENDAI (The Graduate
University for Advanced Studies), 2-21-1 Osawa, Mitaka, Tokyo
181-8588, Japan
        \and
        CSMES, The American University of Paris, 2bis passage Landrieu
75007 Paris, France
        \and
        S. N. Bose National Centre for Basic Sciences, Block JD, Sector III,
Salt Lake, Kolkata 700106, India
        \and
        Instituto Argentino de Radioastronomía (CCT-La Plata, CONICET;
CICPBA), C.C. No. 5, 1894, Villa Elisa, Buenos Aires, Argentina
        \and 
        Department of Astronomy, Yunnan University, Kunming, 650091,
PR China
        \and
        Institute of Astronomy, National Tsing Hua University, Hsinchu
30013, Taiwan
             }

   \date{Received January 1, 2022; accepted January 1, 2022}

 
  \abstract
   {A crucial aspect in addressing the challenge of measuring the core mass function (CMF), that is pivotal for comprehending the origin of the initial mass function (IMF), lies in constraining the temperatures of the cores.}
   {We aim to measure the luminosity, mass, column density and dust temperature of star-forming regions imaged by the ALMA-IMF large program. These fields were chosen to encompass early evolutionary stages of massive protoclusters. High angular resolution mapping is required to capture the properties of protostellar and pre-stellar cores within these regions, and to effectively separate them from larger features, such as dusty filaments.}
   {We employed the point process mapping (PPMAP) technique, enabling us to perform spectral energy distribution fitting of far-infrared and submillimeter observations across the 15 ALMA-IMF fields, at an unmatched 2.5$^{\prime \prime}$ angular resolution. By combining the modified blackbody model with near-infrared data, we derived bolometric luminosity maps. We estimated the errors impacting values of each pixel in the temperature, column density, and luminosity maps. Subsequently, we employed the extraction algorithm \textsl{getsf} on the luminosity maps in order to detect luminosity peaks and measure their associated masses.}
   {We obtained high-resolution constraints on the luminosity, dust temperature, and mass of protoclusters, that are in agreement with previously reported measurements made at a coarser angular resolution. We find that the luminosity-to-mass ratio correlates with the evolutionary stage of the studied regions, albeit with intra-region variability. We compiled a PPMAP source catalog of 313 luminosity peaks using \textsl{getsf} on the derived bolometric luminosity maps. The PPMAP source catalog provides constraints on the mass and luminosity of protostars and cores, although one source may encompass several objects. Finally, we compare the estimated luminosity-to-mass ratio of PPMAP sources with evolutionary tracks and discuss the limitations imposed by the 2.5$^{\prime \prime}$ beam.}
   {}


   \maketitle
%

\section{Introduction}

\begin{SCfigure*}[1][htb]
   \begin{centering}
      \includegraphics[width=0.7\textwidth, trim={5.25cm 3cm 6.5cm 3cm},clip]{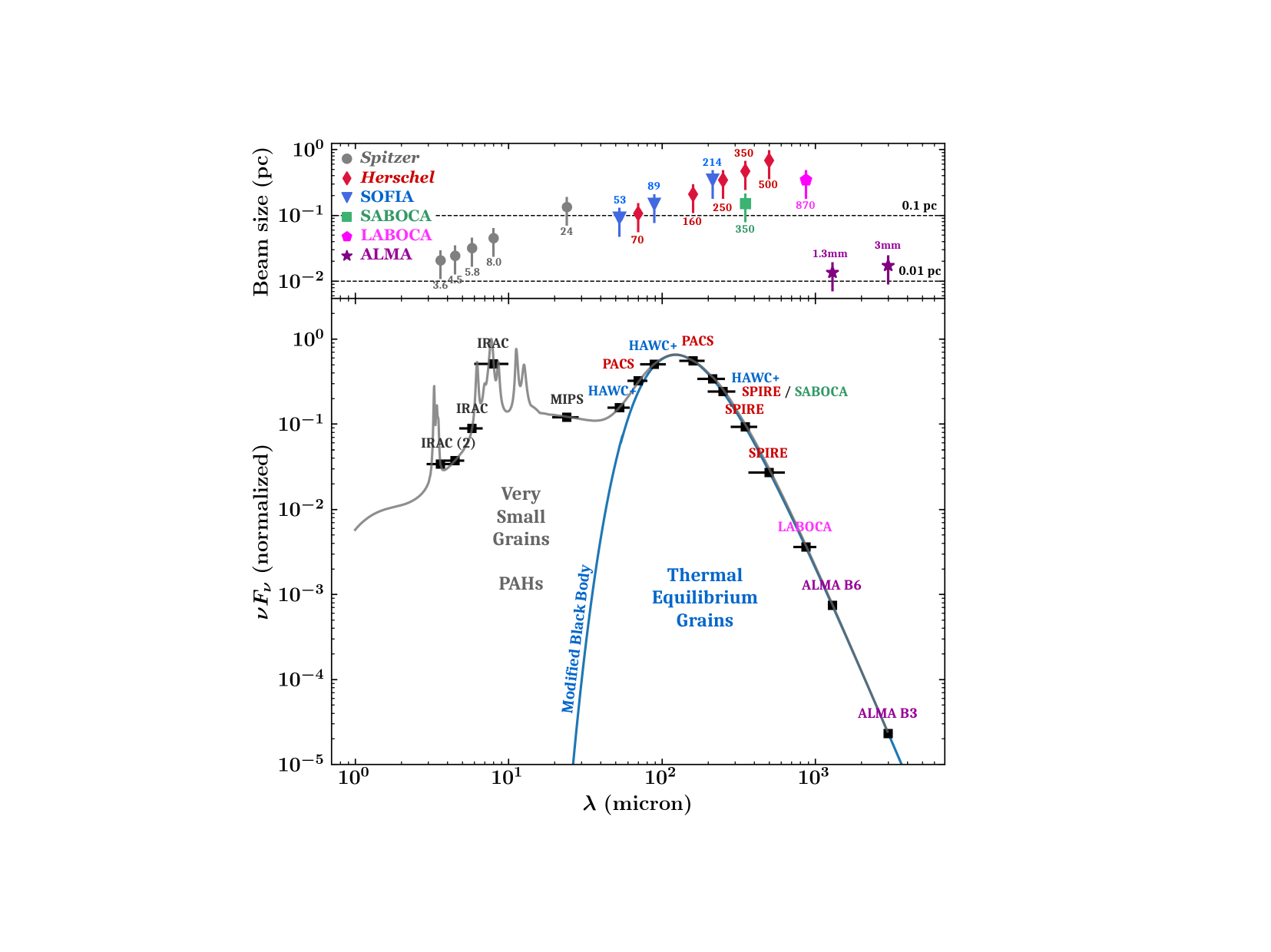}
      \caption{Observational constraints. \textit{Bottom panel}: illustrative spectral energy distribution (SED), produced using a THEMIS grain mixture (\citealt{Jones2017}, gray curve). The blue curve represents a single modified blackbody that best fits the far-infrared and millimeter range of the SED. Black markers are overlaid on the gray curve to indicate the SED coverage enabled by the observations listed in Table~\ref{table:ObservationSummary}, with the horizontal bars representing the bandwidths. \textit{Top panel}: Beam size of the observations used in our analysis, with the wavelength of the observations indicated on top of each marker (in micron, except for ALMA markers). Vertical bars represent the distance-induced variation in physical beam size across the ALMA-IMF sample.}
         \label{fig:SED-coverage}
   \end{centering}
\end{SCfigure*}

The Atacama large millimeter array -- initial mass function (ALMA-IMF\footnote{\url{https://www.almaimf.com/}}) large program surveyed massive protoclusters of the Milky Way ($2.5 - 33 \times 10^3 M_\odot$, see paper I by \citealt{Motte2022}). With distances spanning from 2 to 5.5~kpc, performing interferometric observations with ALMA was paramount to trace the dust and gas emission at a scale that probes the formation and evolution of pre-stellar cores and protostars in the dense gas of dusty filaments (see paper II by \citealt{Ginsburg2022} and paper VII by \citealt{Cunningham2023}). Measuring the mass and thus the temperature of these structures is essential to understand the conditions in which stars form, and to constrain the core mass function (hereafter CMF), since the estimated core masses may vary substantially depending on the adopted temperature. In addition to the spectrum of masses, measuring the bolometric luminosity is indispensable to build the luminosity-to-mass ratio, a fundamental quantity that can be related to the evolutionary stage of protostellar objects (\citealt{Motte2001}; \citealt{Elia2010}; \citealt{Csengeri2016}; \citealt{Mottram2017}). 
One of the challenges faced by the ALMA-IMF program is that millimeter observations, on their own, are insufficient to measure the bolometric luminosity.
This paper addresses all of these issues and provides high-resolution (2.5$^{\prime \prime}$) column density, temperature, and luminosity maps based on a multiwavelength approach.

Constraining the bolometric luminosity, dust column density, and temperature requires the spectral energy distribution (hereafter SED) of dust grains to be modeled. While it is possible to accurately describe the scattering, absorption, and reemission of starlight through the dusty interstellar medium, phenomenological approximations are more practical in most cases (see \citealt{Galliano2018}, and references therein). The modified blackbody (MBB) description, a widely used approximation, allows such measurements to be inferred from an analysis of the far-infrared (hereafter FIR, $ 70 \leq \lambda \leq 500 ~\upmu$m) and millimeter fluxes (see Fig.~\ref{fig:SED-coverage}, bottom panel). Assuming that the majority of the dust mass is in large grains ($r> 0.02~\upmu$m, \citealt{Galliano2018}), and that these grains are in thermal equilibrium (because of their large enthalpy), both the mass $M_\mathrm{dust}$ and temperature $T_\mathrm{dust}$ can be measured through SED fitting based on the following equation:
\begin{equation}
L_\lambda = M_\mathrm{dust} \kappa_0 \left( \frac{\lambda}{ \lambda_0} \right)^{-\beta} 4\pi B_\lambda (T_\mathrm{dust}),
\end{equation}
where $L_\lambda$ ($\mathrm{W~m^{-1}}$) is the monochromatic luminosity, $\kappa_0$ ($\mathrm{kg^{-1}}$) the dust mass absorption coefficient, $\beta$ the opacity index (both tied to the dust grains' physical and chemical properties), $\lambda$ the wavelength, and $B_\lambda$ the Planck function. It should be noted, however, that the MBB description cannot reproduce the dust emission at shorter wavelengths ($\lambda < 70 ~\upmu$m), since the stochastic heating of very small grains and the contribution of aromatic features (\citealt{Duley&Williams1981}; \citealt{Leger&Puget1984}; \citealt{Allamandola1985}) result in a departure from the Planck function (see Fig.~\ref{fig:SED-coverage}, bottom panel). In this paper, we do not attempt to accurately model the mid- and near-infrared domain of the dust SED ($1 \leq \lambda \leq 70 ~\upmu$m), and instead focus on the emission of dust grains in thermal equilibrium.
To this end, a thorough sampling of the SED above 70~$\upmu$m is required to constrain both $M_\mathrm{dust}$ and $T_\mathrm{dust}$ through SED fitting, in particular in the 70-250~$\upmu$m range, since the peak of the SED traces the dust temperature and opacity index. 

We acknowledge that this method is subject to biases, since line-of-sight variations of the temperature and measurement noise can induce a degeneracy between $T_\mathrm{dust}$ and $\beta$ (\citealt{Shetty2009}; \citealt{Kelly2012}; \citealt{Galliano2018}). Including observations longward of 250~$\upmu$m can help to alleviate this issue.
Several observatories can provide such FIR and millimeter measurements, namely \textit{Herschel}, SOFIA, APEX and ALMA (see Sect.~\ref{sect:observations} and Table~\ref{table:ObservationSummary}).
Relying on a diversity of instruments and bands immediately poses a problem, illustrated in the top panel of Fig.~\ref{fig:SED-coverage}: while the angular resolution of ALMA observations such as performed for the ALMA-IMF program lies between $0.29\arcsec \times 0.26$\arcsec and $1.52\arcsec \times 1.30$\arcsec, the angular resolution of \textit{Herschel}/SPIRE ranges from 17.6$^{\prime \prime}$ (at 250~$\upmu$m) to 35.2$^{\prime \prime}$ (at 500~$\upmu$m). The standard procedure (e.g., \citealt{Galametz2012}; \citealt{Aniano2012}; \citealt{Giannetti2013}; \citealt{Kohler2014}; \citealt{Guzman2015}) for SED fitting involves smoothing the observations to the same angular resolution, that is, to the coarsest resolution (35.2$^{\prime \prime}$, in our case). Applying this smoothing procedure would entirely defeat the purpose of high-angular ALMA observations, undermining the immense usefulness of high resolution long wavelength data.
Alternatively, Fourier-space combination of \textit{Herschel} images with ground-based single-dish bolometer data would allow to work at an intermediate resolution, but the improved angular resolution attained with this technique remains coarse with respect to ALMA observations (e.g., $10^{\prime \prime}$, \citealt{Lin2016,Lin2017}; $18^{\prime \prime}$, \citealt{Palmeirim2013}; \citealt{Konyves2020}; \citealt{Ladjelate2020}).

To address this issue and retain the high-angular resolution information from ALMA observations, we employ the point process mapping (PPMAP) algorithm developed by \citet{Marsh2015}. PPMAP allows us to combine and reproduce multiwavelength observations using the MBB description while preserving the spatial information contained in the higher angular scale maps.
Recently, PPMAP was applied to reveal and constrain dust structures in supernova remnants (\citealt{Chawner2019}; \citealt{Chawner2020}), star-forming filaments (\citealt{Howard2019}; \citealt{Howard2021}) and in the Milky Way disk \citep{Bates2023}.
All these studies were based on \textit{Herschel} observations, and some included SCUBA-2 data \citep{Holland2013}.
For the first time, \citet{Motte2018b} advanced PPMAP so far as to simultaneously fit observations from a data set that spanned two orders of magnitudes in angular resolution (from $0.37^{\prime \prime} \times 0.53^{\prime \prime}$ to $35.2^{\prime \prime}$). The results they achieved with a $2.5^{\prime \prime}$ resolution in the massive W43-MM1 protocluster compelled us to apply this novel procedure to the analysis of the 15 ALMA-IMF fields.
Through PPMAP, we perform SED fitting for a large set of continuum observations, while preserving the high-angular resolution capabilities of ALMA, providing a first step toward constraining the luminosity, column density, and temperature of the population of candidate cores and protostars.

In Sect.~\ref{sect:observations}, we present the ALMA and complementary continuum observations toward the ALMA-IMF protoclusters. We then proceed to the analysis in Sect.~\ref{sect:PPMAP}, where we describe the PPMAP algorithm and the methods used to apply it to our specific problem. The derived luminosity, dust temperature and column density maps are presented and compared with previous studies in Sect.~\ref{sect:calibration}. Lastly, the construction of a PPMAP luminosity peaks catalog, encompassing luminosity and mass measurements, is detailed in Sect.~\ref{sect:discussion}, in which we discuss our findings.

\begin{table*}[htb]
\centering
{\caption{Summary of available surveys we used.}       
\label{table:ObservationSummary}}      
{\centering                          
\begin{tabular}{c c c c c c c}        
\hline \hline         \\[-1.0em]
\footnotesize{Survey} & \footnotesize{Wavelength} & \footnotesize{Angular resolution} & \footnotesize{Spatial coverage} & \footnotesize{Reference(s)} \\
\footnotesize{} & \footnotesize{($\upmu$m)} & \footnotesize{(arcsec)} & \footnotesize{} \\
\hline \\[-1.0em]
\footnotesize{ALMA-IMF} & \footnotesize{} & \footnotesize{} \\
\cline{1-1} \\[-1.0em]
\footnotesize{Band 6$^1$} & \footnotesize{1300} & \footnotesize{$0.35 \times 0.27 - 1.09 \times 0.70^2$} & \footnotesize{$\sim 1^\prime \times 1^\prime$} & \footnotesize{Galván-Madrid et al. (in prep.),} \\
\footnotesize{Band 3$^3$} & \footnotesize{3000} & \footnotesize{$0.29 \times 0.26 - 1.52 \times 1.30^2$} & & \footnotesize{and \citet{Diaz2023}} \\
\footnotesize{APEX} & \footnotesize{} & \footnotesize{}\\
\cline{1-1} \\[-1.0em]
\footnotesize{SABOCA} & \footnotesize{350} & \footnotesize{7.8} & \footnotesize{$\sim 2^\prime \times 2^\prime$} & \footnotesize{\citet{Lin2019}}\\
\footnotesize{LABOCA} & \footnotesize{870} & \footnotesize{19.2} & \footnotesize{> $1^\circ \times 1^\circ$} & \footnotesize{\citet{Schuller2009}, \citet{Csengeri2014}}\\
\footnotesize{Hi-GAL$+$HOBYS (\textit{Herschel})} & \footnotesize{} & \footnotesize{}\\
\cline{1-1} \\[-1.0em]
\footnotesize{PACS} & \footnotesize{70} & \footnotesize{5.6} & \footnotesize{> $1^\circ \times 1^\circ$} & \footnotesize{\citet{Molinari2010},}\\
\footnotesize{PACS} & \footnotesize{160} & \footnotesize{10.7} & & \footnotesize{\citet{Motte2010}}\\
\footnotesize{SPIRE} & \footnotesize{250} & \footnotesize{17.6}\\
\footnotesize{SPIRE} & \footnotesize{350} & \footnotesize{23.9}\\
\footnotesize{SPIRE} & \footnotesize{500} & \footnotesize{35.2}\\
\footnotesize{SOFIA} & \footnotesize{} & \footnotesize{}\\
\cline{1-1} \\[-1.0em]
\footnotesize{HAWC+} & \footnotesize{53} & \footnotesize{4.85} & \footnotesize{$\sim 3^\prime \times 3^\prime$} & \footnotesize{\citet{Vaillancourt2016},}\\
\footnotesize{HAWC+} & \footnotesize{89} & \footnotesize{7.8} & & \footnotesize{\citet{Pillai2023}}\\
\footnotesize{HAWC+} & \footnotesize{214} & \footnotesize{18.2} & \footnotesize{}\\
\footnotesize{MIPSGAL (\textit{Spitzer})} & \footnotesize{} & \footnotesize{}\\
\cline{1-1} \\[-1.0em]
\footnotesize{MIPS} & \footnotesize{24} & \footnotesize{5.6} & \footnotesize{> $1^\circ \times 1^\circ$} & \footnotesize{\citet{Carey2009}} \\
\footnotesize{GLIMPSE (\textit{Spitzer})} & \footnotesize{} & \footnotesize{}\\
\cline{1-1} \\[-1.0em]
\footnotesize{IRAC} & \footnotesize{3.6} & \footnotesize{1.7} & \footnotesize{> $1^\circ \times 1^\circ$} & \footnotesize{\citet{Benjamin2003}}\\
\footnotesize{IRAC} & \footnotesize{4.5} & \footnotesize{1.7}\\
\footnotesize{IRAC} & \footnotesize{5.8} & \footnotesize{1.9}\\
\footnotesize{IRAC} & \footnotesize{8.0} & \footnotesize{2.0}\\
\hline
\end{tabular}}
\begin{tabular}{l}
\footnotesize{$^1$As reported in Sect.~\ref{subsect:ALMAobs}, Band 6 observations decontaminated from free-free emission were used in our analysis (Galván-Madrid et al., in prep.).} \\
\footnotesize{$^2$We indicate both the best (first) and coarsest (second) angular resolutions obtained with the ALMA-IMF observations.} \\
\footnotesize{$^3$ALMA Band 3 observations were not used in the PPMAP analysis, in order to prevent contamination by free-free emission (see Sect.~\ref{sect:free-free}).}
\end{tabular}
\end{table*}

\section{Observations}\label{sect:observations}

\begin{figure}[htb]
   \begin{centering}
      \includegraphics[width=\hsize, trim={0.5cm 1.5cm 1.25cm 0.25cm},clip]{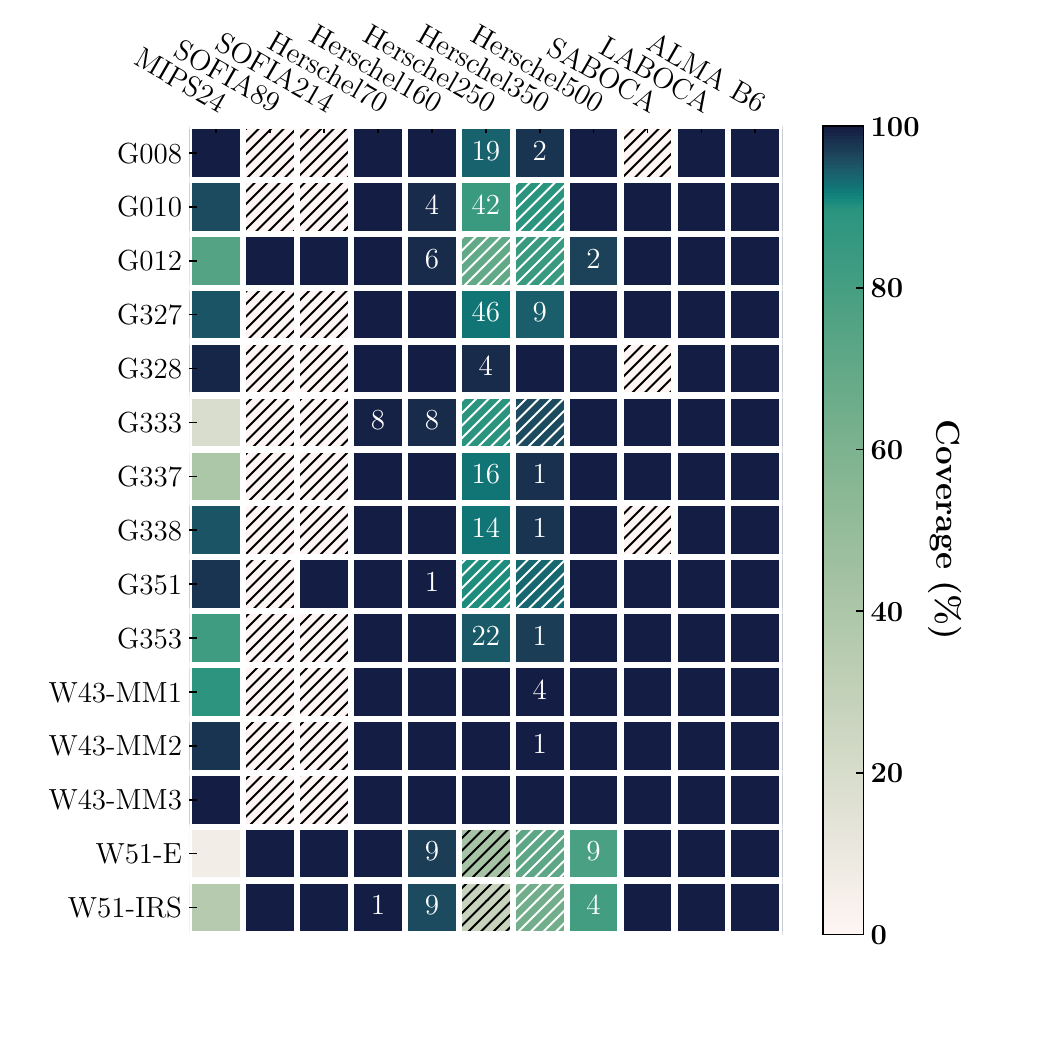}
      \caption{Data coverage chart. The color scale represents the percentage of observed and unsaturated pixels in each pair of region and map. \textit{Spitzer}/IRAC data, which has no saturated pixels in the regions studied, is not shown here. The numbers account for the amount of pixels replaced through \texttt{astrofix} (example: in G012.80, the Hi-GAL 160 $\upmu$m map has 6 saturated pixels that were interpolated). Maps that are either missing or discarded from the analysis are hatched.}
         \label{fig:datacompleteness}
   \end{centering}
\end{figure}

In addition to the new observations obtained with the ALMA, the modified blackbody SED fitting analysis requires far-infrared data. To cover the necessary wavelength range between 70~$\upmu$m and 870~$\upmu$m, we have gathered observations from five different instruments. Additional mid- and near-infrared observations are required to derive bolometric luminosities. The details of these observations, along with the corresponding instruments, are provided in Table~\ref{table:ObservationSummary} and shown in Fig.~\ref{fig:datacompleteness}.
To complete the sampling of the SED below 70~$\upmu$m and estimate the bolometric luminosity (see Sect.~\ref{subsect:luminosity}), we have also collected archival observations between 3.6~$\upmu$m and 24~$\upmu$m.
In the following subsections, we present these different observations and the corresponding instruments in detail.

\subsection{ALMA-IMF images}\label{subsect:ALMAobs}

We used the continuum images presented in \citet{Diaz2023}. These images are the result of combining the 3 mm (Band 3) and 1 mm (Band 6) ALMA-IMF continuum images presented in Paper I \citep{Motte2022} and Paper II \citep{Ginsburg2022} with the pilot of the Mustang-2 Galactic plane survey (MGPS90, \citealt{Ginsburg2020}) at 3 mm and the Bolocam Galactic plane survey (BGPS, \citealt{Aguirre2011}, \citealt{Ginsburg2013}). The more evolved ALMA-IMF fields have a significant contribution of free-free emission in the continuum images. To aid our photometry measurements, we also used the estimates of pure dust emission presented in Galván-Madrid et al. (in prep.), which subtract the free-free contribution to the continuum at 1~mm using the H$41\alpha$ recombination line within the ALMA-IMF data set. Section~\ref{sect:free-free} provide further details about this procedure. The absolute flux calibration for the ALMA observations is estimated to have an uncertainty of 10\% as reported by \citet{Ginsburg2022}.

\subsection{ATLASGAL \& APEX/SABOCA observations}

The ALMA-IMF regions were mapped by the APEX telescope large area survey of the Galaxy (ATLASGAL, \citealt{Schuller2009}). APEX/LABOCA observations provide a 870 $\upmu$m data point at an angular resolution similar to \textit{Herschel}/SPIRE observations at 250 $\upmu$m (see Table~\ref{table:ObservationSummary}). Furthermore, 12 out of 15 ALMA-IMF regions were mapped by APEX/SABOCA at 350~$\upmu$m \citep{Lin2019}.
The absolute flux calibration uncertainties on SABOCA and LABOCA observations are estimated to be 20\% and 15\%, respectively (\citealt{Lin2019}; \citealt{Contreras2013}).

\subsection{Hi-GAL survey}

The ALMA-IMF regions have also been extensively mapped by the \textit{Herschel} infrared Galactic plane survey (Hi-GAL, \citealt{Molinari2010}). Moreover, W43 was also imaged in the high-gain mode of PACS and SPIRE by HOBYS, a key imaging survey with \textit{Herschel} (\citealt{Motte2010}; \citealt{NguyenLuong2013}), to correct the field for saturation. The combination of PACS and SPIRE observations provides data points from 70~$\upmu$m to 500~$\upmu$m. However, it is worth noting that some Hi-GAL maps contain \lq \lq NaN'' (Not-a-Number) values, that correspond to pixels that are saturated. For instance, in the PSW band (250~$\upmu$m) observations of the G012.80 (W33) region, there are 154 saturated pixels according to Table B.1 in \citet{Molinari2016}. To address this issue, we have applied interpolation to replace the values of these saturated pixels. This interpolation process was performed using Gaussian process regression, as implemented in the \texttt{astrofix} \texttt{Python} package (\citealt{Zhang2021}), described in Appendix~\ref{appendix:PPMAP}. The extent of pixel replacement through interpolation is visualized in Fig.~\ref{fig:datacompleteness}.
When observations reach saturation levels that preclude interpolation, an alternative approach is to substitute them with SABOCA and/or SOFIA observations (Sect.~\ref{sect:SOFIA}). This solution is generally applicable, except in the case of G333.60, where the \textit{Herschel}/SPIRE image at 250 $\upmu$m is saturated, but there are not SOFIA observations at 214 $\upmu$m (as shown in Fig.~\ref{fig:datacompleteness}, sixth row). This makes G333 the least constrained region within our study.
For \textit{Herschel}/PACS and \textit{Herschel}/SPIRE observations, the absolute flux calibration uncertainties are estimated to be 10\% and 7\%, respectively, as reported by \citet{Galametz2014}.

\subsection{SOFIA/HAWC+ observations}\label{sect:SOFIA}
G012.80, G351.77, W51-E, and W51-IRS underwent observations at 53, 89, and 214 $\upmu$m, conducted by \citet{Vaillancourt2016} and \citet{Pillai2023}, employing the 2.7~m stratospheric observatory for infrared astronomy (SOFIA) telescope \citep{Temi2018}. These observations used the High-resolution Airborne Wideband Camera-plus (HAWC+; \citealt{Harper2018}). In our analysis of the G012.80, G351.77, W51-E, and W51-IRS dust emission, the SOFIA/HAWC+ data were employed to better sample the mid-infrared portion of the SED and to replace saturated \textit{Herschel}/SPIRE 250~$\upmu$m maps. The absolute flux calibration uncertainties for SOFIA observations are estimated at 15\% for 53~$\upmu$m and 89~$\upmu$m, and 20\% for 214~$\upmu$m \citep{Chuss2019}.

\subsection{\textit{Spitzer} surveys}
The ALMA-IMF protoclusters were imaged with \textit{Spitzer} instruments. The MIPSGAL survey \citep{Benjamin2003} covered the Galactic plane with the \textit{Spitzer}/MIPS camera, while the GLIMPSE survey \citep{Carey2009} provides \textit{Spitzer}/IRAC observations.
The absolute flux calibration uncertainties for \textit{Spitzer} observations are estimated at 4\% for MIPS at 24~$\upmu$m \citep{Engelbracht2007}, and 2\% for IRAC at 3.6, 4.5, 5.8 and 8.0~$\upmu$m \citep{Reach2005}.

\section{PPMAP description and analysis}\label{sect:PPMAP}

\subsection{Point process mapping}\label{sect:PPMAPdescription}

The point process mapping procedure, denoted as PPMAP (\citealt{Marsh2015}; \citealt{Marsh2017}), is an iterative Bayesian SED fitting algorithm that allows to account for the mixing of physical conditions along the line of sight (dust temperature gradients and variations of the opacity index $\beta$). 
PPMAP is grounded in the point process formalism (\citealt{Richardson&Marsh1987}; \citealt{Richardson&Marsh1992}, \citealt{Marsh2006}). 
The point process formalism represents complex astrophysical systems as an arrangement of individual components, referred to as \lq \lq points'' (e.g., \citealt{Marsh2015}). Points do not correspond to individual astrophysical objects in the image, nor to pixels; rather, they serve as elements that facilitate image representation. These elements are defined by a set of physical parameters, corresponding to a specific position within a state space. For our present purposes, the state space includes the position on the celestial sphere ($x$, $y$) and the dust temperature $T_\mathrm{dust}$. Hence, the state space has a dimensionality $N_\mathrm{states} = N_\mathrm{temp} \times N_x \times N_y$, where $N_\mathrm{temp}$, $N_x$ and $N_y$ are the number of temperature and positional cells that the points may occupy.
The system can then be characterized by a vector containing the occupation number of each cell within the state space, denoted $\mathbf{\Gamma}$. 
In essence, the local density of points corresponds to the density of dust at a specific sky position and temperature. Therefore, the dust column density within the $n^\mathrm{th}$ cell is determined by the corresponding occupation number $\Gamma_n$, and the set of occupation numbers $\mathbf{\Gamma}$ can also be viewed as a probability density function. The underlying PPMAP measurement model is expressed by the equation:
\begin{equation}
\mathbf{d} = \mathbf{A \Gamma} + \mu .
\end{equation}
Here, $\mathbf{d}$ is the vector of observational measurement. The $m^\mathrm{th}$ component of $\mathbf{d}$ pertains to the pixel value at the coordinates ($x_m$, $y_m$) within the observed image at the wavelength $\lambda_m$. The term $\mu$ represents the measurement noise, assumed to be a Gaussian random process. Lastly, $\mathbf{A}$ denotes the system response matrix. Within this matrix, the $mn^\mathrm{th}$ element corresponds to the response of the $m^\mathrm{th}$ measurement to a point situated in the $n^\mathrm{th}$ cell of the state space (characterized by spatial position $x_n$, $y_n$, and temperature $T_n$).
The PPMAP algorithm aims to solve this equation for $\mathbf{\Gamma}$, given a set of observations $\mathbf{d}$ and an a priori distribution of points. This is achieved by minimizing the mean square error, ensuring that the best estimate is the a posteriori expectation value of $\mathbf{\Gamma}$.
The initial distribution (or \lq\lq prior'') across all positions and temperatures is a Gaussian random process, expressed as:
\begin{equation}\label{eq:prior}
P(\Gamma_n) = \frac{1}{\sigma \sqrt{2 \pi}} \mathrm{exp} \left( \frac{-(\Gamma_n -\eta)^2}{2 \sigma^2} \right),
\end{equation}
where $\Gamma_n$ is the occupation number in the $n^\mathrm{th}$ element of the state space, and $\sigma = \sqrt{\eta (1 - 1 / N_\mathrm{states})}$. The quantity $\eta$ is referred to as the \lq \lq dilution'', since it controls the number of points relatively to the number of cells in state space. The initial distribution expressed by Eq.~(\ref{eq:prior}) means that points are equally likely to occupy any $x$, $y$ position, and any given value in the user-defined log$(T)$ temperature distribution.

\begin{figure}[htb]
   \begin{centering}
      \includegraphics[width=\hsize, trim={0cm 0cm -0.5cm 0cm},clip]{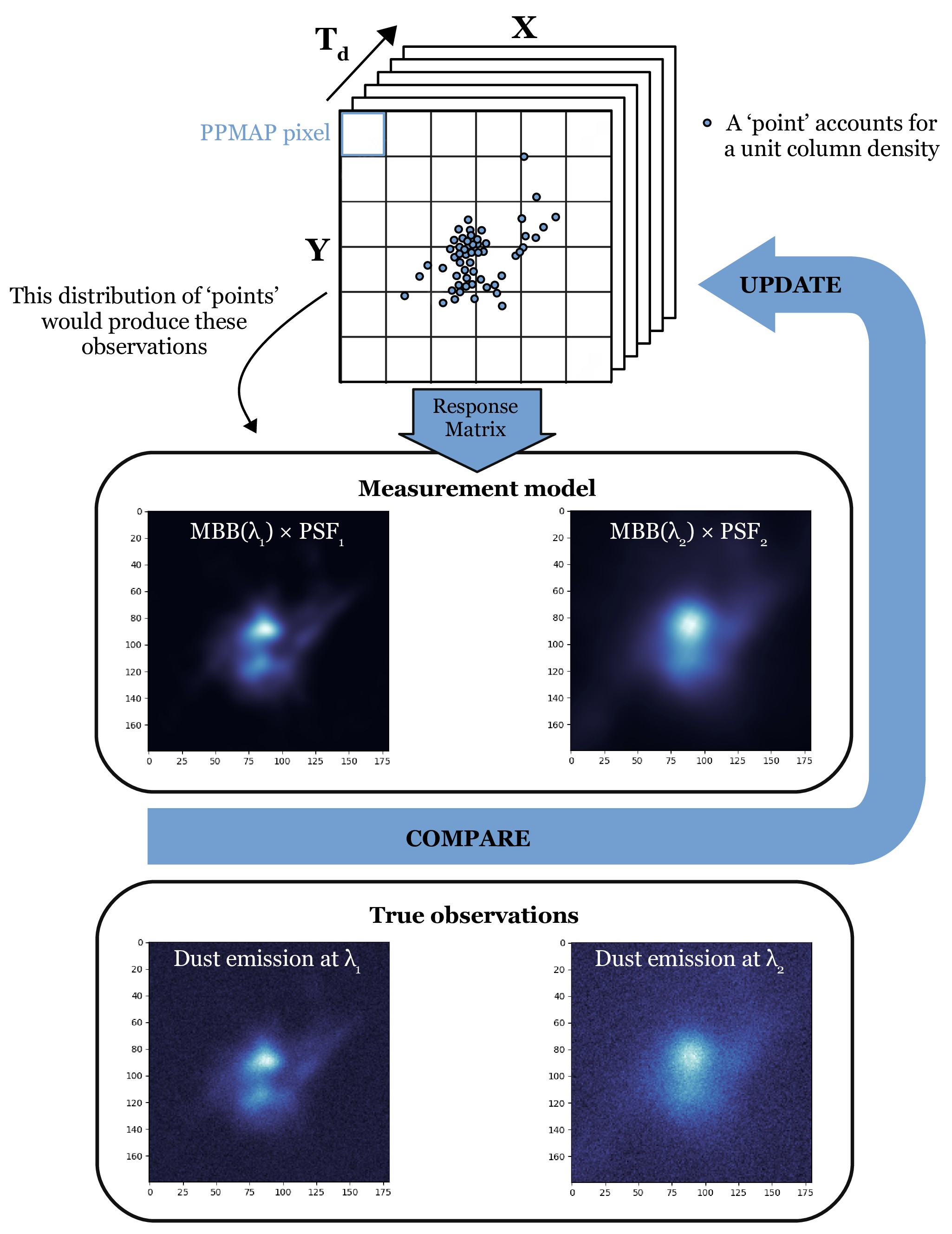}
      \caption{Schematic representation of the PPMAP iterative process. G012.80 images (at $\lambda_1 = 350~\upmu$m and $\lambda_2 = 870~\upmu$m) are used for illustrative purpose. In the upper part, we represent how PPMAP distributes \lq \lq points'' in a continuous state space ($X$, $Y$, $T_\mathrm{dust}$) that can be divided into finite cells (corresponding to PPMAP pixels, that is, with a size fixed by the user, independent of the pixel size of the observed images). This distribution is then translated into a synthetic continuum emission map through the MBB description, taking into account the PSF of the instruments. Synthetic observations are finally compared to true observations, allowing to update the distribution of points. These iterative steps are repeated until the model converges.}
         \label{fig:PPMAP-process}
   \end{centering}
\end{figure}  

PPMAP operates under the assumption that the radiation emitted by dust across observed wavelengths is optically thin. Consequently, the system response matrix $\textbf{A}$ takes the form of the MBB approximation, expressed as:
\begin{equation}\label{eq:PPMAPresponsematrix}
A_{mn} = H_{\lambda_m} K_{\lambda_m}(T_n) B_{\lambda_m}(T_n) \kappa_{\lambda_m}\Delta \Omega_m .
\end{equation}
Here, $H_\lambda$ represents the convolution operator associated with the point spread function (PSF) at wavelength $\lambda$, $K_\lambda (T)$ accounts for a possible color correction, pertaining to the finite bandwidth of observations (cf. \ref{appendix:colorcorrection}), $B_\lambda (T)$ denotes the Planck function, $\kappa _{\lambda}$ corresponds to the dust opacity law, and $\Delta \Omega$ denotes the solid angle corresponding to a specific pixel in the output map. The $H_\lambda$ operator enables PPMAP to function without downgrading the spatial resolution of input maps, provided that the model benefits from accurate beam profiles.
The PPMAP dust opacity law, $\kappa_\lambda$, exhibits a wavelength dependence parametrized by the opacity index $\beta$:
\begin{equation}
\beta = - \frac{d \mathrm{ln} (\kappa_\lambda)}{d \mathrm{ln} (\lambda)} .
\end{equation}
Thus, for any given $\lambda$, the dust opacity law $\kappa_\lambda$ can be represented as:
\begin{equation}\label{eq:Kappa300}
\kappa_\lambda = \kappa_{300} \left( \frac{\lambda}{300~\upmu \mathrm{m}} \right) ^{-\beta} .
\end{equation}
Here, $\beta$ denotes the opacity power-law index, and $\kappa_{300} = 0.1~\mathrm{cm^2 g^{-1}}$ represents the reference opacity, measured at $\lambda_0 = 300$ $\upmu$m, encompassing both dust and gas mass contributions. The selection of the dust absorption coefficient employed by PPMAP ($\kappa_{300} = 0.1~\mathrm{cm^2 g^{-1}}$) is in line with a gas-to-dust mass ratio of 100 \citep{Hildebrand1983}.
The value of $\kappa_{300}$ is identical for all points, and remains unchanged during the iterative process.
Employing Eq.~(\ref{eq:Kappa300}) with $\beta=1.8$ yields $\kappa_{1.3\mathrm{mm}} = 0.007~\mathrm{cm^2 g^{-1}}$, which is consistent with the value adopted by \citet{Armante2024} following \citet{Ossenkopf&Henning1994} ($\kappa_{1.3\mathrm{mm}} = 0.01 ^{+0.005}_{-0.0033}~\mathrm{cm^2 g^{-1}}$).
The reference opacity is in fact not well known, and may vary across the protoclusters.
Depending on the size distribution and the composition of the dust, a range $\kappa_{1.3\mathrm{mm}} = 0.002 - 0.03~\mathrm{cm^2 g^{-1}}$ is predicted by \citet{Ysard2019} for the diffuse interstellar medium (cf. \lq \lq Mix~1'' and \lq \lq Mix~2'', power-law size distributions). Aggregated grain models better represent denser media, and in that case the reference opacity is predicted to increase by a factor 3 to 7, depending on the addition of ice mantles into the models \citep{Koehler2015}.

\begin{figure*}[htb]
   \begin{centering}
      \includegraphics[width=\hsize, trim={4.5cm 1.75cm 4.85cm 1.75cm},clip]{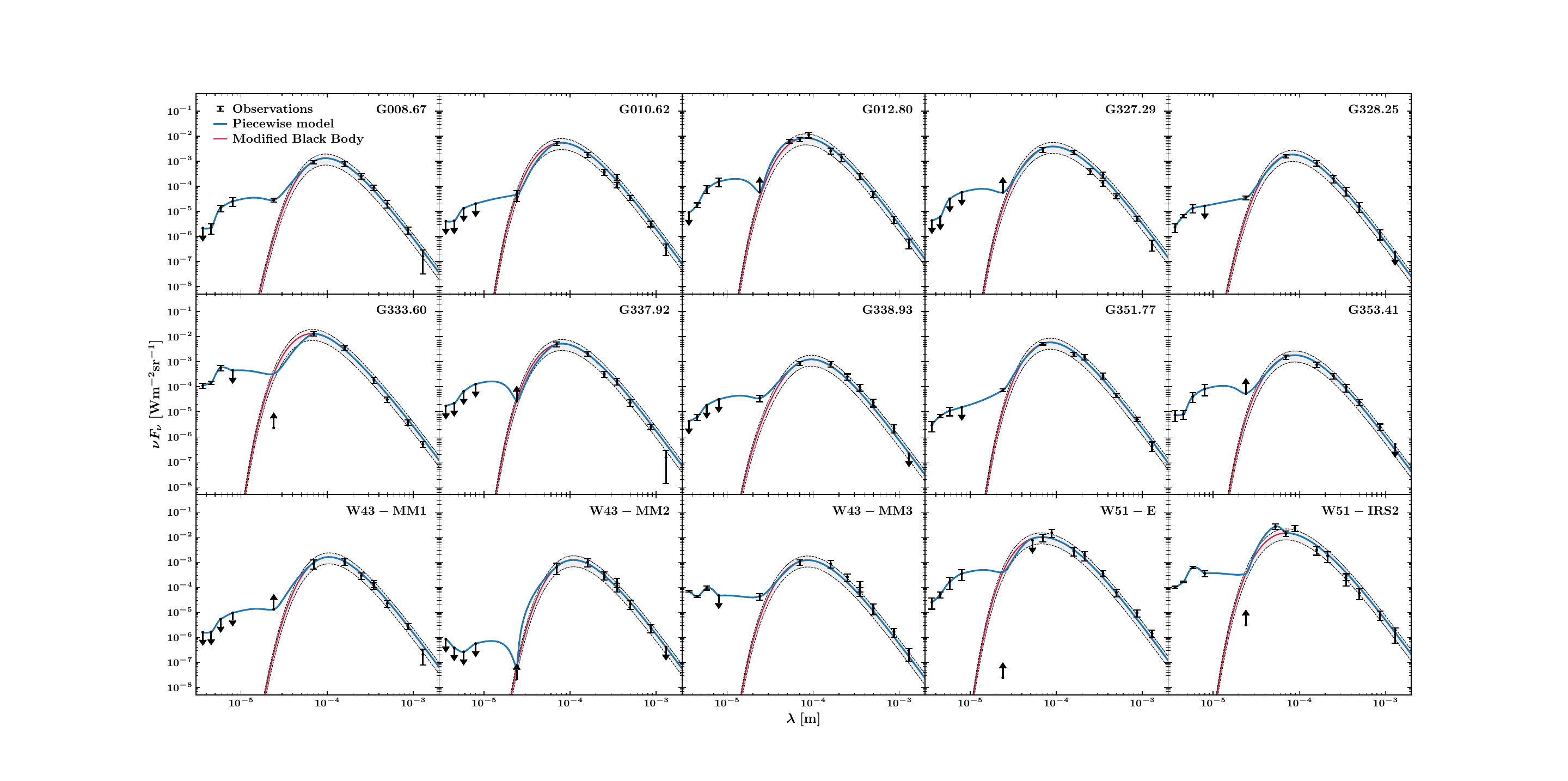}
      \caption{SEDs extracted from the ATLASGAL sources' footprints (\citealt{Contreras2013}, \citealt{Urquhart2014}, see Sect.~\ref{subsect:luminosity}) corresponding to the protoclusters mapped by ALMA-IMF. The actual observations are represented by black points, while the red curve depicts the PPMAP MBB that provides the best fit to the data, with the gray shaded area representing the $\pm2 \sigma$ standard deviation of the best fit. The blue curve represents the total piecewise model described in Sect.~\ref{subsect:luminosity}. Downward and upward arrows respectively correspond to lower limits and saturated observations.}
         \label{fig:SEDs}
   \end{centering}
\end{figure*}

Unlike conventional modified blackbody fitting approaches, PPMAP circumvents the need to homogenize the input observational data to a common resolution. Instead, when provided with a collection of observational data pertaining to dust continuum emission at varying instrumental resolutions, PPMAP produces maps of column density and temperature that are simultaneously optimized with respect to each specific PSFs associated with each dataset.
Figure~\ref{fig:PPMAP-process} provides a schematic illustration of the stepwise approach employed by PPMAP. In practical terms, PPMAP initializes an array given an a priori distribution of points. Subsequently, it generates a corresponding synthetic map for comparison with actual maps, accounting for synthetic noise.
In each pixel of the maps, PPMAP minimizes the reduced-$\chi^2$ metric, calculated from the deviations between components of the measurement model $\textbf{d}$ and the corresponding observations, that are initially resampled to a common pixel size. Employing a truncated hierarchy of integro-differential equations (described in details by \citealt{Marsh2015}, Sect.~2.3), the distribution of points in the state space is iteratively updated until the model converges to match the observations.
Upon completion of the process, the dust column density and temperature are given by the expectation value $E(\Gamma_n \lvert \mathbf{d})$. The H$_2$ column density is derived assuming a reference opacity of $\kappa_{300}=0.1$~cm$^2$~g$^{-1}$, fractional abundance by mass of hydrogen and molecular hydrogen $X_\mathrm{H}=0.7$, $X_\mathrm{H_2}=1$, and fractional abundance by mass of dust $Z_\mathrm{D}=0.01$ \citep{Howard2019, Howard2021}. The differential column density cube $N_\mathrm{H_2}(T_\mathrm{dust})$ represents the H$_2$ column density at different dust temperatures, such that:
\begin{equation}\label{eq:PPMAP-N}
N_\mathrm{H_2}(i, j) = \sum_{k=1}^{N_\mathrm{temp}} N_\mathrm{H_2}(i, j, T_{\mathrm{dust}, k}) .
\end{equation}
Similarly, the density-weighted dust temperature map is defined by the following average quantity, based on the differential column density cube:
\begin{equation}\label{eq:PPMAP-T}
T_\mathrm{dust}(i, j) = \frac{1}{N_\mathrm{H_2}(i, j)} \sum_{k=1}^{N_\mathrm{temp}} \left\lbrace T_{\mathrm{dust}, k} N_\mathrm{H_2}(i, j, T_{\mathrm{dust}, k}) \right\rbrace .
\end{equation}
Lastly, in accordance with Eq.~(\ref{eq:PPMAPresponsematrix}), the synthetic intensity produced by PPMAP at wavelength $\lambda$ in any pixel is determined by the following sum:
\begin{equation}\label{eq:PPMAPmod}
I_\lambda(i, j) = \sum_{k=1}^{N_\mathrm{temp}} \left\lbrace \left( \frac{N_\mathrm{H_2}(i, j, T_{\mathrm{dust}, k})}{2.1 \times 10^{24}\mathrm{cm^{-2}}} \right) \left( \frac{\lambda}{300\mathrm{\upmu m}} \right)^{-\beta} B_\lambda(T_{\mathrm{dust}, k}) \right\rbrace .
\end{equation}
Here, $N_\mathrm{H_2}(T_{\mathrm{d}, k}) / 2.1 \times 10^{24}~\mathrm{cm^{-2}} = \tau_{300}$ represents the optical depth at 300 $\upmu$m. The numerical constant depends on the adopted values of $\kappa_{300}$, $X_\mathrm{H}$, $X_\mathrm{H_2}$ and $Z_\mathrm{D}$.

\subsection{PPMAP analysis}\label{sect:PPMAPanalysis}

We apply the PPMAP procedure to the continuum data set detailed in Sect.~\ref{sect:observations}. A comprehensive account of the methodology employed for implementing PPMAP to the data is provided in Appendix~\ref{appendix:PPMAP}. Here, we present an overview of the principles adopted throughout our analysis.
To prevent contamination by free-free emission, we exclude the ALMA Band 3 data from our analysis (that is, the ALMA 3~mm continuum map, see also Sect.~\ref{sect:free-free} on the impact of free-free emission on PPMAP products). Consequently, the input images for PPMAP encompass the wavelength range from 70~$\upmu$m (\textit{Herschel}/PACS) to 1.3 mm (ALMA Band 6). 

The opacity indices predicted by dust models describing diffuse and dense interstellar media are $\beta = 1.5$ and $\beta = 1.8$ \citep{Koehler2015}, respectively. There is in fact a range of plausible values, and $\beta$ may vary across the field of observations, but we adopt a fixed value to minimize effects from the degeneracy between the temperature and opacity index and thus better constrain the temperature (\citealt{Shetty2009}; \citealt{Kelly2012}; \citealt{Galliano2018}).
We therefore fix the opacity index to $\beta = 1.8$, and we employ 8 MBB components with temperature values ranging from 10~K to 50~K, consistent with prior PPMAP applications (e.g., \citealt{Marsh2017}; \citealt{Howard2019}; \citealt{Whitworth2019}; \citealt{Chawner2020}; \citealt{Howard2021}). Adopting a number of temperature components higher than 8 would result in larger uncertainties without improving the temperature sampling significantly. We experimented with larger temperature ranges and found that $10 - 50$~K is sufficient to reproduce the observations. The final temperature estimate can exceed this temperature range following on the a posteriori temperature correction (see Sect.~\ref{sect:opacity-correction}). The pixel sizes of the Nyquist-sampled PPMAP arrays are 1.25$^{\prime \prime}$, corresponding to an angular resolution of 2.5$^{\prime \prime}$.
We execute the PPMAP analysis twice. The initial run (\lq \lq Run1'') uses the ALMA B6 data decontaminated from free-free emission (as released by Galván-Madrid et al., in prep.), primarily created for deriving column density and dust temperature. The subsequent run (\lq \lq Run2'') employs the standard ALMA B6 data (as released by \citealt{Diaz2023}), that is more optimal for accurately determining the luminosity by taking into account the millimeter excess tied to free-free emission.
Column density and temperature estimates are directly obtained through the application of PPMAP, but deriving the bolometric luminosity requires additional steps.

\subsection{Bolometric luminosity measurements}\label{subsect:luminosity}

\begin{table*}[htb]
\centering
{\caption{Results of the PPMAP analysis of ALMA-IMF protoclusters.}
\label{table:Results}}      
{\centering                          
\begin{tabular}{l c c c c c c c c c}        
\hline  \hline       \\[-1.0em]
\footnotesize{Region} & \footnotesize{Distance$^a$} & \footnotesize{$\overline{T_\mathrm{dust}}$$^b$} & \footnotesize{$N_\mathrm{peak}$$^c$} & \multicolumn{3}{c}{\footnotesize{ALMA footprint$^d$}} & \multicolumn{3}{c}{\footnotesize{ATLASGAL footprint$^e$}} \\
\cmidrule(lr{.75em}){5-7}
\cmidrule(lr{.75em}){8-10}
\footnotesize{} & \footnotesize{(kpc)} & \footnotesize{(K)} & \footnotesize{($10^{24}$ cm$^{-2}$)} & \footnotesize{$M$ ($10^2 M_{\odot}$)} & \footnotesize{$L_\mathrm{bol}$ ($10^3 L_\odot$)} & \footnotesize{$L / M$ ($ L_\odot / M_{\odot}$)} & \footnotesize{$M^\prime$ ($10^2 M_{\odot}$)} & \footnotesize{$L_\mathrm{bol}^\prime$ ($10^3 L_\odot$)} & \footnotesize{$L^\prime/M^\prime$ ($ L_\odot / M_{\odot}$)} \\ \\[-1.0em]
\hline \\[-1.0em]
\footnotesize{W43-MM1} & \footnotesize{$5.5\pm 0.4$} & \footnotesize{27 $\pm$ 5} & \footnotesize{12 $\pm$ 3} & \footnotesize{170 $\pm$ 60} & \footnotesize{250 $\pm$ 90} & \footnotesize{15} & \footnotesize{110 $\pm$ 30} & \footnotesize{80 $\pm$ 30} & \footnotesize{7}\\
\footnotesize{W43-MM2} & \footnotesize{$5.5\pm 0.4$} & \footnotesize{25 $\pm$ 5} & \footnotesize{9 $\pm$ 3} & \footnotesize{150 $\pm$ 60} & \footnotesize{180 $\pm$ 60} & \footnotesize{12} & \footnotesize{50 $\pm$ 20} & \footnotesize{30 $\pm$ 10} & \footnotesize{6}\\
\footnotesize{G338.93} & \footnotesize{$3.9 \pm 1.0$} & \footnotesize{25 $\pm$ 5} & \footnotesize{1.7 $\pm$ 0.5} & \footnotesize{40 $\pm$ 30} & \footnotesize{100 $\pm$ 60} & \footnotesize{25} & \footnotesize{40 $\pm$ 20} & \footnotesize{90 $\pm$ 30} & \footnotesize{23}\\
\footnotesize{G328.25} & \footnotesize{$2.5 \pm 0.5$} & \footnotesize{25 $\pm$ 5} & \footnotesize{1.9 $\pm$ 0.5} & \footnotesize{10 $\pm$ 5} & \footnotesize{50 $\pm$ 30} & \footnotesize{50} & \footnotesize{8 $\pm$ 3} & \footnotesize{30 $\pm$ 10} & \footnotesize{38}\\
\footnotesize{G337.92} & \footnotesize{$2.7 \pm 0.7$} & \footnotesize{22 $\pm$ 4} & \footnotesize{8 $\pm$ 2} & \footnotesize{30 $\pm$ 20} & \footnotesize{130 $\pm$ 80} & \footnotesize{43} & \footnotesize{20 $\pm$ 10} & \footnotesize{70 $\pm$ 30} & \footnotesize{35}\\
\footnotesize{G327.29} & \footnotesize{$2.5 \pm 0.5$} & \footnotesize{25 $\pm$ 5} & \footnotesize{60 $\pm$ 20} & \footnotesize{80 $\pm$ 40} & \footnotesize{110 $\pm$ 60} & \footnotesize{14} & \footnotesize{50 $\pm$ 20} & \footnotesize{50 $\pm$ 20} & \footnotesize{10}\\
\hline \\[-1.0em]
\footnotesize{G351.77} & \footnotesize{$2.0\pm 0.7$} & \footnotesize{26 $\pm$ 5} & \footnotesize{19 $\pm$ 5} & \footnotesize{20 $\pm$ 10} & \footnotesize{80 $\pm$ 60} & \footnotesize{40} & \footnotesize{13 $\pm$ 5} & \footnotesize{60 $\pm$ 20} & \footnotesize{46}\\
\footnotesize{G008.67} & \footnotesize{$3.4 \pm 0.3$} & \footnotesize{21 $\pm$ 4} & \footnotesize{4 $\pm$ 1} & \footnotesize{30 $\pm$ 10} & \footnotesize{70 $\pm$ 30} & \footnotesize{23} & \footnotesize{20 $\pm$ 10} & \footnotesize{40 $\pm$ 10} & \footnotesize{20}\\
\footnotesize{W43-MM3} & \footnotesize{$5.5\pm 0.4$} & \footnotesize{25 $\pm$ 5} & \footnotesize{3 $\pm$ 1} & \footnotesize{80 $\pm$ 30} & \footnotesize{180 $\pm$ 70} & \footnotesize{23} & \footnotesize{20 $\pm$ 10} & \footnotesize{50 $\pm$ 20} & \footnotesize{25}\\
\footnotesize{W51-E} & \footnotesize{$5.4\pm 0.3$} & \footnotesize{23 $\pm$ 4} & \footnotesize{30 $\pm$ 10} & \footnotesize{240 $\pm$ 70} & \footnotesize{1200 $\pm$ 400} & \footnotesize{50} & \footnotesize{170 $\pm$ 50} & \footnotesize{500 $\pm$ 200} & \footnotesize{29}\\
\footnotesize{G353.41} & \footnotesize{$2.0\pm 0.7$} & \footnotesize{21 $\pm$ 4} & \footnotesize{7 $\pm$ 2} & \footnotesize{20 $\pm$ 10} & \footnotesize{60 $\pm$ 50} & \footnotesize{30} & \footnotesize{12 $\pm$ 5} & \footnotesize{30 $\pm$ 10} & \footnotesize{25}\\
\hline \\[-1.0em]
\footnotesize{G010.62} & \footnotesize{$5.0 \pm 0.5$} & \footnotesize{24 $\pm$ 5} & \footnotesize{5 $\pm$ 1} & \footnotesize{110 $\pm$ 40} & \footnotesize{500 $\pm$ 200} & \footnotesize{45} & \footnotesize{80 $\pm$ 20} & \footnotesize{300 $\pm$ 100} & \footnotesize{38}\\
\footnotesize{W51-IRS2} & \footnotesize{$5.4\pm 0.3$} & \footnotesize{29 $\pm$ 5} & \footnotesize{50 $\pm$ 20} & \footnotesize{250 $\pm$ 80} & \footnotesize{1300 $\pm$ 500} & \footnotesize{52} & \footnotesize{190 $\pm$ 60} & \footnotesize{300 $\pm$ 100} & \footnotesize{16}\\
\footnotesize{G012.80} & \footnotesize{$2.4 \pm 0.2$} & \footnotesize{28 $\pm$ 5} & \footnotesize{3 $\pm$ 1} & \footnotesize{120 $\pm$ 40} & \footnotesize{300 $\pm$ 100} & \footnotesize{25} & \footnotesize{20 $\pm$ 10} & \footnotesize{180 $\pm$ 60} & \footnotesize{90}\\
\footnotesize{G333.60} & \footnotesize{$4.2 \pm 0.7$} & \footnotesize{27 $\pm$ 5} & \footnotesize{3 $\pm$ 1} & \footnotesize{130 $\pm$ 60} & \footnotesize{1400 $\pm$ 700} & \footnotesize{108} & \footnotesize{50 $\pm$ 20} & \footnotesize{800 $\pm$ 300} & \footnotesize{160}\\
\hline
\end{tabular}}
\\~\\
\raggedright
\footnotesize{$^a$From Table 1 in \citet{Motte2022}.\\
$^b$Spatially-averaged density-weighted dust temperature, measured in the corrected PPMAP temperature map (cf Sect.~\ref{sect:opacity-correction}). \\ 
$^c$ Peak H$_2$ column density, measured in the PPMAP column density map.\\
$^d$ Mass of gas, derived from the PPMAP column density map; bolometric luminosity, derived from the PPMAP MBB in addition to observations by \textit{Spitzer} and SOFIA (that is, between 3.6~$\upmu$m and 1.3~mm); and luminosity-to-mass ratio, measured in the ALMA footprint (cf. Sect.~\ref{subsect:luminosity}).\\
$^e$ Mass, bolometric luminosity and luminosity-to-mass ratio measured in the ATLASGAL source footprint (\citealt{Contreras2013}, \citealt{Urquhart2014}, cf Sect.~\ref{subsect:luminosity}).\\}
\end{table*}

Using the outcomes of PPMAP Run2 (in order to take into account the free-free contribution to the luminosity), we computed the luminosity for all observed ALMA-IMF fields. The \lq \lq PPMAP Luminosity'' ($L_\mathrm{MBB}$) is defined as the integral $4\pi d^2 \int I_\nu \mathrm{d}\nu$, where $I_\nu$ is defined by Eq.~(\ref{eq:PPMAPmod}), and $d$ denotes the distance to the observed star-forming region. On the other hand, the bolometric luminosity ($L_\mathrm{bol}$) accounts for the additional near-infrared flux estimated from background-subtracted \textit{Spitzer}/IRAC, \textit{Spitzer}/MIPS, and SOFIA/HAWC+ observations, extracted from the MIPSGAL, GLIMPSE, and SOFIA archives (\citealt{Carey2009}; \citealt{Benjamin2003}; \citealt{Vaillancourt2016}; \citealt{Pillai2023}; also see Table ~\ref{table:ObservationSummary}).
We performed a pixel-per-pixel merge between MIPSGAL, GLIMPSE, and HAWC+ observations and the PPMAP output SED using a piecewise cubic Hermite interpolating polynomial from the \texttt{scipy} package in \texttt{Python}. This results in a composite SED model as follows:
\begin{equation}
\begin{split}
I_\lambda^\prime(\lambda < 70~\mathrm{\upmu m}) = \mathrm{PCHI}(\lambda) , \\
I_\lambda^\prime(\lambda \geq 70~\mathrm{\upmu m}) = \mathrm{MBB}(\lambda) .
\end{split}
\end{equation}
Here,\lq \lq PCHI'' denotes the cubic spline function employed for interpolating near-infrared data points, and \lq \lq MBB'' represents the best-fit PPMAP model. The bolometric luminosity is then defined as $L_\mathrm{bol} = 4 \pi d^2 \int_{3.6~\upmu \mathrm{m}}^{1.3~\mathrm{mm}} I_\lambda^\prime d\lambda$.
The presence of saturation in the \textit{Spitzer}/MIPS band occasionally resulted in lower limits on the infrared flux at 24~$\upmu$m (refer to the first column in Fig.~\ref{fig:datacompleteness}). Additionally, our integration approach effectively merges 2.5$^{\prime \prime}$ products with observations acquired at coarser resolutions (\textit{Spitzer}/MIPS at 5.6$^{\prime \prime}$ and SOFIA/HAWC+ at 4.85$^{\prime \prime}$), resulting in a composite angular resolution. These two limitations have a relatively low impact on the outcome, given that the dominant contribution to the luminosity arises from the MBB emission ($L_\mathrm{MBB} / L_\mathrm{bol} \geq 0.77$ across the 15 star-forming regions studied, where $L_\mathrm{MBB}$ and $L_\mathrm{bol}$ are respectively the modified blackbody and bolometric luminosities integrated over the field of observations).

Figure~\ref{fig:SEDs} displays a representative sample of illustrative spectral energy distributions (SEDs) extracted from the ATLASGAL sources' footprints encompassing the protoclusters (\citealt{Contreras2013}, \citealt{Urquhart2014}). All of the observations align within $\pm 2$ standard deviation ($\sigma$) of the PPMAP MBB model, and 86\% of the observations maintain agreement within $\pm 1 \sigma$. We note that the ALMA error bars appear larger as a result of the integration area being large with respect to the 1.3~mm sources sizes.

\subsection{Temperature correction and final products}\label{sect:opacity-correction}

\begin{figure}[htb]
   \begin{centering}
      \includegraphics[width=\hsize, trim={1.15cm 4.25cm 9.5cm 4.75cm},clip]{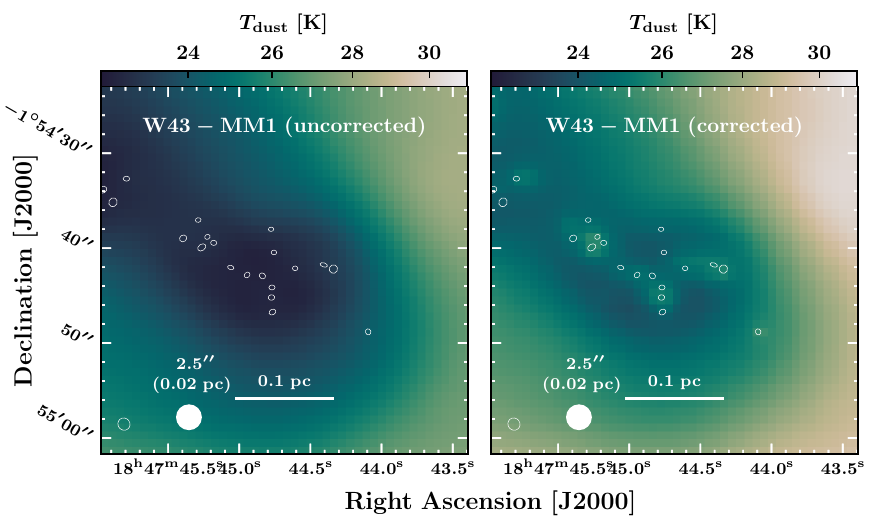}
      \caption{Temperature correction of the PPMAP images illustrated. The left panels displays the W43-MM1 Main-West image before correction, while the right panels presents the post-corrected map. White ellipses outline continuum cores identified by \citet{Louvet2024} in the ALMA 1.3~mm images at 0.4-0.9$^{\prime \prime}$ angular resolution.}
         \label{fig:MM1-tempchange}
   \end{centering}
\end{figure}

As reported in Sect.~\ref{sect:PPMAPdescription}, PPMAP assumes that the observed astrophysical object is optically thin to the thermal radiation emitted by dust across all input wavelengths.
This primarily leads to a bias in the estimate of the dust temperature of deeply embedded sources, since the infrared fluxes at $\lambda \leq 250$ $\upmu$m may significantly deviate from the MBB shape \citep{Menshchikov2016}. To mitigate this limitation, in our analysis we have systematically applied an a posteriori correction to the PPMAP-derived temperature maps.
After running PPMAP on synthetic observations generated with a dust model that incorporates the effect of the optical depth, the PPMAP outcome is compared with the input model dust parameters. We derive a correction table from this comparison, that can then be applied to PPMAP products (the dust model used to build this correction table is described in Appendix~\ref{appendix:tempcorr}). 
Figure~\ref{fig:MM1-tempchange} illustrates the change in temperature following the correction of the W43-MM1 temperature image.
The outcome of the temperature correction is evident in localized areas, where the temperature is increased. For instance, in G012.80 the 99$^\mathrm{th}$ percentile temperature raises to 38.0~K from 33.2~K after the opacity correction. The magnitude of this correction scales with the column density estimated by PPMAP, as shown in Fig.~\ref{fig:tcorr}, therefore the high-density pixels benefit the most from it. As a result, the correction allows a better representation of embedded protostars and hot cores.

The final products delivered with this study are the maps of the H$_2$ column density ($N_\mathrm{H_2}$, in cm$^{-2}$), bolometric luminosity ($L_\mathrm{bol}$, in $L_\mathrm{sun} / \mathrm{px}$) and dust temperature ($T_\mathrm{dust}$, in Kelvin) (see Fig.~\ref{fig:PPMAPflowchart} for reference). The dust temperature maps are declined in two versions:
\begin{enumerate}
\item The direct output of PPMAP, denoted hereafter as $T_\mathrm{dust}^\prime$, represents the best-fit MBB temperature derived under the assumption of optically thin emission.
\item The a posteriori correction of the temperature yields a new estimate, denoted hereafter as $T_\mathrm{dust}$.
\end{enumerate}
The opacity-corrected temperature $T_\mathrm{dust}$ generally provides a better representation of the dust temperature, except in instances where the foreground is heated. Therefore, the temperature of internally heated, optically thick dust cores are best estimated by the second version of the temperature map, that we hereafter consider the default. Impending studies will attempt to determine the best combination of both maps based on the identification of prestellar cores and candidate protostars in the ALMA-IMF fields (Motte et al., in prep.).

An other caveat tied to PPMAP-derived temperatures is the impact of the $2.5^{\prime \prime}$ angular resolution. Emission of the ALMA-IMF cores with a median size of $\sim 2100$ au \citep{Motte2022} is diluted in a PPMAP beam of physical size 6000 -- 14~000 au, depending on the distance of the region. This dilution of the signal may result in underestimating the temperature of protostellar cores and overestimating that of prestellar cores, since cold and warm dust are mixed in the $2.5^{\prime \prime}$ beam, as well as along the stratified line of sight. 
Therefore, additional processing should be applied to PPMAP temperature maps prior to their use in constraining core temperatures. This endeavor is also being undertaken by Motte et al. (in prep.), to which we refer the reader for a detailed account of the methodology.

\section{Results, validation, and caveats}\label{sect:calibration}

Following the procedures described in Sect.~\ref{sect:PPMAP} and expanded upon in Appendix~\ref{appendix:PPMAP}, we have applied PPMAP to multiwavelength observations and obtained luminosity, column density and temperature maps at a 2.5$^{\prime \prime}$ angular resolution. We here present these outputs, evaluate their reliability against an accepted reference and then discuss their uncertainties.

\subsection{Results}\label{subsect:maps}

\begin{figure}[htb]
   \begin{centering}
      \includegraphics[width=\hsize, trim={1.15cm 4.25cm 9.5cm 4.75cm},clip]{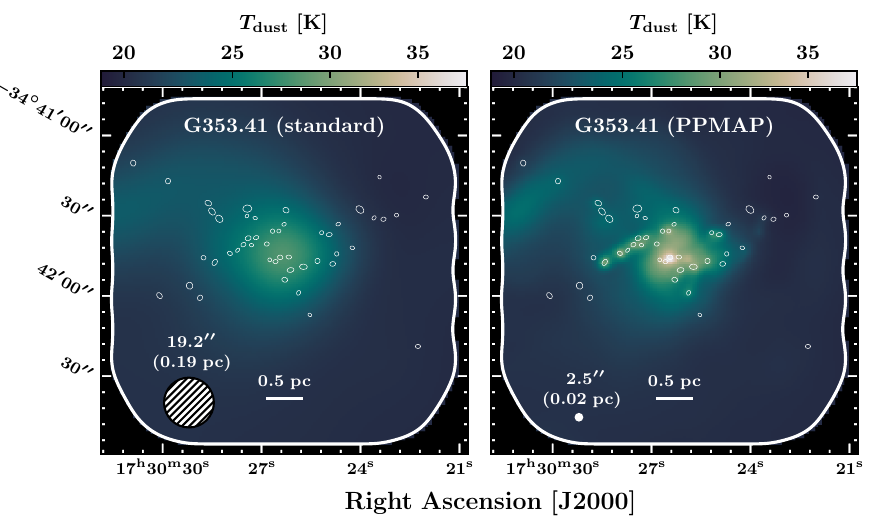}
      \includegraphics[width=\hsize, trim={1.15cm 4.25cm 9.5cm 4.75cm},clip]{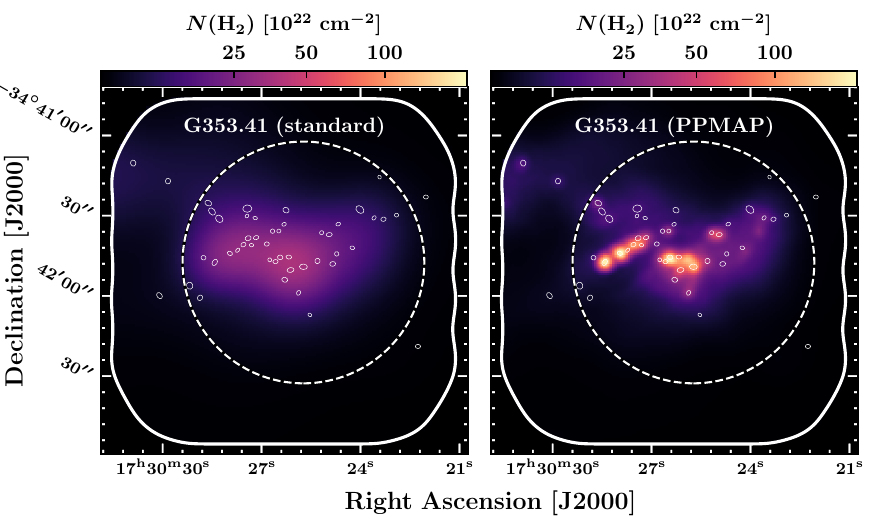}
      \caption{Resolution enhancement of the dust temperature (\textit{top panels}) and column density (\textit{bottom panels}) images, achieved through the application of PPMAP to the G353.41 dataset. The left panels display the maps derived by smoothing all input images to a uniform resolution of 19.2$^{\prime \prime}$, while the right panels represent the PPMAP images at an angular resolution of 2.5$^{\prime \prime}$. White ellipses outline continuum cores identified by \citet{Louvet2024} in the ALMA 1.3~mm images at 0.4-0.9$^{\prime \prime}$ angular resolution. The larger dashed circles represent the footprint of the ATLASGAL source AGAL353.409-00.361 (\citealt{Contreras2013}, \citealt{Urquhart2014}).}
         \label{fig:PPMAP-resolution}
   \end{centering}
\end{figure}

\begin{figure*}[htbp!]
   \begin{centering}
      \includegraphics[width=\hsize, trim={0.25cm 4.25cm 1cm 4.75cm},clip]{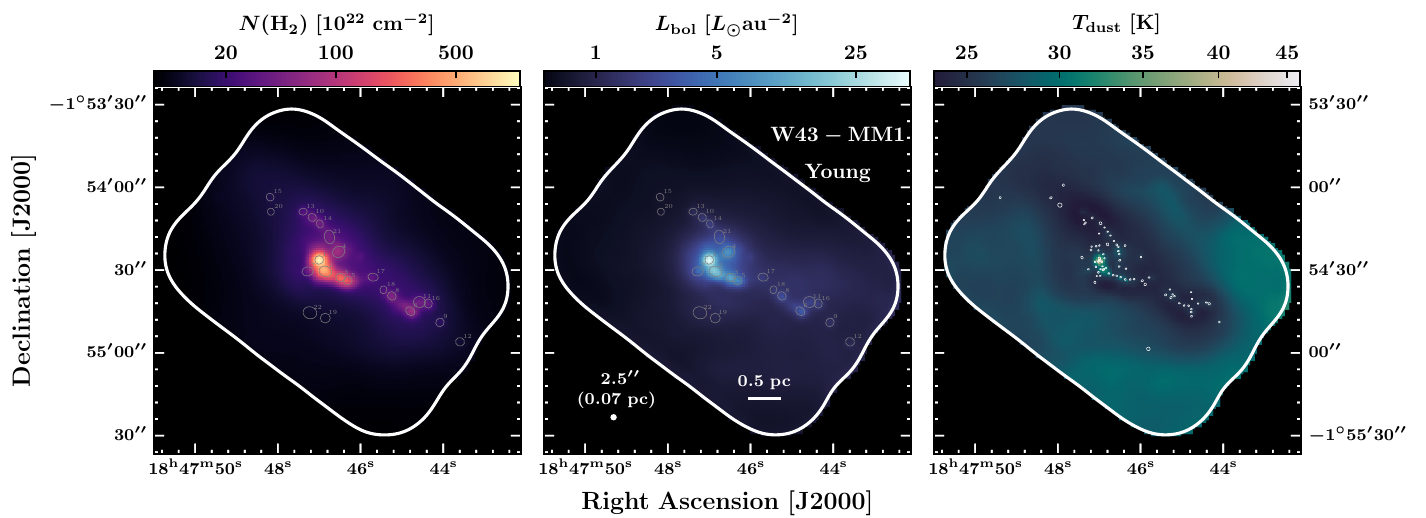}
      \includegraphics[width=\hsize, trim={0.25cm 4.25cm 1cm 4.75cm},clip]{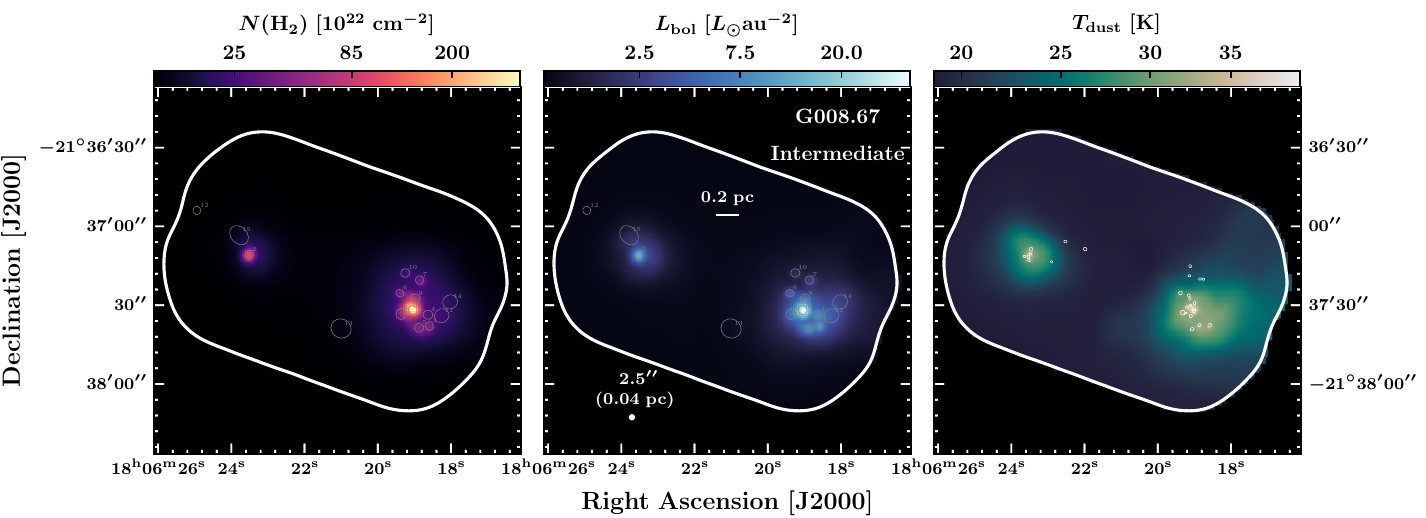}
      \includegraphics[width=\hsize, trim={0.25cm 4.25cm 1cm 4.75cm},clip]{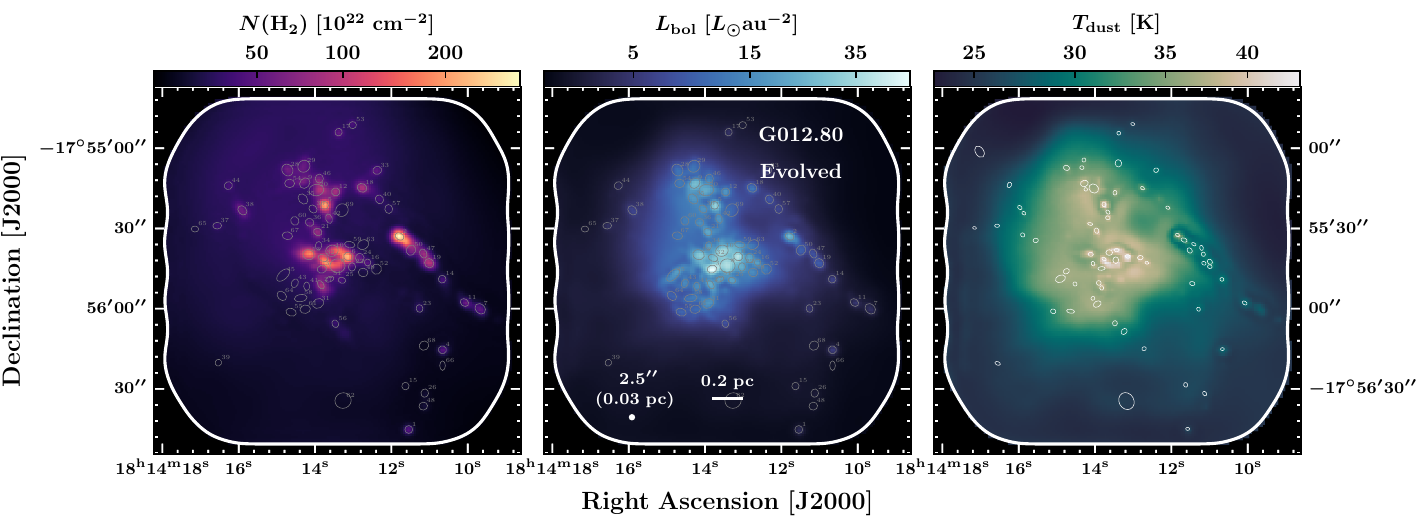}
      \caption{PPMAP products illustrated for three example regions: the young W43-MM1 (top), intermediate G008.67 (center), and evolved G012.80 (bottom) protoclusters. From left to right: column density map ($N(\mathrm{H_2}$)), bolometric luminosity ($L_\mathrm{bol}$), dust temperature ($T_\mathrm{dust}$). White continuous contours outline the ALMA 1.3~mm mosaic areas. The luminosity peaks extracted from the PPMAP luminosity maps (see Sect.~\ref{sect:GETSF}) are overlaid in gray. The continuum cores identified by \citet{Louvet2024} in the ALMA 1.3~mm images are overlaid in white. The size of the ellipses reflects the FWHM of the sources.}
         \label{fig:PPMAPs}
   \end{centering}
\end{figure*}  

Figure~\ref{fig:PPMAP-resolution} illustrates the improvement in angular resolution achieved through PPMAP's application. We used the temperature and column density images of G353.41 as examples to illustrate the importance of gaining angular resolution for the ALMA-IMF studies.
In this figure, we contrast the outcomes of the PPMAP approach at a 2.5$^{\prime \prime}$ resolution with the more typical approach, that requires smoothing all input images to the same angular resolution. To make this illustration, we first substituted the \textit{Herschel}/SPIRE image at 350~$\upmu$m with the SABOCA image and excluded the \textit{Herschel}/SPIRE image at 500~$\upmu$m, thus preventing further smoothing to a 35.2$^{\prime \prime}$ resolution. We then smoothed all continuum images to the coarsest angular resolution, which is that of LABOCA observations, 19.2$^{\prime \prime}$. Finally we performed SED fitting using PPMAP.

Complete representations of PPMAP luminosity, column density and temperature maps obtained for all regions are shown in Fig.~\ref{fig:PPMAPs-supplementary}. The spatial variations of the reduced $\chi^2$ square metric are also shown in Fig.~\ref{fig:PPMAPs-chi2} for all regions.
We here present an overview of these PPMAP data products for a subset of regions, selected to provide one example for each evolutionary stage (young, intermediate, evolved, as outlined by \citealt{Motte2022}).
In Fig.~\ref{fig:PPMAPs}, we present the column density, bolometric luminosity and dust temperature maps obtained for this specific subset. These temperature images correspond to those corrected for the opacity because we consider them to best represent the dust temperature in dense regions. In the following, we simply call them the temperature images (see Sect.~\ref{sect:opacity-correction} and Appendix~\ref{appendix:tempcorr} for a description of the temperature correction procedure).

With a 2.5$^{\prime \prime}$ angular resolution, PPMAP captures the morphology of filamentary structures, pinpoint the location of cores, protostars, \HII regions, and constrain their surrounding physical conditions.
The angular resolution provided by PPMAP offers an insight into the relationships between the continuum cores identified by \citet{Louvet2024} in the ALMA-IMF 1.3~mm continuum images and the column density and dust temperature maps. These cores, with typical sizes of 0.4-0.9$^{\prime \prime}$ (equivalent to $\sim$2000-4000 au), align with filaments  and aggregate within central hubs, a trend depicted in Figs.~\ref{fig:PPMAP-resolution}--\ref{fig:PPMAPs}. Additionally, within the temperature images we observe correlations between massive, hot protostars and warmer spots, while \HII regions appear as extended areas of enhanced temperature (as evidenced in the evolved protocluster G012.80, shown in the bottom panel of Fig.~\ref{fig:PPMAPs}, also see \citealt{Armante2024}). This underscores the pivotal role of PPMAP's resolution-optimization capacity in achieving our goals.

Table~\ref{table:Results} presents the luminosity measurements using two distinct approaches:
\begin{enumerate}
\item The bolometric luminosity ($L_\mathrm{bol}$) measured in the primary beam response of the ALMA 12 m array mosaics (hereafter referred to as the \lq \lq ALMA-IMF mosaic footprint'', as outlined in Paper I, \citealt{Motte2022}, Fig.~1).\\[-1.0em]
\item The bolometric luminosity ($L_\mathrm{bol}^\prime$) measured in the ATLASGAL sources' footprints, as outlined in \citet{Contreras2013} and \citet{Urquhart2014}. These sources were extracted from the ATLASGAL survey \citep{Schuller2009} using the source extraction routine SEXtractor. Protoclusters imaged by ALMA-IMF are associated with either one or two ATLASGAL sources (two sources in the case of G008.67), that are always smaller than the ALMA-IMF mosaic footprints. An example of single ATLASGAL source is shown in the bottom panel of Fig.~\ref{fig:PPMAP-resolution}.
\end{enumerate}
Table~\ref{table:Results} also provides a compilation of average dust temperatures, peak column densities and total masses. The regions are classified according to their evolutionary stages, as defined in \citet{Motte2022}, categorized as \lq \lq young'', \lq \lq intermediate'', and \lq \lq evolved'', in descending order in the table. The luminosity-to-mass ratio, $L / M$, exhibits a noticeable trend aligned with this classification: younger regions generally correspond to lower ratios ($L / M = 20$ on average for the young protoclusters listed in Table~\ref{table:Results}), and vice versa ($L / M = 76$ on average for the evolved protoclusters listed in Table~\ref{table:Results}). There is also a trend of increasingly high temperatures across the evolutionary stages, with a 99$^\mathrm{th}$ percentile temperature of 33.7~K,  35.4~K and 38.7~K respectively for the young, intermediate and evolved regions.
The average temperatures presented in the table, ranging from 21~K to 29~K, are representative of broader regions rather than locally heated zones (see Sect.~\ref{subsect:maps}). The $99^\mathrm{th}$ percentile temperature ranges from 28~K to 42~K, depending on the specific region. As such, the temperatures inferred by PPMAP within the vicinity of massive cores are higher than those measured by \citet{Wienen2012, Wienen2018}, \citet{Koenig2017} and used in \citet{Motte2022} as the means to estimate core masses ($T_\mathrm{dust} = 20 - 30$~K). The average column density measured in the 2.5$^{\prime \prime}$ maps ranges from $2.5 \times 10^{22}$ to $2.5 \times 10^{23}$ cm$^{-2}$, about one to two decades higher than those measured in the wide Herschel images of nearby star-forming regions (\citealt{Arzoumanian2011}, \citealt{Hill2011}).

\subsection{Comparison to results from ATLASGAL}\label{subsect:validation}

\begin{figure}[htb]
   \begin{centering}
      \includegraphics[width=0.95\hsize, trim={0cm 0cm 0cm 2.5cm},clip]{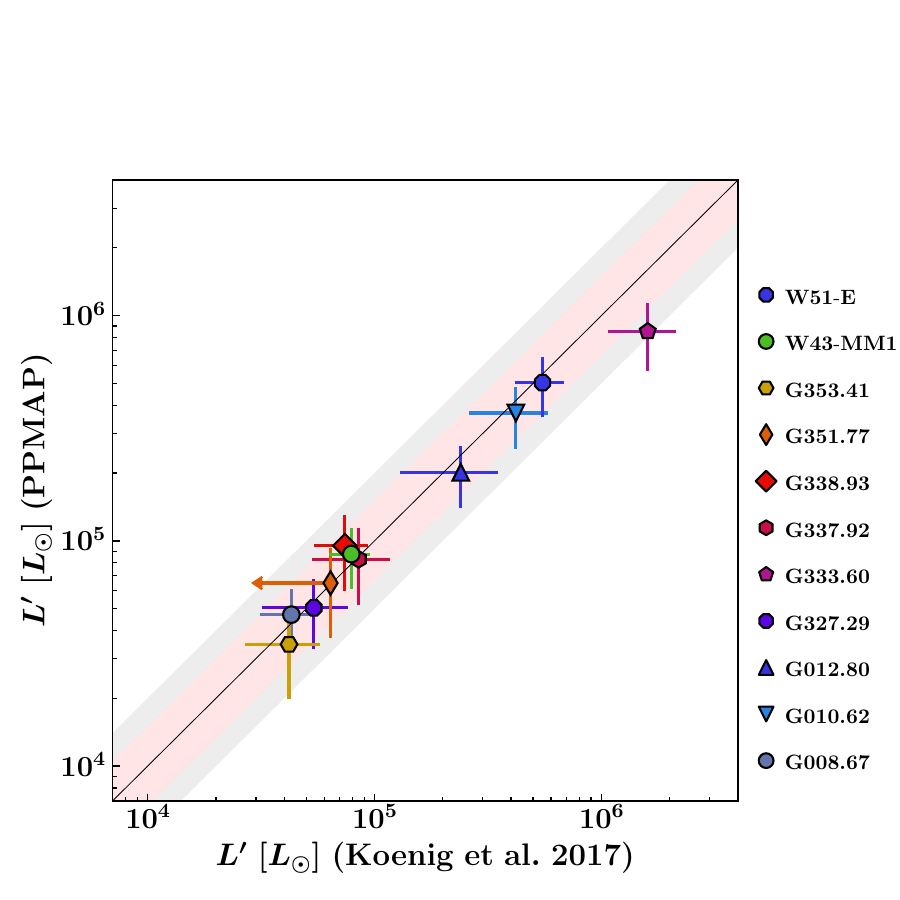}
      \includegraphics[width=0.95\hsize, trim={0cm 0cm 0cm 2.5cm},clip]{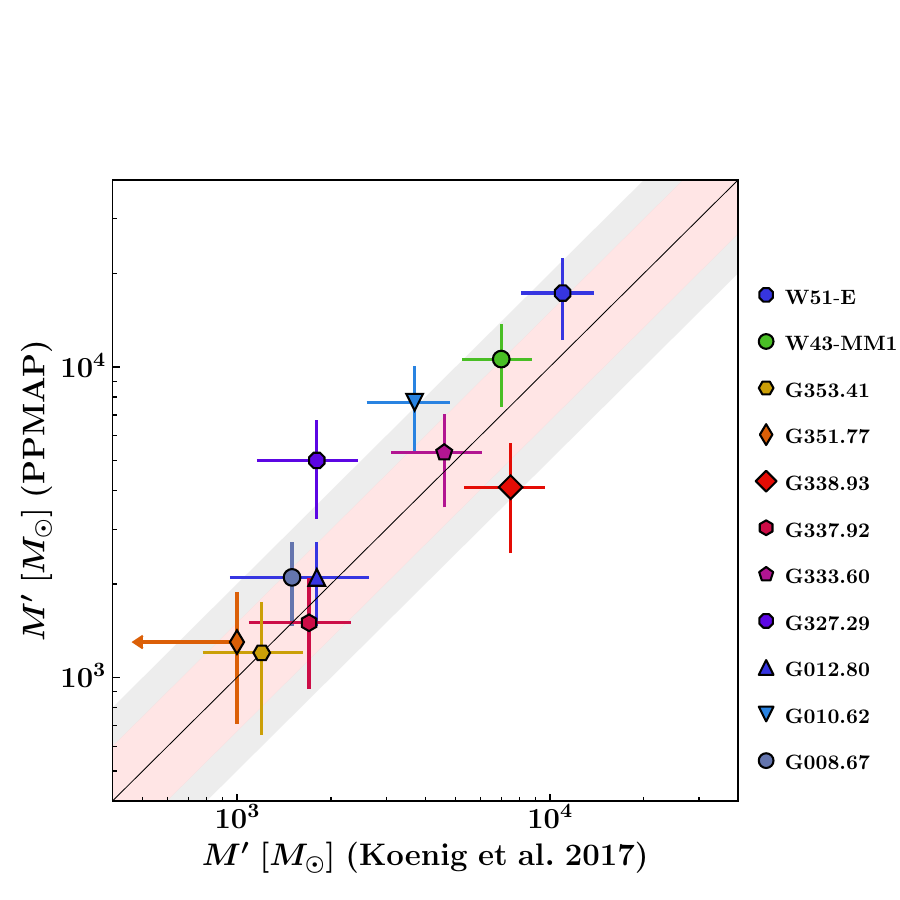}
      \caption{Accuracy of PPMAP measurements. Comparison of estimates made by PPMAP (see Tab.~\ref{table:Results}) and \citet{Koenig2017} for the bolometric luminosity $L_\mathrm{bol}^\prime$ (\textit{top panel}) and mass $M^\prime$ (\textit{bottom panel}), integrated in the aperture defined by \citet{Koenig2017} for 11 of the ALMA-IMF regions covered by both studies. The red and gray shades respectively indicate a 50\% and 100\% uncertainty range around the identity curve, represented by a solid black line.}
         \label{fig:Koenig2017Benchmark}
   \end{centering}
\end{figure}

In order to benchmark our results, Fig.~\ref{fig:Koenig2017Benchmark} compares the bolometric luminosity and mass measured in the PPMAP images to those obtained by \citet{Koenig2017}. Based on a two-temperatures MBB description, they inferred bolometric luminosities and masses for a selected sample of 110 ATLASGAL sources through SED fitting of their mid-infrared to submillimeter flux densities, between 8 and 870 $\upmu$m, that is, a slightly narrower range than ours (3.6~$\upmu$m - 1.3~mm). We performed aperture photometry on the 2.5$^{\prime \prime}$ PPMAP-derived bolometric luminosity and column density maps using the exact same apertures. The \citet{Koenig2017} apertures were designed to ensure consistent flux extraction over the same area from the mid-infrared to the submillimeter range. Throughout the comparison we introduced a systematic correction to account for variations in the adopted distances for the ALMA-IMF regions\footnote{\citet{Koenig2017} assumed 4.8, 5, 3.1, 3.6, 3.2, 4.4, 1.0, 3.4, and 4.9~kpc respectively for G008.67, G010.62, G327.29, G333.60, G337.92, G338.92, G351.77, G353.41, and W43-MM1.}.
The comparison is made on a subset of 10 regions observed by both studies: G008.67, G010.62, G012.80, G327.29, G333.60, G337.92, G338.92, G351.77, G353.41, W43-MM1, and W51-E. 

On the one hand, our comparison (cf. Fig.~\ref{fig:Koenig2017Benchmark}) indicates that our bolometric luminosity estimates are generally consistent with the measurements made by \citet{Koenig2017}. The only noticeable discrepancy, exceeding a standard deviation, is observed for G333.60. This discrepancy is likely due to the saturation of \textit{Spitzer}/MIPS observations in our study, a limitation that was circumvented by \citet{Koenig2017} using MSX observations \citep{Egan2003}. Incorporating MSX observations into our study was not feasible, as our focus is to achieve the most accurate representation of the finer scales in the images (below $5^{\prime \prime}$). This objective contrasts with the relatively coarse angular resolution of MSX (18$^{\prime \prime}$), and the fact that MSX observations can only be added to the SED model a posteriori, since PPMAP cannot reproduce the emission from out-of-equilibrium dust grains.

On the other hand, our mass estimates exhibit more significant discrepancies with \citet{Koenig2017}'s results (see Fig.~\ref{fig:Koenig2017Benchmark}). Because the derived mass depends on the estimated temperature, using the MBB description can yield larger discrepancies in the mass estimates compared to the luminosity estimates. Measuring luminosities involves a robust and straightforward measurement of the area below the data points, whereas the derived mass depends on the estimated temperature. As a consequence, it is anticipated that a more substantial variability may arise in mass, in contrast to luminosity (see Fig.~\ref{fig:Koenig2017Benchmark}).
Furthermore, \citet{Galliano2018} reported that mass estimates may depend on the spatial resolution, since the temperature structure can be hidden in poorly resolved images, while it is accounted for in higher-resolution observations (e.g., Fig.~14 in \citealt{Aniano2012}).
This interpretation is consistent with the fact the more distant protoclusters display larger mass discrepancies, while the less distant protoclusters better align with the identity curve, with the only exception of G327.29.

Additionally, our use of 8 dust temperatures for reproducing the observations contrasts with \citet{Koenig2017} use of at most two temperatures. When using more temperature components, a small amount of warm dust can largely contribute to the 70~$\upmu$m emission, thereby recovering an amount of cold dust that would otherwise would be missed, and consequently increasing the total mass (e.g., \citealt{Aniano2012}). 
Moreover, an other effect is the fact that the longest wavelength used by \citet{Koenig2017} is 870~$\upmu$m, whereas we also used 1.3~mm observations. 
We checked for these two effects and found that reducing the number of temperature components to two does result in higher density-weighted temperatures, while removing the 1.3~mm data generally results in recovering less mass. Taking into account these two effects combined, 14 out of 15 protoclusters masses are consistent with \citet{Koenig2017}'s measurement within $1\sigma$ (with the exception of G338.93, that is consistent within $2 \sigma$).
Finally, the use of a slightly different opacity index ($\beta = 1.75$) and larger reference opacity ($\kappa_{300} \simeq 0.125$~cm$^2$g$^{-1}$) by \citet{Koenig2017} marginally accounts for these mass discrepancies.

\subsection{Uncertainties}\label{sect:caveats}

\subsubsection{Description}

We estimate here the uncertainties inherent to the PPMAP-derived products and measurements. While the MBB description itself constitutes an approximation, our results are influenced by additional complexities. Table~\ref{table:systematic-errors} provides an estimate of the mean values of these uncertainties across the 15 protoclusters studied. The primary sources of errors within the PPMAP process, that impact the determination of luminosity, mass, column density and dust temperatures, are as follows:

\begin{enumerate}
\item Saturated pixels: saturation in the continuum observations used as PPMAP inputs. Most importantly, the saturation of the \textit{Spitzer}/MIPS map at 24~$\upmu$m (see Fig.~\ref{fig:datacompleteness}) may impact the bolometric luminosity estimate, like suggested for that of G333.60 in Fig.~\ref{fig:Koenig2017Benchmark}. In contrast, the saturation of far-infrared and submillimeter images and its effect on PPMAP products are mitigated by the Gaussian process regression we applied (cf. Appendix~\ref{appendix:desaturation}).\\[-1.0em]
\item Free-free emission: the MBB description does not account for the free-free emission that contributes to the ALMA 1.3~mm image of evolved and intermediate regions. We used images approximately corrected for contamination by free-free emission, as described in Sect.~\ref{sect:free-free}.\\[-1.0em]
\item Noise estimates in the FIR to millimeter maps: they determine the uncertainty in estimating the measurement error for the input maps, that in turn determines the relative weights of data points used throughout the PPMAP SED fitting process. All the PPMAP results are thus sensitive to the methods used to determine the noise level of input maps (Sect.~\ref{sect:noise}).\\[-1.0em]
\item PPMAP SED fitting: the uncertainties associated with the PPMAP fitting process for determining dust parameters. Includes systematic errors arising from the adopted opacity index ($\beta = 1.8\pm0.2$).\\[-1.0em]
\item  Correction of the optically thick emission: while this correction is crucial to estimate the dust temperature maps of ALMA-IMF protoclusters, it relies on a model of extinction that introduces uncertainties (cf. Sect.~\ref{sect:opacity-correction}).\\[-1.0em]
\end{enumerate}

Finally, we identified four minor sources of error.

\begin{enumerate}
\item Pointing errors inherent to any map sets taken with different observatories, here leading to relative shifts between \textit{Herschel}, APEX, SOFIA, and ALMA data, could bias the SED fitting (cf. Table~\ref{table:systematic-errors}).\\[-1.0em]
\item Uncertainties on the distance to the Sun of ALMA-IMF protoclusters (see Table~\ref{table:Results}) lead to errors on their luminosity and mass images.\\[-1.0em]
\item Errors introduced by the splines interpolation of the near-infrared observations, below 70~$\upmu$m.
\item Finally, ring-like artifacts around sources are expected to have an impact on the PPMAP-derived measurements (cf. Appendices~\ref{appendix:convergence} and \ref{appendix:post-processing} for a description of these artifacts).
\end{enumerate}

\subsubsection{Quantification of errors}\label{subsect:systematic-errors}

We here quantify the mean uncertainty of each sources of error mentioned above.
Systematic and random errors are listed and quantified in Table~\ref{table:systematic-errors}. The methods we employed to estimate the errors are the following:
\begin{enumerate}
\item To estimate the uncertainties caused by the saturation, free-free emission, noise estimates, opacity index, artifacts, and pointing errors, we ran PPMAP with modified input parameters and maps. We then inferred the errors from the discrepancies between the outcomes. For instance, the uncertainty on the column density originating from the saturation of input maps is defined as $\sigma(N_\mathrm{H_2})= \Delta N_\mathrm{H_2}(\mathrm{RUN_a}, \mathrm{RUN_b}) = \lvert N_\mathrm{H_2}(\mathrm{RUN_a}) - N_\mathrm{H_2}(\mathrm{RUN_b}) \lvert $, where $N_\mathrm{H_2}(\mathrm{RUN_a})$ and $N_\mathrm{H_2}(\mathrm{RUN_b})$ are respectively the column density maps obtained with and without including the saturated images (see Fig.~\ref{fig:datacompleteness}). The characteristics of the different PPMAP runs performed to derive uncertainties are described below Table~\ref{table:systematic-errors}.
\item For the uncertainty inherent to the PPMAP SED fitting process, we used the uncertainty output, \lq \lq \texttt{sigtdens.fits}'' (in which the random errors obtained from the SED fitting are stored), to derive the column density and temperature uncertainties following Eq.~(\ref{eq:PPMAP-N}) and Eq.~(\ref{eq:PPMAP-T}).
\end{enumerate}

\begin{table}[htb]
\centering
{\caption{Empirically derived errors associated with PPMAP measurements (see Sect.~\ref{subsect:systematic-errors}).}       
\label{table:systematic-errors}}      
{\centering                          
\begin{tabular}{l l c c c c}        
\hline \hline         \\[-1.0em]
\footnotesize{uncertainty $u_X$} & \footnotesize{label} & \footnotesize{$\sigma N_\mathrm{H_2} (u_X)$} & \footnotesize{$\sigma T_\mathrm{dust} (u_X)$} \\
\hline \\[-1.0em]
\footnotesize{$u_\mathrm{saturation}$$^a$} & \footnotesize{input saturation} & \footnotesize{9\%} & \footnotesize{3\%}\\
\footnotesize{$u_\mathrm{free-free}$$^b$} & \footnotesize{free-free emission} & \footnotesize{8\%} & \footnotesize{1\%}\\
\footnotesize{$u_\mathrm{PPMAP}$$^c$} & \footnotesize{PPMAP SED fitting} & \footnotesize{8\%} & \footnotesize{11\%}\\
\footnotesize{$u_\mathrm{\texttt{getnoise}}$$^d$} & \footnotesize{noise estimates} & \footnotesize{14\%} & \footnotesize{7\%}\\
\footnotesize{$u_\beta$$^e$} & \footnotesize{opacity index} & \footnotesize{21\%} & \footnotesize{9\%}\\
\footnotesize{$u_\mathrm{artifacts}$$^f$} & \footnotesize{ring-like artifacts} & \footnotesize{3\%} & \footnotesize{$\pm 1~$K}\\
\footnotesize{$u_\mathrm{\Delta\theta = 1^{\prime \prime}}$$^g$} & \footnotesize{\textit{Herschel} pointing (1)} & \footnotesize{0.4\%} & \footnotesize{0.2\%}\\
\footnotesize{$u_\mathrm{\Delta\theta = 2.5^{\prime \prime}}$$^g$} & \footnotesize{\textit{Herschel} pointing (2)} & \footnotesize{0.5\%} & 
\footnotesize{0.8\%}\\ \\[-1.0em]
\hline \\[-1.0em]
\footnotesize{} & \footnotesize{Combined uncertainty} & \footnotesize{29\%} & 
\footnotesize{16\% $\pm$ 1~K}\\
\hline
\end{tabular}}
\\~\\
\raggedright
\footnotesize{
The numbers compiled above are the mean values (across the 15 protoclusters) of the mean errors measured in the uncertainty maps.\\
$^a$Error stemming from the saturation of input continuum maps, assessed by comparing the outcomes obtained with and without the inclusion of saturated maps in the PPMAP analysis.\\
$^b$Systematic error induced by free-free emission. Estimated from the comparison between Run1 and Run2 (cf. Sect.~\ref{sect:PPMAPanalysis}).\\
$^c$PPMAP SED fitting errors, derived from the \texttt{sigtdens.fits} file.\\
$^d$Error related to the noise estimates used to weight the data points fed to PPMAP. Empirically derived by comparing PPMAP products obtained with three distinct noise estimate methods (cf. Sect.~\ref{sect:noise}).\\
$^e$Error caused by the uncertainty on the opacity index, $\beta = 1.8 \pm 0.2$.\\
$^f$Error caused by ring-like artifacts (cf. Appendix~\ref{appendix:post-processing}).\\
$^g$Estimate of the error that would be produced by a 1$^{\prime \prime}$ or 2.5$^{\prime \prime}$ relative shift in the \textit{Herschel} maps with respect to the ALMA observations.
}
\end{table}

We found that the uncertainty caused by potential \textit{Herschel} pointing errors are negligible ($< 1$\%). Meanwhile, the primary contributors to uncertainties in constraining the dust temperature include the errors associated with PPMAP's SED fitting, noise estimates, the choice of the opacity index ($\beta$), and the influence of ring-like artifacts.
Uncertainties are in fact variable across the field of observations, thus we provide uncertainty maps corresponding to the relevant data products. Values in Table~\ref{table:systematic-errors} are an account of the spatially averaged uncertainties (measuring the mean value across the error maps).
The determination of the total uncertainties for column density ($N_\mathrm{H_2}$) and temperature ($T_\mathrm{dust}$) employs the combined standard uncertainty ($u_\mathrm{tot}^2 = \sum_i u_i^2$, that is, a quadratic sum over the uncertainties listed in Table~\ref{table:systematic-errors}), where $u$ pertains to the errors enumerated above. The resulting total uncertainties are $u_{N_\mathrm{H_2}} = 0.29 N_\mathrm{H_2}$ and $u_{T_\mathrm{dust}}= 0.16 T_\mathrm{dust} \pm 1$~K. It should be noted that these total uncertainties do not include \textit{i.)} the uncertainty on $\kappa_{300}$, due to the unknown composition and size distribution of dust (\citealt{Koehler2015}, \citealt{Ysard2019}, \citealt{Schirmer2020}); \textit{ii.)} the bias induced by the PPMAP assumption of optical thinness, that may significantly affect the temperature and mass estimates, in particular toward high column density pixels. These considerations, along with the beam dilution bias mentioned earlier, should be treated as additional uncertainties if their relevance arises in the context of using our PPMAP estimates.
Finally, we propagated the quadratic sum of errors for parameters such as $\beta$, $N$, and $T_\mathrm{dust}$ to infer errors for luminosity and mass estimates (as defined by Eq.~(\ref{eq:PPMAPmod})), employing the same combined standard uncertainty approach. Variable errors associated with distances, as outlined in the second column of Table~\ref{table:Results}, are also factored into the mass and luminosity uncertainties.

\section{Discussion}\label{sect:discussion}

The PPMAP products presented in this paper have the potential for further analyses.
Firstly, high-resolution dust temperature maps are an essential prerequisite for deriving core masses and constructing core mass functions (CMFs). Although the PPMAP 2.5$^{\prime \prime}$ beam is roughly five-fold larger than that of ALMA observations, our temperature maps currently offer the most comprehensive coverage and the best resolution available for the ALMA-IMF survey. In fact, ongoing studies by \citet{Louvet2024} and \citet{Armante2024} are employing these PPMAP-derived temperature maps for CMF investigations.
Moreover, our column density maps provide a means to characterize the structure of the protoclusters (as illustrated in Fig.~\ref{fig:PPMAP-resolution}), and could be used along with different molecular tracers (e.g., N$_2$H$^+$) to derive abundances in different parts of the same protocluster, or between protoclusters, that has a potential use as an evolutionary indicator. Finally, the luminosity-to-mass ratio may also constitute a tracer of the evolutionary stage of these regions, even enabling the discernment of subregions within the ALMA-IMF fields. These tools open up new possibilities for further exploration within the ALMA-IMF survey, a topic we discuss in the following sections, featuring a selection of examples. More comprehensive analyses will be the focus of future studies.

\begin{figure}[htb]
   \begin{centering}
      \includegraphics[width=\hsize, trim={0.5cm 0.5cm 2.25cm 1.5cm},clip]{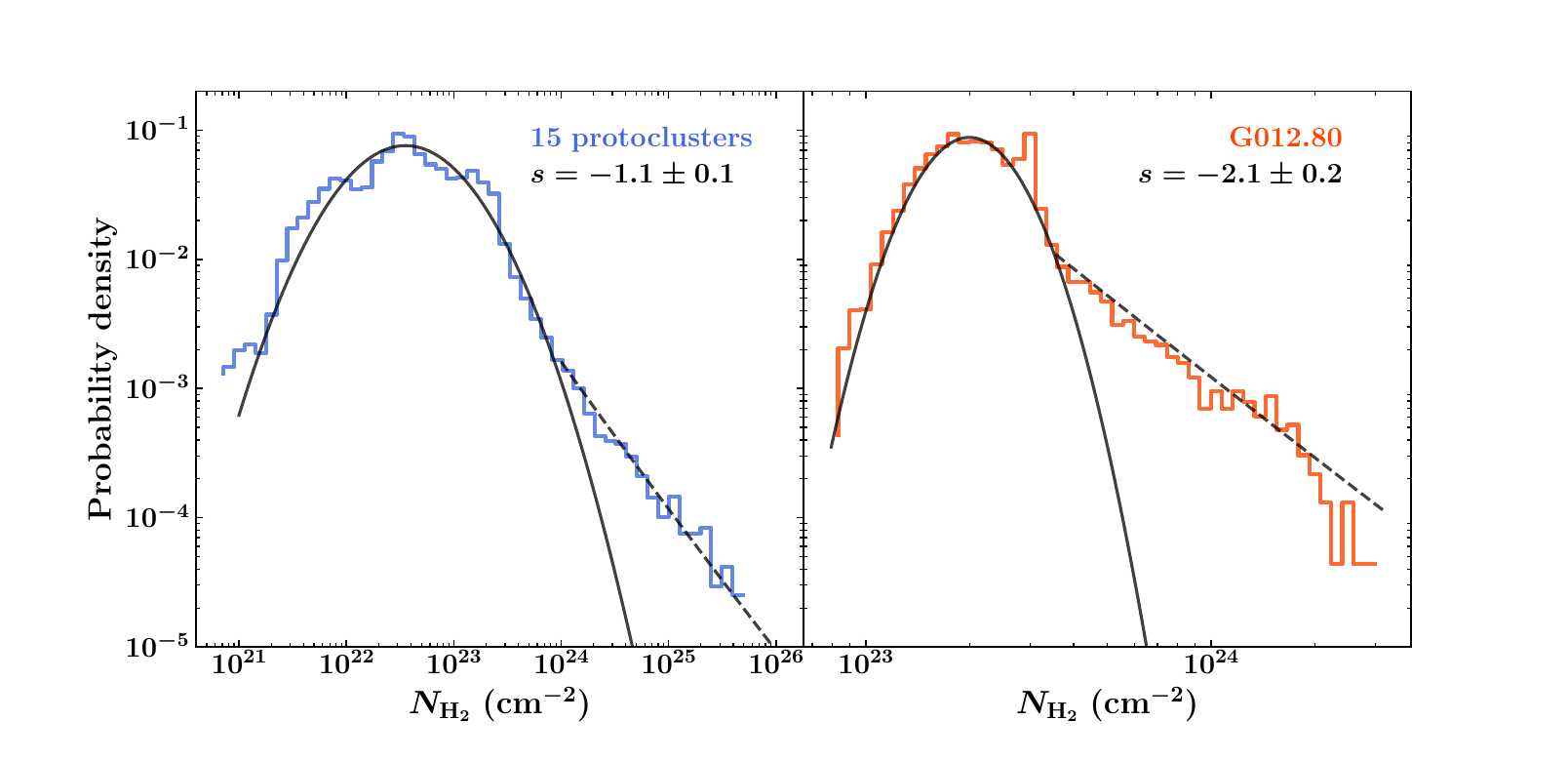}
      \caption{Probability density functions (PDFs) of the PPMAP-derived column density, normalized with respect to the area. The cumulative PDF across the 15 ALMA-IMF regions is shown on the left, and the PDF measured in the evolved G012.80 protocluster is shown on the right. Solid black lines represent the lognormal distribution that best fits the PDFs, while the dashed black lines correspond to the power-law tail, with the power-law index $s$ indicated in the upper-right corner.}
         \label{fig:cdens-PDFs}
   \end{centering}
\end{figure}

\begin{figure*}[htb!]
   \begin{centering}
      \includegraphics[width=\hsize, trim={0.25cm 4.25cm 1cm 4.75cm},clip]{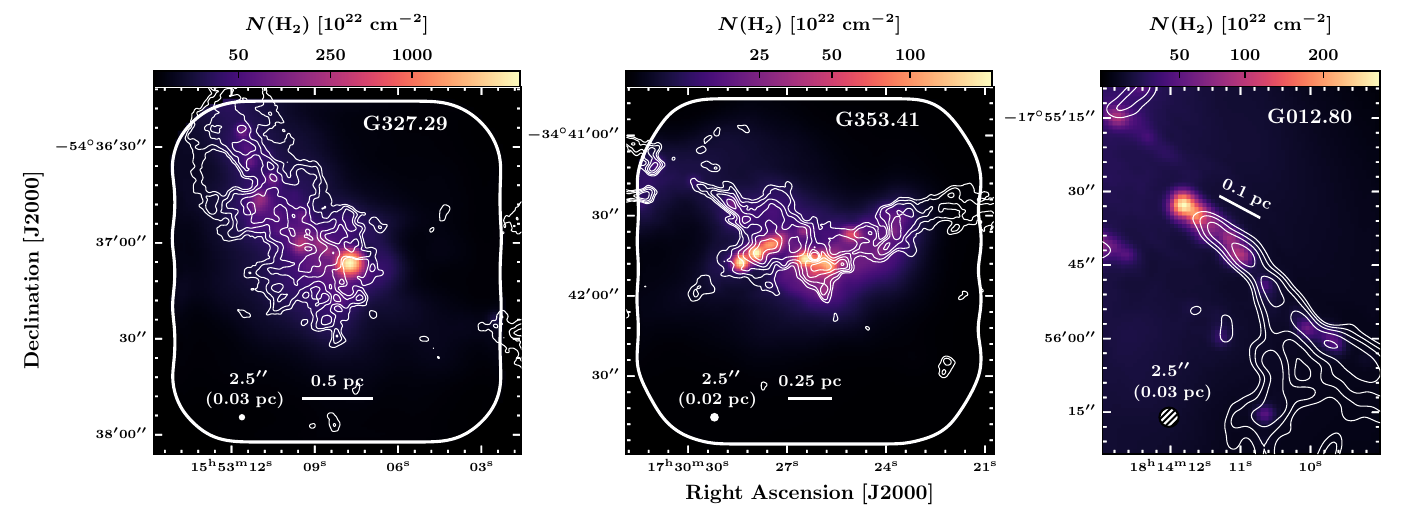}
      \caption{PPMAP column density maps compared with the N$_2$H$^+$ integrated intensity map (overlaid in white contours), for three specific regions (from left to right: G327.29, G353.41, and G012.80). Contour levels: logarithmically spaced between 50 to 225 K km s$^{-1}$ (left panel), 50 to 150 K km s$^{-1}$ (center panel), and 50 to 150 K km s$^{-1}$ (right panel).}
         \label{fig:PPMAP-N2H+}
   \end{centering}
\end{figure*} 

\subsection{PPMAP-derived column density}

\subsubsection{Probability density functions}

Figure~\ref{fig:cdens-PDFs} presents the mean probability density function (hereafter PDF) of the PPMAP-derived column density across the 15 regions studied. Column density PDFs of molecular cloud are well described by a lognormal distribution in addition to a power-law tail (\citealt{Kainulainen2009}; \citealt{Schneider2015}; \citealt{Schneider2022}, refer to \citealt{Pouteau2022} for an analysis of column density PDFs in W43-MM2\&MM3). The low-density tail is always limited by the included sky area, therefore the lognormal shape and position may be biased by the relatively small size of ALMA fields (typically $1^\prime \times 1 ^\prime$).
Even though they span a wide range in column densities (from $10^{21}$ cm$^{-2}$ to $10^{25}$ cm$^{-2}$), distances to the Sun and evolutionary stages, the cumulative PDF of the 15 regions shown in the left panel of Fig.~\ref{fig:cdens-PDFs} can be roughly described by a lognormal distribution, although substructures are seen.
As an example of an individual PDF measured over the extent of the ALMA footprint, we show the PDF of the evolved G012.80 protocluster in the right panel of Fig.~\ref{fig:cdens-PDFs}. In this individual case the  best-fit lognormal distribution better matches the measurements, and for higher column densities the departure from the lognormal distribution can be clearly defined. Deviations from the lognormal shape are predicted for gas structures governed by self-gravity (\citealt{Schneider2015} and references therein). The flattening of the distribution at higher column densities is described by a power-law, $p \propto (N_\mathrm{H_2}) ^{-s}$. We measure $s=2.1 \pm 0.2$ in G012.80, a value that is consistent with gravitational collapse of an isothermal sphere \citep{Schneider2015}.
A comprehensive analysis of the column density PDFs is not in the scope of this study, we refer to  \citet{Diaz2023} for a systematic, high-resolution study of column density PDFs inferred with PPMAP temperature maps.

\subsubsection{Comparison of the PPMAP column density maps with a N$_2$H$^+$ line}

From the comparison with \citet{Koenig2017}'s measurements described in section~\ref{subsect:validation}, we established that the PPMAP products, including the column density and temperature maps, are robust in terms of large of the large scale measurements.
These parameters ($N_\mathrm{H_2}$, $T_\mathrm{dust}$) are vital to understanding the chemistry in massive star-forming protoclusters (column density to measure abundances, temperature in relation to potential energy barriers), and to constraining dust simulations.
However, the comparison presented in section~\ref{subsect:validation} pertains to mean values measured within large apertures (on the order of $10^{\prime \prime}$). To assess our results at the precise angular resolution enabled by PPMAP (2.5$^{\prime \prime}$), we made further comparisons with observations of comparable resolution, such as the ALMA-IMF spectral data \citep{Cunningham2023}. Through this comparison we aim to check that small features ($\sim 2.5^{\prime \prime}$) found in the ALMA high-resolution data are reproduced to some extent in the PPMAP products.

\begin{figure}[htb]
   \begin{centering}
      \includegraphics[width=0.9\hsize, trim={0.25cm 0.5cm 1.5cm 2cm},clip]{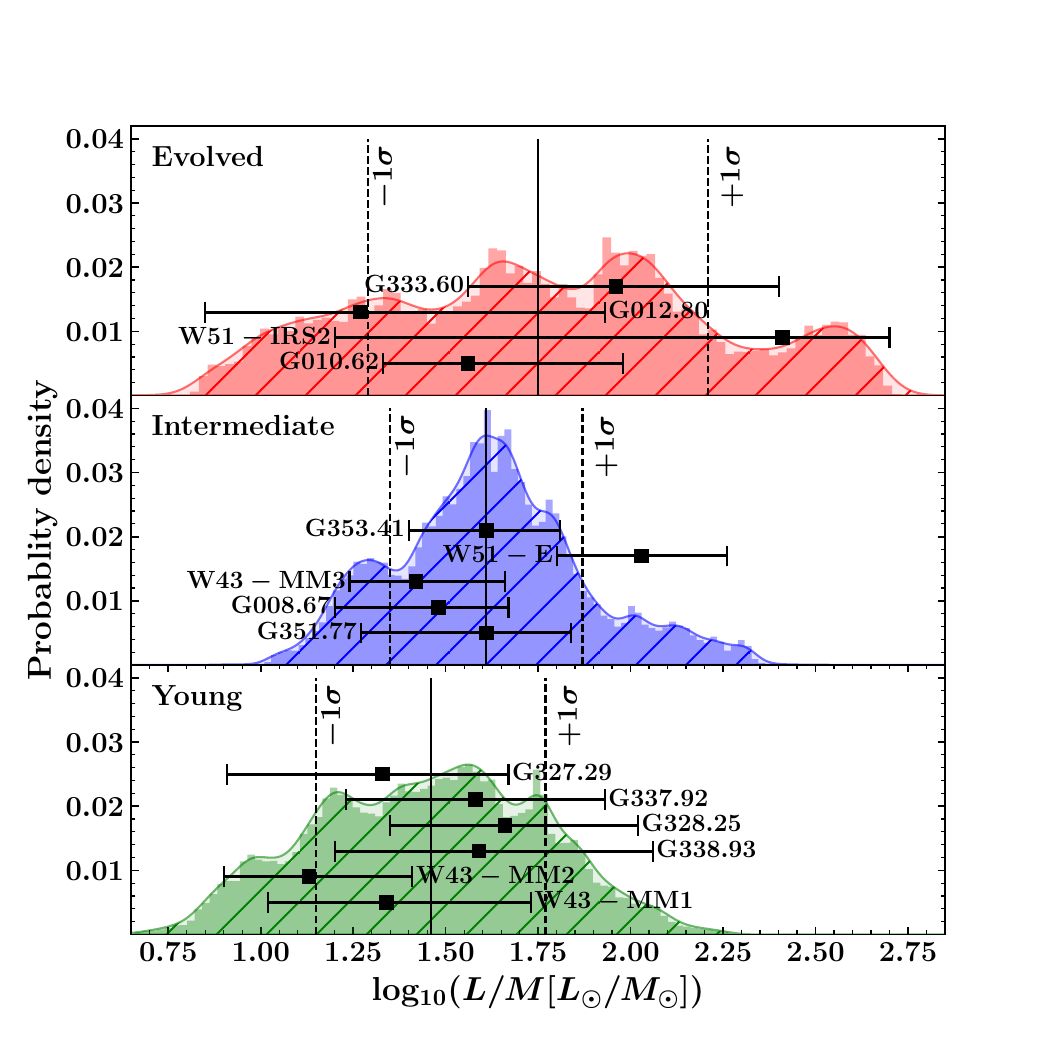}
      \caption{Probability density functions (PDFs) of the luminosity-to-mass ratio, normalized with respect to the area. Cumulative PDFs are shown for the evolved (\textit{top panel}: G010.62, W51-IRS2, G012.80, G333.60), intermediate (\textit{central panel}: G351.77, G008.67, W43-MM3, W51-E, G353.41) and young (\textit{bottom}: W43-MM1, W43-MM2, G338.93, G328.25, G337.92, G327.29) protoclusters. Black markers and horizontal bars represent the median, first and last decile $L/M$ measurements within individual protoclusters.}
         \label{fig:LM-PDFs}
   \end{centering}
\end{figure}

Figure~\ref{fig:PPMAP-N2H+} compares our column density maps with the N$_2$H$^{+}$ J=1--0 integrated line emission (Stutz et al. in prep; Alvarez-Gutierrez et al. in prep), a tracer of the dense and cold medium \citep{Pety2017} that has the potential to correlate with dusty filaments. Our analysis across regions in different evolutionary stages (G327.29: young, G353.41: intermediate, G012.80: evolved) reveals a general consistency between the PPMAP-derived column density and the N$_2$H$^{+}$ integrated intensity map. 
Toward G327.29 and G353.41, both the global filamentary morphology and a fraction of the local emission peaks peaks are coherent between the dust column density and N$_2$H$^{+}$ maps.
Particularly remarkable is the image in the right panel of Fig.~\ref{fig:PPMAP-N2H+}, displaying W33 Main-West filament (\citealt{Immer2014}; \citealt{Armante2024}). Along the filament, faint column density peaks align with local maxima of N$_2$H$^{+}$ emission, while N$_2$H$^{+}$ fades toward the central, higher column density peak. Local protostellar heating could account for the absence of N$_2$H$^{+}$ emission within the central source, where a heightened gas temperature must result in its chemical destruction following the desorption of CO from grain mantles (\citealt{Lee2004}; \citealt{Busquet2011}; \citealt{Sanhueza2012}). A hot core was detected at this location, (\citealt{Armante2024}; \citealt{Bonfand2024}), and the PPMAP-derived temperature does register a local increase in this specific area, up to 37~K (see Fig.~\ref{fig:PPMAP-N2H+}), a measurement that may be consistent with a localized temperature of a hundred Kelvin, if we account for beam dilution (Motte et al., in prep.).
Consistent associations between the positions of hot cores and local temperature increases in the dust temperature maps strengthen the case that the PPMAP products are reliable at their native angular scale of 2.5$^{\prime \prime}$, and demonstrate that they offer opportunities for interpretations of the physical and chemical mechanisms at work in the ALMA-IMF fields of observations.

\begin{figure}[htb]
   \begin{centering}
      \includegraphics[width=0.9\hsize, trim={0.25cm 0cm 0cm 1cm},clip]{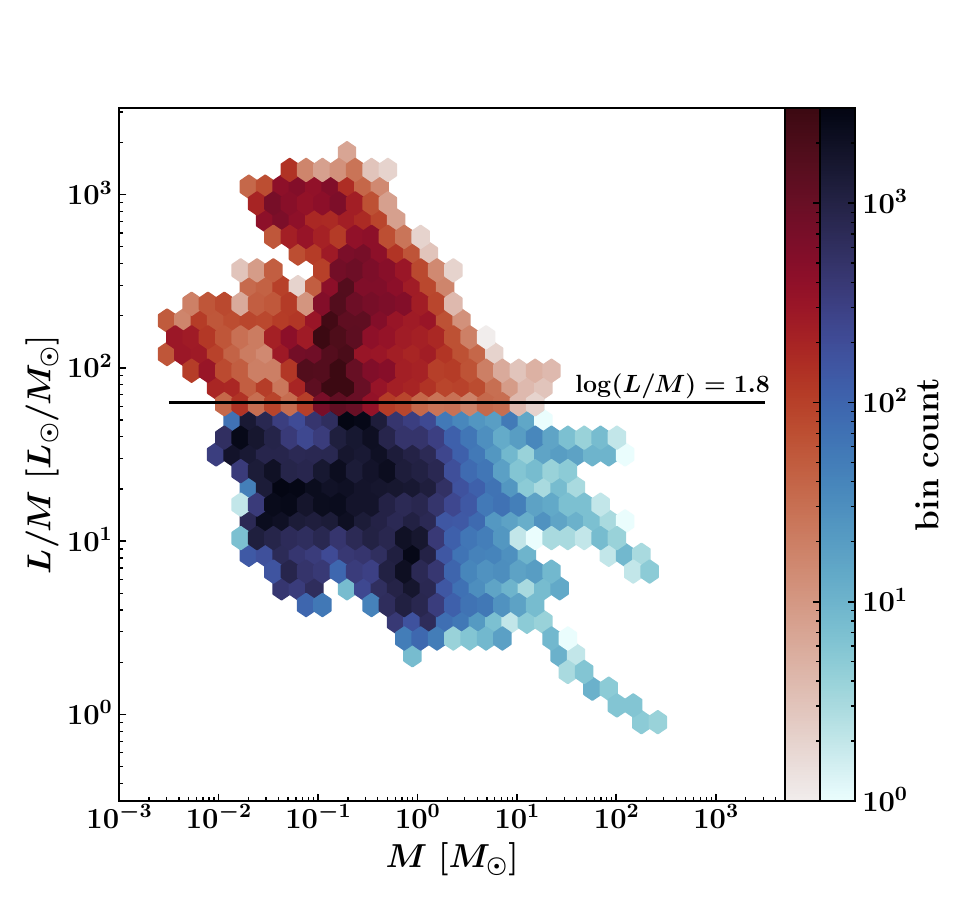}
      \includegraphics[width=0.9\hsize, trim={1cm 4cm 14.5cm 5cm},clip]{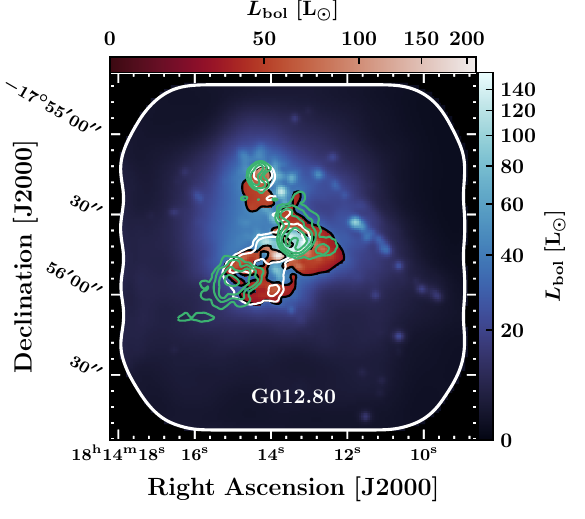}
      \caption{Luminosity-to-mass ratio unveiled by PPMAP. \textit{Top panel}: we show the complete histogram of $L / M$ across the 15 ALMA-IMF regions studied, separated into two samples by the equation $\mathrm{log}(L / M) = 1.8$. \textit{Bottom panel}: as an example, a decomposition of the evolved G012.80 protocluster's luminosity map is performed. Pixels with higher luminosity-to-mass ratio ($\mathrm{log}(L / M) \geq 1.8$ are plotted with a red colormap, while their counterpart ($\mathrm{log}(L / M) \leq 1.8$) are shown with a blue colormap. Superimposed white contours illustrate the H41$\alpha$ line emission, that traces regions dominated by free-free emission (contour levels: logarithmically spaced between 0.025 and 0.075 Jy beam$^{-1}$). Green contours illustrate the \NeII line emission (contour levels: logarithmically spaced between 0.005 and 0.04 erg s$^{-1}$ cm$^{-2}$ sr$^{-1}$).}
         \label{fig:PPMAP-decomposition}
   \end{centering}
\end{figure}

\subsection{Luminosity and mass of PPMAP luminosity peaks}\label{sect:GETSF}

Here we discuss the (bolometric) luminosity-to-mass ratio measured over the full extent of the protoclusters, before we delve into the luminosity-to-mass ratio of smaller sources mapped at the 2.5$^{\prime \prime}$ angular resolution. The bolometric luminosity is directly measured from the bolometric luminosity maps described in Sect.~\ref{subsect:luminosity} (hence, including free-free emission), and the total gas mass is derived from the H$_2$ column density maps.
Figure~\ref{fig:LM-PDFs} shows the pixel-per-pixel PDFs of the luminosity-to-mass ratio partitioned with respect to the evolutionary stage proposed by \citet{Motte2022}.

As star-forming protoclusters evolve, the contribution of \HII regions to the luminosity is expected to gradually increase, thus the luminosity-to-mass ratio should be enhanced in the more evolved regions. This trend is found, with the respective distributions shifting toward a higher $L/M$ ratio across the \lq \lq young'', \lq \lq intermediate'' and \lq \lq evolved'' protoclusters, although a significant dispersion is measured: for any pair of regions the mean values of $L / M$ are consistent with each other at a $\pm 1 \sigma$ level, where $\sigma$ is the standard deviation of the distribution.
The large dispersion may be primarily attributed to intra-region variability: regions within the ALMA-IMF survey, despite their relatively small sizes of a few parsecs, can encompass subregions with differing characteristics. This may result in a blend of lower and higher $L / M$ ratios across the field of observations, as we observe in Fig.~\ref{fig:LM-PDFs}.
This interpretation is reinforced by the fact that evolved regions display the largest dispersions. Indeed, evolved protoclusters may harbor a combination of young and more evolved subregions, pertaining to inhomogeneous initial conditions.
Consequently, the spatially averaged $L / M$ ratio (over the extent of the ALMA footprint) may not systematically serve as a robust indicator of the global evolutionary stage of one protocluster.

In Fig.~\ref{fig:PPMAP-decomposition}, we illustrate the pixel-per-pixel spatial variations of the luminosity-to-mass ratio, specifically in the evolved G012.80 protocluster. Through a luminosity map clustering approach based on an arbitrary threshold ($\mathrm{log}(L / M) = 1.8$), we discern a broad correlation between areas exhibiting higher $L / M$ ratios and \HII regions, that are traced by H41$\alpha$ and \NeII line emissions \citep{Beilis2022}. This illustrates the capability to map luminosity-to-mass ratio variations at a 2.5$^{\prime \prime}$ angular resolution and to constrain the nature and evolutionary stage of resolved structures.
In the subsequent subsection, we outline a method for estimating the luminosity-to-mass ratios of resolved luminosity peaks that may be extracted as individual sources.

\subsubsection{Source extraction with \textsl{getsf}}

Here we aim to compile a systematic catalog encompassing the luminosity and mass estimates of luminosity peaks found in the PPMAP dataset. With a limited angular resolution of 2.5$^{\prime \prime}$, it is conceivable that certain sources may not correspond to single entities, but rather to clusters of luminous sources, that could include a mixture of protostars, pre-stellar cores, and ultra- or hyper-compact \HII regions. Impending studies will attempt to perform cross-identifications based on already established core and protostar catalogs (Motte et al., in prep.). 
To identify sources within the luminosity maps, we employed the \textsl{getsf} method described by \citet{Menshchikov2021}. We direct interested readers to that publication for a detailed exposition of the procedure. Below, the underlying principles and application criteria pertinent to our dataset are briefly summarized.

The \textsl{getsf} method entails a spatial deconstruction of observed images, effectively separating structural constituents and their background. The technique aims to parse distinct spatial scales and segregating sources and filaments from both one another and the background. Characterized by a single parameter, namely an approximate maximum size of sources to be extracted, detection yields initial approximations of source footprints, dimensions, and fluxes. Subsequently, more precise measurements of the source sizes and fluxes are conducted on background-subtracted images and, if warranted, on auxiliary images.

We executed the \textsl{getsf} algorithm on the PPMAP luminosity maps generated from Run2, that is, without taking into account the free-free subtraction performed by Galvan-Madrid et al. (in prep.). This choice is rooted in our goal to ensure the best representation of the bolometric luminosities of the PPMAP luminosity peaks, by including the contribution of free-free emission.
In addition, PPMAP column density maps resulting from Run1 were simultaneously given to \textsl{getsf} as auxiliary data, in order to measure the mass of the sources.
Our initial step involved a resampling of the luminosity maps targeted at achieving a three-pixel sampling of the PPMAP beam (as elaborated in Appendix~\ref{appendix:resampling}), because it improves the detection of luminosity peaks associated with protostars.
Furthermore, we fixed the maximum source size to 5$^{\prime \prime}$, a value equivalent to twice the dimensions of the PPMAP beam.

\subsubsection{Results of the \textsl{getsf} extraction}

The outcomes of the \textsl{getsf} extraction process are presented in Table~\ref{table:compact-sources}. This table provides details including celestial coordinates, angular and spatial full width at half maximum (FWHM), background-subtracted luminosities and masses, as well as the luminosity-to-mass ratio of the PPMAP luminosity peaks.
We estimated the completeness level of the catalog of 313 luminosity peaks to be $\sim 60$~$L_\odot$, with a tendency for a better completeness level in young regions ($\sim 30$~$L_\odot$) than in evolved ones ($\sim 100$~$L_\odot$).
In some instances (18\% of the peaks), the luminosity peaks have no counterparts in the column density maps, resulting in a mass below our detection threshold. These mass measurements are discarded and signalled by the symbol \lq \lq -'' (see Table~\ref{table:compact-sources}). We interpret the luminosity peaks with no massive counterpart as diffuse areas heated by evolved protostars or \HII regions.

The spatial distribution of the PPMAP luminosity peaks is superimposed on the column density and luminosity maps, in Figs.~\ref{fig:PPMAPs} and~\ref{fig:PPMAPs-supplementary}. Upon visual examination, we observe a general correspondence between the PPMAP luminosity peaks and features such as dusty filaments, \HII regions, clusters of cores and protostars extracted from the ALMA images.
Correlations are also observed between PPMAP luminosity peaks and individual cores and/or protostars extracted from the ALMA images.
A detailed cross-analysis of the PPMAP source catalog and the \textsl{getsf} core catalog extracted from ALMA continuum images falls beyond the scope of this paper and will be covered in a subsequent work by Motte et al. (in prep.).

The relation between the bolometric luminosity and mass is a useful metric to constrain the evolutionary stage of star forming objects.
The PPMAP source sample spans a considerable range of luminosity-to-mass ratios, across four orders of magnitude (from $L / M \simeq 10^{-1}$ to $L / M \simeq 10^{3}$, in solar units). This range underscores the wide spectrum of physical conditions and object types encompassed within the sample, ranging from (bright) ultra- and hyper-compact \HII regions to (faint) cold, massive star-forming cores. Due consideration must be given to the fact that a single PPMAP source may correspond to several blended objects, possibly of different nature, because of the limited angular resolution (2.5$^{\prime \prime}$).

\begin{figure}[htb]
   \begin{centering}
      \includegraphics[width=\hsize, trim={0cm 0.25cm 0.75cm 1cm},clip]{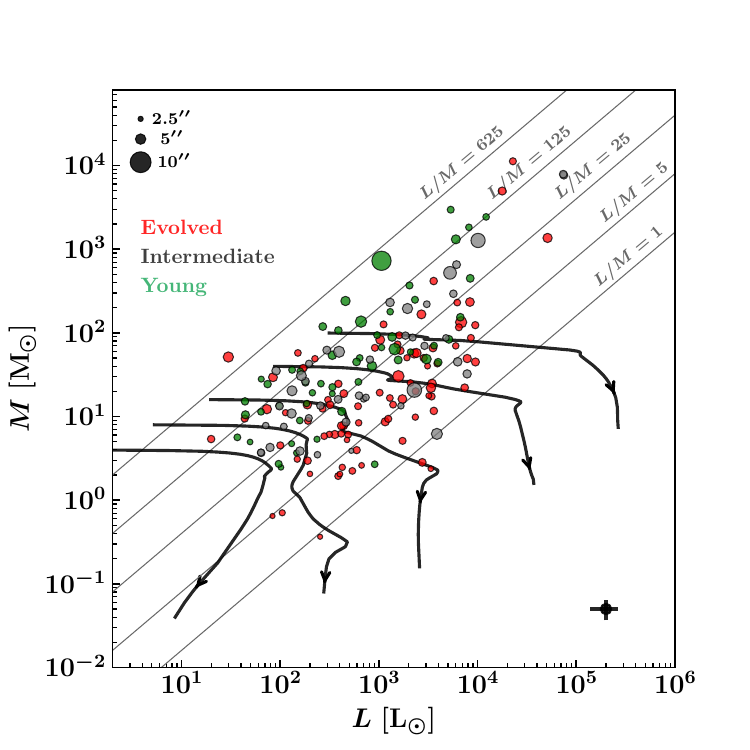}
      \caption{Mass and luminosity distribution of PPMAP luminosity peaks extracted by \textsl{getsf}. Caveat: with a 2.5$^{\prime \prime}$ angular resolution and at a distance of 2-5.5~kpc, PPMAP sources may correspond to several cores and/or protostars. The size of the markers reflects the FWHM of the sources, while the color indicates the evolutionary stage of the protoclusters they belong to. Typical error bars are shown in the bottom-right corner of the diagram. Solid black lines represent the evolutionary tracks from \citet{Motte2018a} and \citet{Duarte-Cabral2013} for final stellar masses of 2, 4, 8, 20 and 50 $M_\sun$.}
         \label{fig:LM}
   \end{centering}
\end{figure}

In Fig.~\ref{fig:LM}, we present the distribution of masses and luminosities for the PPMAP source catalog, and its comparison with evolutionary tracks from \citet{Motte2018a} and \citet{Duarte-Cabral2013}. 
The path of individual objects within the $M$-$L_\mathrm{bol}$ diagram can be predicted by accretion models. From a given initial envelope mass, the mass is expected to decrease at a given rate through material accretion onto the central star, in addition to material ejection by molecular outflows. The luminosity, on the other hand, is a function of stellar mass, hence it grows over time. These mechanisms steer the progression within the $M$-$L_{\mathrm{bol}}$ diagram from the top-left extremity to the end of the evolutionary tracks, allowing to follow the time evolution of cores and protostars.
The positions of luminosity peaks shown in Fig.~\ref{fig:LM} are generally consistent with the accretion models presented by \citet{Motte2018a} and \citet{Duarte-Cabral2013}, although approximately 20\% of the sources are above the 50~$M_\odot$ final stellar mass track. We interpret these massive sources as clusters of unresolved cores and/or protostars.
Furthermore, we observe that although the distributions of luminosity peaks pertaining to evolved and young protoclusters are overlapping, their centroid diverge within the diagram, aligning with different segments of the evolutionary tracks.
Indeed, the median values of the luminosity-to-mass ratio ($L / M$) of luminosity peaks demonstrate a trend across regions when classified based on their evolutionary stages, as outlined by \citet{Motte2022}. Specifically, we measure mean $L / M$ values of 253.5, 12.8, and 12.3 $L_\sun / M_\sun$ for the luminosity peaks of evolved, intermediate, and young regions, respectively (with a respective dispersion around the mean of 100.0, 4.2 and 8.0 $L_\sun / M_\sun$, estimated by the mean absolute deviation).

\subsection{Summary \& perspectives}


Using the multiwavelength, multiresolution Bayesian algorithm PPMAP, we performed a SED analysis of the dust emission in the 15 ALMA-IMF protoclusters. Near-infrared to millimeter observations from 8 instruments were included in the MBB analysis, spanning angular resolutions from subarcsecond (ALMA) to 35.2$^{\prime \prime}$ (\textit{Herschel}). Our results are the following:
\begin{enumerate}
\item We present new measurements of the bolometric luminosity, column density and dust temperature toward the 15 massive ALMA-IMF protoclusters (Tab.~\ref{table:Results}, Fig.~\ref{fig:SEDs}). The PPMAP estimates are consistent with previous measurements at a coarser angular resolution (\citealt{Koenig2017}, Fig.~\ref{fig:Koenig2017Benchmark}), and thus constitute a benchmarked mapping of the dust parameters at the best angular resolution currently attainable (2.5$^{\prime \prime}$).
\item We compared our column density and dust temperature maps with the continuum cores identified by \citet{Louvet2024}, the hot cores identified by \citet{Bonfand2024} and the N$_2$H$^+$ J=1--0 line, showing that the 2.5$^{\prime \prime}$ features found in the PPMAP products are consistent with sources and structures mapped at ALMA's native resolution, 0.3$^{\prime \prime}$-0.9$^{\prime \prime}$ (Fig.~\ref{fig:PPMAPs}, Fig.~\ref{fig:PPMAPs-supplementary}, Fig.~\ref{fig:PPMAP-N2H+}).
\item The pixel-per-pixel analysis of the luminosity-to-mass ratio shows that more evolved regions have, on average, a larger luminosity-to-mass ratio, although intra-region variability is observed (Fig.~\ref{fig:LM-PDFs}). We show, with an example in the G012.80 protocluster, that subregions pertaining to different evolutionary stages can be separated by setting a luminosity-to-mass ratio threshold (Fig.~\ref{fig:PPMAP-decomposition}).
\item Using \textsl{getsf} \citep{Menshchikov2021} on the PPMAP bolometric luminosity maps, we have extracted a catalog of 313 PPMAP sources, with the associated bolometric luminosity and mass measurements (Tab.~\ref{table:compact-sources}). We find that the luminosity-to-mass ratio of PPMAP sources tends to be consistent with the evolutionary stage attributed to the different protoclusters (\citealt{Motte2022}, Galván-Madrid et al., in prep.). We compared the PPMAP sources with evolutionary tracks from \citet{Duarte-Cabral2013} and \citet{Motte2018a}, although the comparison is biased by the fact that one PPMAP source may correspond to several unresolved cores and/or protostars (Fig.~\ref{fig:LM}).
\end{enumerate}

Numerous potential directions stem from our current findings.
Our analysis resulted in the determination of luminosity-to-mass ratios across ALMA footprints, ATLASGAL sources' footprints, and PPMAP luminosity peaks extracted via \textsl{getsf}. Our catalog of 313 luminosity peaks presents a resource for forthcoming investigations. The cross-identification of PPMAP luminosity peaks with \HII regions offers a pathway to estimate their luminosities. The dataset of column density, temperature, and luminosity maps we provide opens doors to investigate relationships among these quantities. 
Furthermore, the PPMAP-derived column density, in conjunction with PPMAP source and core catalogs, lays the groundwork for examining core formation efficiency and its dependence with the local gas density at a 2.5$^{\prime \prime}$ angular resolution, following the original study by \citet{Louvet2014} in W43-MM1.

Finally, despite potential challenges related to associating single PPMAP source with multiple protostars, our catalog may help constraining the temperatures and luminosity-to-mass ratios of young stellar objects. Assuming that a robust cross-identification can be completed, the catalog enables a direct comparison with accretion models, based on the $M_\mathrm{core}$ versus $L_\mathrm{bol}$ evolutionary diagram (e.g., \citealt{Duarte-Cabral2013}).
One such study, underway by Motte et al. (in prep.), focuses on cross-identification and further processing of PPMAP-derived temperatures to estimate core and protostar temperatures and luminosities.

\begin{acknowledgements}
This paper makes use of the following ALMA data: ADS/JAO.ALMA\#2017.1.01355.L, \#2013.1.01365.S, and \#2015.1.01273.S. ALMA is a partnership of ESO (representing its member states), NSF (USA) and NINS (Japan), together with NRC (Canada), MOST and ASIAA (Taiwan), and KASI (Republic of Korea), in cooperation with the Republic of Chile. The Joint ALMA Observatory is operated by ESO, AUI/NRAO and NAOJ. This project has received funding from the European Research Council (ERC) via the ERC Synergy Grant ECOGAL (grant 855130), from the French Agence Nationale de la Recherche (ANR) through the project COSMHIC (ANR-20- CE31-0009), and the French Programme National de Physique Stellaire and Physique et Chimie du Milieu Interstellaire (PNPS and PCMI) of CNRS/INSU (with INC/INP/IN2P3).
MB is a postdoctoral fellow in the University of Virginia’s VICO collaboration and is funded by grants from the NASA Astrophysics Theory Program (grant number 80NSSC18K0558) and the NSF Astronomy \& Astrophysics program (grant number 2206516).
AS gratefully acknowledges support by the Fondecyt Regular (project
code 1220610), and ANID BASAL project FB210003.
\end{acknowledgements}

\nocite{Louvet2024}
\bibliographystyle{aa}
\bibliography{biblio}

\appendix

\section{Technical insights into the PPMAP implementation}\label{appendix:PPMAP}

\begin{figure*}[htb]
   \begin{centering}
      \includegraphics[width=\hsize, trim={0.5cm 8.5cm 9.5cm 1cm},clip]{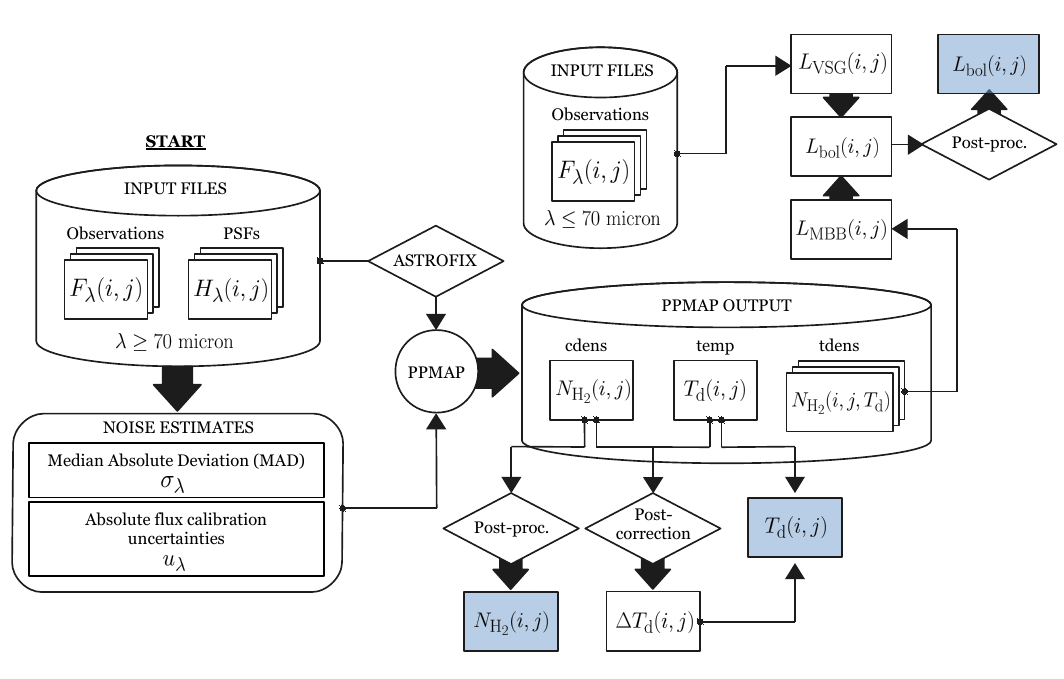}
      \caption{Flow chart of the data preprocessing, PPMAP calculations, and products' post-processing steps applied in this study. \lq \lq Post-correction'' and \lq \lq Post-proc.'' respectively refer to the treatments described in Appendix~\ref{appendix:tempcorr} and Appendix~\ref{appendix:post-processing}. \lq \lq ASTROFIX'' is the Gaussian process regression algorithm described in Appendix~\ref{appendix:desaturation}. \lq \lq VSG'' stands for very small grains. The light blue boxes correspond to the final products released with this study.}
         \label{fig:PPMAPflowchart}
   \end{centering}
\end{figure*} 

In this appendix, we provide a comprehensive overview of the systematic steps employed while using the PPMAP software for the data analysis. The following sections outline the process, starting from the preparation of input continuum maps and point spread functions (PSFs) to the post-correction procedures applied to the PPMAP data products. Fig~\ref{fig:PPMAPflowchart} represents a flow chart of these procedures.

\subsection{Preprocessing}

\subsubsection{Correction of saturated pixels}\label{appendix:desaturation}

Several \textit{Herschel}/PACS and \textit{Herschel}/SPIRE maps exhibit saturation in the regions around the brightest sources, particularly within the peak of the spectral energy distribution (around 160-250~$\upmu$m). As a result, some pixels in these regions contain \lq \lq NaN'' values, rendering the use of PPMAP impossible for these affected maps. To address this issue, we took the following steps:

\begin{enumerate}
    \item In severe cases of saturation (11 out of 124 images in our study), where a significant portion of the map was affected ($\geq 25$\%) and the data quality was compromised, we chose to entirely remove these maps from the pool of input data.

    \item In cases where the saturation level was lower (less than 25\% of the map is saturated), we applied interpolation to estimate the missing values. We used the \texttt{astrofix} Python package \citep{Zhang2021}, that employs Gaussian process regression for this purpose. In this process, the algorithm used the unsaturated areas of the available input images as a training set to learn and determine the missing values, using an optimized Gaussian interpolation kernel (as shown in Fig.~\ref{fig:datacompleteness}, 25 out of 124 images underwent interpolation for the present study, with a median of 6 interpolated pixels per image).
\end{enumerate}

Figure~\ref{fig:datacompleteness} illustrates the extent of the pixels replaced using interpolation in the affected \textit{Herschel} maps.
We acknowledge that interpolating missing pixels, especially in regions of high intensity peaks, can be an uncertain procedure. To quantify the uncertainties associated with this treatment, we performed two iterations of the PPMAP analysis: one with the interpolated maps and the other without any interpolation (excluding the saturated maps from the analysis). By comparing the results between these two iterations, we measured the discrepancies introduced by the interpolation process. These discrepancies were then propagated into the final uncertainty maps released with this study (see Sect.~\ref{subsect:systematic-errors} and Table~\ref{table:systematic-errors}).

\subsubsection{Preparation of point spread functions}

To combine a set of input images with different angular resolutions, PPMAP requires accurate Point Spread Function (PSF) profiles. The current PPMAP version\footnote{\url{https://github.com/ahoward-cf/ppmap}} is shipped with \textit{Herschel} PSFs projected on $256 \times 256$ frames. We replaced these files with new ones projected on a larger $2048 \times 2048$ frame. This adjustment was necessary to ensure correct sampling of both the higher-resolution (ALMA) and lower-resolution (\textit{Herschel}) PSFs while using a single array size for all PSFs, as required by PPMAP. The PSFs we used for \textit{Herschel}/SPIRE (at 250, 350, and 500 $\upmu$m) were downloaded from the \textit{Herschel} science archive\footnote{\url{http://archives.esac.esa.int/hsa/legacy/ADP/PSF/SPIRE/SPIRE-P/}}, while those for \textit{Herschel}/PACS (at 70 and 160 $\upmu$m) were obtained from \citet{Bocchio2016}. \citet{Traficante2011} reported that the beams in the Hi-GAL images are elongated because of the sampling procedure adopted in the Galactic plane survey, with an ellipticity lower than 15\% across all bands.
We produced our own synthetic beam profile for SABOCA, LABOCA, and ALMA observations using the \texttt{Gaussian2DKernel} function from the \texttt{astropy.convolution} package in Python.

\begin{figure}[htb]
   \begin{centering}
      \includegraphics[width=\hsize, trim={1.15cm 4.25cm 9.5cm 4.75cm},clip]{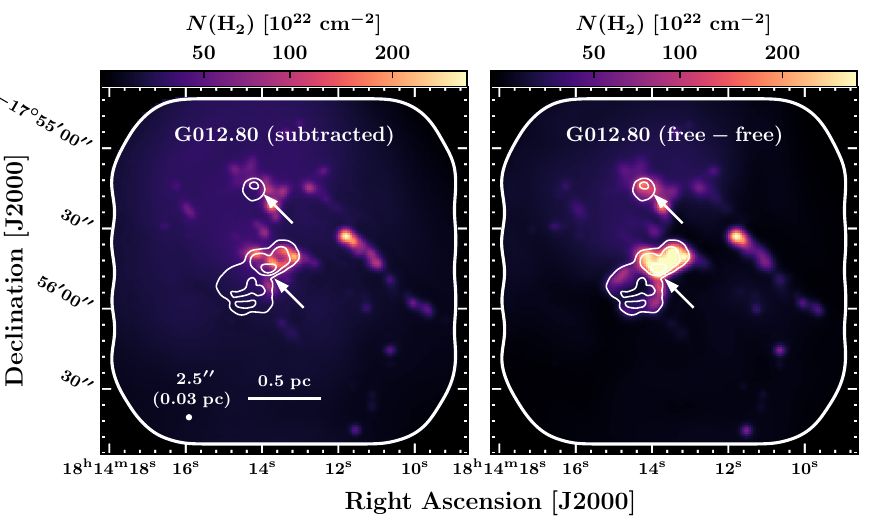}
      \includegraphics[width=\hsize, trim={1.15cm 4.25cm 9.5cm 4.75cm},clip]{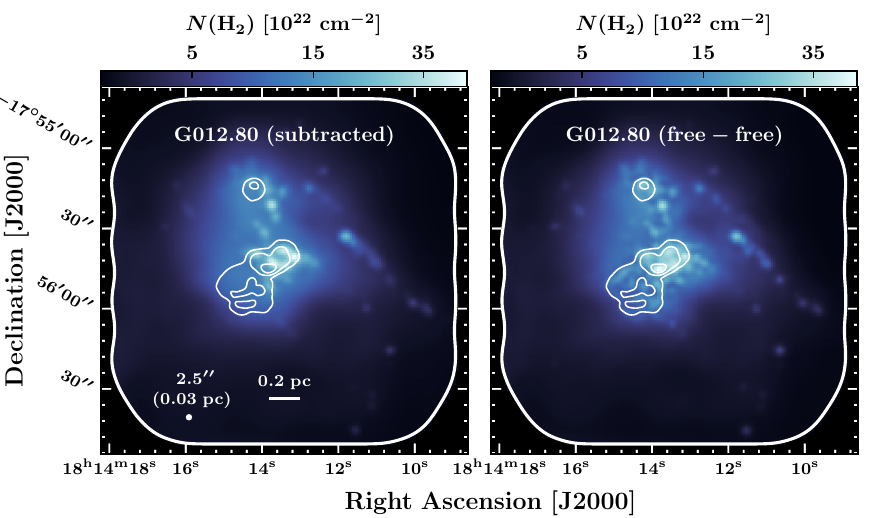}
      \caption{Demonstration of the impact of free-free subtraction, carried out by Galván-Madrid et al. (in prep.), on the PPMAP-derived column density (top panel) and luminosity (bottom panel) maps of G012.80. The left panel displays the map obtained from the subtracted data (Run1), while the right panel depicts the outcomes achieved with the standard ALMA data (Run2). Superimposed white contours illustrate the H41$\alpha$ line emission, that traces regions dominated by free-free emission (contour levels: logarithmically spaced between 0.045 to 0.5 Jy beam$^{-1}$). Arrows indicate the positions of the main \HII regions.}
         \label{fig:free-free}
   \end{centering}
\end{figure}  

\subsubsection{Free-free emission in the ALMA 1.3~mm emission map}\label{sect:free-free}

As introduced in Sect.~\ref{sect:observations}, to mitigate the contamination by free-free emission, we have excluded the ALMA Band 3 data from the PPMAP inputs. In addition, we employed two versions of ALMA Band 6 observations : the \lq \lq native'' version, and a free-free subtracted version, following a process detailed by Galván-Madrid et al. (in prep.). The subtraction method relied on the H41$\alpha$ line emission to disentangle the thermal emission of dust grains from the free-free emission. This was achieved by assuming an electron temperature of $7000$~K and variable helium abundances, depending on the specific region under consideration. Throughout the PPMAP analysis, we either used the free-free subtracted maps to obtain the dust temperature and column density maps (refered to as the PPMAP \lq \lq Run1''), or the native ALMA Band 6 maps to estimate the luminosity (refered to as the PPMAP \lq \lq Run2''). The motivations for these two runs are described in Sect.~\ref{sect:PPMAPanalysis}.
In Fig.~\ref{fig:free-free}, we emphasize the distinctions in the PPMAP-derived column density between the Run1 (without free-free emission) and Run2 (with free-free emission) outcomes.
While the overall column density within the ALMA footprint registers a mere 12.5\% increment between the two instances, the column density estimate is significantly enhanced toward localized areas (up to a factor 20). This increase is particularly pronounced around the G012.80 Main-Central and Main-North \HII regions as outlined by \citet{Armante2024}. Conversely, the impact of the free-free removal process on the Main-West filament is minimal (relative difference below 10\%), reinforcing its classification as a primarily dusty filament. The disparity between Run1 and Run2 underscores that the free-free subtraction is necessary to precisely estimate the column density in evolved regions.

\subsection{PPMAP SED fitting}

\subsubsection{Determination of measurement errors}\label{sect:noise}

As discussed in Sect.~\ref{sect:PPMAPdescription}, the PPMAP Bayesian fitting procedure necessitates accurate error estimates, in order to attribute a correct weight to the input observations. We explored three distinct methods to determine the noise:
\begin{enumerate}
    \item First method (applied to ALMA, LABOCA, SABOCA observations): Assuming a Gaussian distribution for the sky noise, we conducted a Gaussian fit of the data histograms to extract the standard deviation (referred to as $\sigma_1$).

    \item Second method (applied to \textit{Herschel} observations): due to the high signal-to-noise sky background associated with \textit{Herschel} maps, the first method was not applicable. Instead, we employed the built-in PPMAP noise estimate function, \texttt{getnoise}, by setting the \texttt{sigobs} input parameter to 0 in the \texttt{premap} routine \citep{Marsh2015}. Using Gaussian kernels of variable scales, this function facilitates the measurement of sky noise through iterative subtraction of a smoothed background (referred to as $\sigma_2$).

    \item Finally, in a third attempt to estimate the noise, we calculated the median absolute deviation (referred to as \lq \lq MAD'', and denoted as $\sigma_3$) in empty areas of the observed fields (searching for such empty areas in a 30$^\prime$ radius around the \textit{Herschel} field center coordinates). MAD is a robust statistic that provides an estimate of the image noise, based on the median of the absolute deviations from the median pixel value in the image \citep{Leys2013}. 
\end{enumerate}

Based on our investigation, when applied to ALMA, LABOCA, and SABOCA observations, all three noise measurement methods exhibited a general agreement, with relative differences within the range of 30-60\% when compared to $\sigma_2$ (the noise estimate obtained from the PPMAP built-in subroutine). The specific values of these relative differences depended on the regions being analyzed.
We always observed that $\sigma_2 > \sigma_1 > \sigma_3$, indicating that the third method consistently yields lower noise estimates compared to the other two methods. This outcome is particularly beneficial since it effectively increases the weight of the 1.3~mm image, thus allowing a better representation of the low signal-to-noise features in the observed protoclusters (such as faint cores).

Therefore, we incorporated the median absolute deviation measurement, along with the systematic error on the absolute flux calibration of the sky background $F_\mathrm{\lambda}^\mathrm{sky}$, to estimate the total errors for each input map at a given wavelength $\lambda$ (e.g., $\sigma_{70\mathrm{\upmu m}}$, $\sigma_{350\mathrm{\upmu m}}$, $\sigma_{870\mathrm{\upmu m}}$, $\sigma_{1.3\mathrm{mm}}$, etc). The error estimates $\sigma_\lambda$ were computed as follows:

\begin{equation}\label{eq:error-estimate}
\sigma_\lambda = \mathrm{MAD}(F_\lambda) + u_\lambda F_\mathrm{\lambda}^\mathrm{sky} ,
\end{equation}
where, MAD$(F_\lambda)$ represents the median absolute deviation measured in the map $F_\lambda$, and $u_\lambda$ denotes the absolute flux calibration error (e.g., 15\% for LABOCA, as discussed in Sect.~\ref{sect:observations}). The resulting values of $\sigma_\lambda$ were subsequently assigned to the input parameter \texttt{sigobs} for the corresponding wavelengths. The relative contribution of the two terms in Eq.~(\ref{eq:error-estimate}) varies significantly depending on the map. In typical low signal-to-noise pixels (dominated by Gaussian noise in the ALMA observations), the error estimates for \textit{Herschel} maps are primarily dominated by the sky background component, while the MAD component is the largest contributor to the ALMA Band 6 errors.

\subsubsection{\textit{Herschel} color correction}\label{appendix:colorcorrection}

We applied color corrections to the \textit{Herschel}/PACS and SPIRE observations (see \citealt{Balog2014}; \citealt{Valtchanov2018} and references therein). These corrections are taken directly from the PACS and SPIRE user manuals, and consider the observed flux in each band as an integral over the product of the instrumental bandpass and the source's spectrum, assuming a MBB profile with $\beta = 2$. The correction tables used in our analysis are consistent with the correction factors used in \citet{Motte2018b} for the PPMAP analysis of W43-MM1. These color corrections are provided to PPMAP as input tables and then automatically applied during the analysis. While it is ideally more accurate to include the color corrections, their effect on the PPMAP outcome is minimal ($< 1$\% in relative variation of the column density and temperature estimates) and can be considered negligible in our analysis, with respect to other sources of uncertainties.

\subsubsection{PPMAP parameters}

\begin{table}[htb]
\centering
{\caption{PPMAP input parameters.}       
\label{table:PPMAPfixedparams}}      
{\centering                          
\begin{tabular}{l l c c c c}        
\hline \hline         \\[-1.0em]
\footnotesize{Input parameter} & \footnotesize{Description} & \footnotesize{Adopted value} \\
\hline \\[-1.0em]
\footnotesize{\texttt{pixel}} & \footnotesize{sampling ($^{\prime \prime}$)} & \footnotesize{1.25 (a)}\\
\footnotesize{\texttt{ncells}} & \footnotesize{cell size (px)} & \footnotesize{60 (a)}\\
\footnotesize{\texttt{noverlap}} & \footnotesize{cell overlap (px)} & \footnotesize{40, 50 (a)}\\
\footnotesize{\texttt{kappa}300} & \footnotesize{ref. opacity (cm$^2$ g$^{-1}$)} & \footnotesize{0.1 (b)}\\
\footnotesize{\texttt{eta}} & \footnotesize{$\eta$ (\textit{a priori} dilution)} & \footnotesize{0 (c)}\\
\footnotesize{\texttt{maxiterat}} & \footnotesize{maximum nb. of iterations} & \footnotesize{10~000}\\
\footnotesize{\texttt{beta}} & \footnotesize{$\beta$ (dust opacity)} & \footnotesize{1.8}\\
\footnotesize{\texttt{Nt}} & \footnotesize{nb. of MBB components} & \footnotesize{8}\\
\footnotesize{\texttt{temprange}} & \footnotesize{$T_\mathrm{dust}$ (K)} & \footnotesize{10 -- 50}\\
\footnotesize{\texttt{sigobs}} & \footnotesize{measurement error} & \footnotesize{see Sect.~\ref{sect:noise}}\\
\footnotesize{\texttt{highpass}} & \footnotesize{high-pass filtering} & \footnotesize{\lq \lq yes'' (d)}\\
\hline
\end{tabular}}
\\~\\
\raggedright
\footnotesize{
The parameters listed above are entered in the input file \texttt{premap.inp}. The reference opacity at 300 $\upmu$m is defined with respect to the total mass \citep{Marsh2015}.\\
(a) The maps are Nyquist-sampled with pixels of size 1.25$^{\prime \prime}$, hence the synthetic resolution is 2.5$^{\prime \prime}$. \texttt{noverlap} $\geq 0.5$ \texttt{ncells} is recommended by the PPMAP user manual. With \texttt{ncells} $=60$ and \texttt{noverlap} $=40$, all ALMA-IMF regions are contained either in a $3 \times 3$ or $5 \times 5$ mosaic. In some cases we increased \texttt{noverlap} to 50 in order to enhance the quality of the subfields seaming, thereby using $7 \times 7$ mosaics. \\
(b) The reference opacity $\kappa_{300}$ is defined by Eq.~(\ref{eq:Kappa300}).\\
(c) The \textit{a priori} dilution input parameter, \lq \lq \texttt{eta}'', characterizes the degree to which the procedure is forced to represent the data with the least number of \lq \lq points'’ (that is, with the least number of MBB components with distinct dust temperatures $T_\mathrm{dust}$ ). Values in the range 0.01 -- 1 are recommended, but setting the input to \lq \lq 0'' causes an internal PPMAP subroutine to adopt a default value, defined as $\eta = N_\mathrm{prior} / N_\mathrm{states}$, where $N_\mathrm{prior}$ is the maximum number of components that the data may constrain, and $N_\mathrm{states} = N_X \times N_Y \times N_T \times N_\beta$ is the number of states that a \lq \lq point'' can occupy (see Sect.~\ref{sect:PPMAPdescription}).\\
(d) Enables a PPMAP subroutine to subtract mean value from model and ground-based observed images during the SED fitting process.\\
}\end{table}

The fixed PPMAP parameters adopted in our initial analysis are shown in Table~\ref{table:PPMAPfixedparams}. Following previous applications of PPMAP (e.g., \citealt{Marsh2017}; \citealt{Howard2019}; \citealt{Whitworth2019}; \citealt{Chawner2020}; \citealt{Howard2021}), we assumed that we can reproduce the observations in the FIR to submm (70 $\upmu$m to 1.3 mm) with 8 MBB components between 10~K and 50~K (using a logarithmic spacing). We have systematically checked that allowing PPMAP to employ lower or higher temperature components does not provide a substantial change in column density $N_\mathrm{H_2}$ and luminosity (within $\pm$5\%).
Ground-based input images inherently suppress low spatial frequencies. To address this, we enabled the input parameter \texttt{highpass} (see note on Table~\ref{table:PPMAPfixedparams} footnote (d)), allowing PPMAP to subtract a constant background from the internal model images. This subtraction is applied at wavelengths corresponding to the ground-based data (SABOCA, LABOCA, and ALMA). The goal is to enhance the alignment between the model and the spatial characteristics of these particular observations, ultimately improving the fitting process.
We adopted a standard reference opacity $\kappa_{300}=0.1~\mathrm{cm^2g^{-1}}$, defined with respect to the total mass and corresponding to a gas-to-dust ratio of 100 (\citealt{Marsh2015}; \citealt{Hildebrand1983}). Based on previous studies (\citealt{Planck2014}; \citealt{Juvela2015}; \citealt{Koehler2015} and references therein for more diffuse gas), in a first approach we fixed the dust opacity to $\beta = 1.8$ for all regions. If we were to assume $\beta = 2$, the estimated dust temperatures would likely be lower (see Table~\ref{table:systematic-errors}). Finally, to balance between our high-resolution requirements and potential artifact-inducing effects, we adopted a synthetic angular resolution of 2.5$^{\prime \prime}$ for the analysis. In fact, extensive testing of PPMAP indicated that further reducing the angular resolution could lead to stronger ring-like artifacts (see Appendix~\ref{appendix:post-processing}).

\subsubsection{Temperature and opacity index optimization}
Based on our first series of PPMAP runs, performed with a fixed temperature range ($T_\mathrm{dust} = 10-50$~K, using 8 MBB components) and fixed opacity index ($\beta=1.8$), we refined the input temperature range. From the temperature distributions measured in the zero-order PPMAP output, we systematically lowered the amplitude of the temperature range (e.g., from $T_\mathrm{dust} = 10-50$~K to $T_\mathrm{dust} = 21-39$~K in G012.80). This adjustment allowed us to obtain more precise temperature estimates for the sources, by decreasing the $\Delta T$ interval between the temperature components.
Additionally, we explored the possibility of making the opacity index, $\beta$, a free parameter in the PPMAP analysis. To achieve this, we conducted iterative runs, modifying the value of $\beta$ in increments of $\Delta \beta = 0.1$, while measuring the corresponding change in chi-square ($\chi^2$). Following this procedure, we found a best-fit value of the opacity index, that could be different from $\beta=1.8$ (between $\beta=1.4$ and $\beta=2.1$, depending on the specific fields of observation). However, we acknowledge that the degeneracy between the temperature and opacity index makes our attempt at simultaneously constraining them uncertain.
The variations in $\chi^2$ associated with the modification of $\beta$ were negligible compared to the expected $\chi^2$ variation at a $1 \sigma$ level. More precisely, for any $\beta_i$ and $\beta_j$, we found that $\Delta \chi^2 (\beta_i, \beta_j) \ll [(1 + 1 / (N - p)] \mathrm{min}(\chi^2)$, where $N$ is the number of data points used to fit the modified blackbody and $p$ is the number of parameters. As a result, we decided to fix $\beta = 1.8$ for all regions, and focused on the determination of the temperature. We acknowledge the potential uncertainty in the opacity index, that may differ significantly from the assumed value of $\beta = 1.8$ across the ALMA-IMF protoclusters. To explore the impact of different opacity indices, we assessed variations in the PPMAP-derived temperature. For instance, we observed average changes of $\Delta T_\mathrm{dust} = + 3$~K and $\Delta T_\mathrm{dust} = + 12$~K when adopting $\beta = 1.5$ and $\beta = 1$ (representative of more evolved dust in denser media), respectively.

\subsubsection{Criteria to stop PPMAP iterations}\label{appendix:convergence}
The quality of PPMAP outputs underwent a systematic validation process through an a posteriori chi-square test, where we manually compared the synthetic fluxes obtained from the PPMAP model to the actual observed fluxes.
Based on this a posteriori validation of the product's reliability, we monitored the convergence of the PPMAP iterative process to prevent over-fitting, that can lead to increasingly severe ring-like artifacts in the column density and temperature maps (Sect.~\ref{appendix:post-processing}). We refer to this particular effect as \lq \lq over-fitting'',  since the associated change in synthetic flux is minimal with respect to measurement errors. We systematically controlled for and prevented over-fitting through minor tweaks of the PPMAP iteration number, while validating the outcome through the aforementioned chi-square test. These adjustments were necessary to strike a balance and ensure the best representation of the observed data without introducing excessive artifacts. Most importantly, we note that although it does mitigate the growth of the spurious features, this validation process does not prevent their development.

\begin{figure}[htb]
   \begin{centering}
      \includegraphics[width=0.95\hsize, trim={0.5cm 0.5cm 1.5cm 2cm},clip]{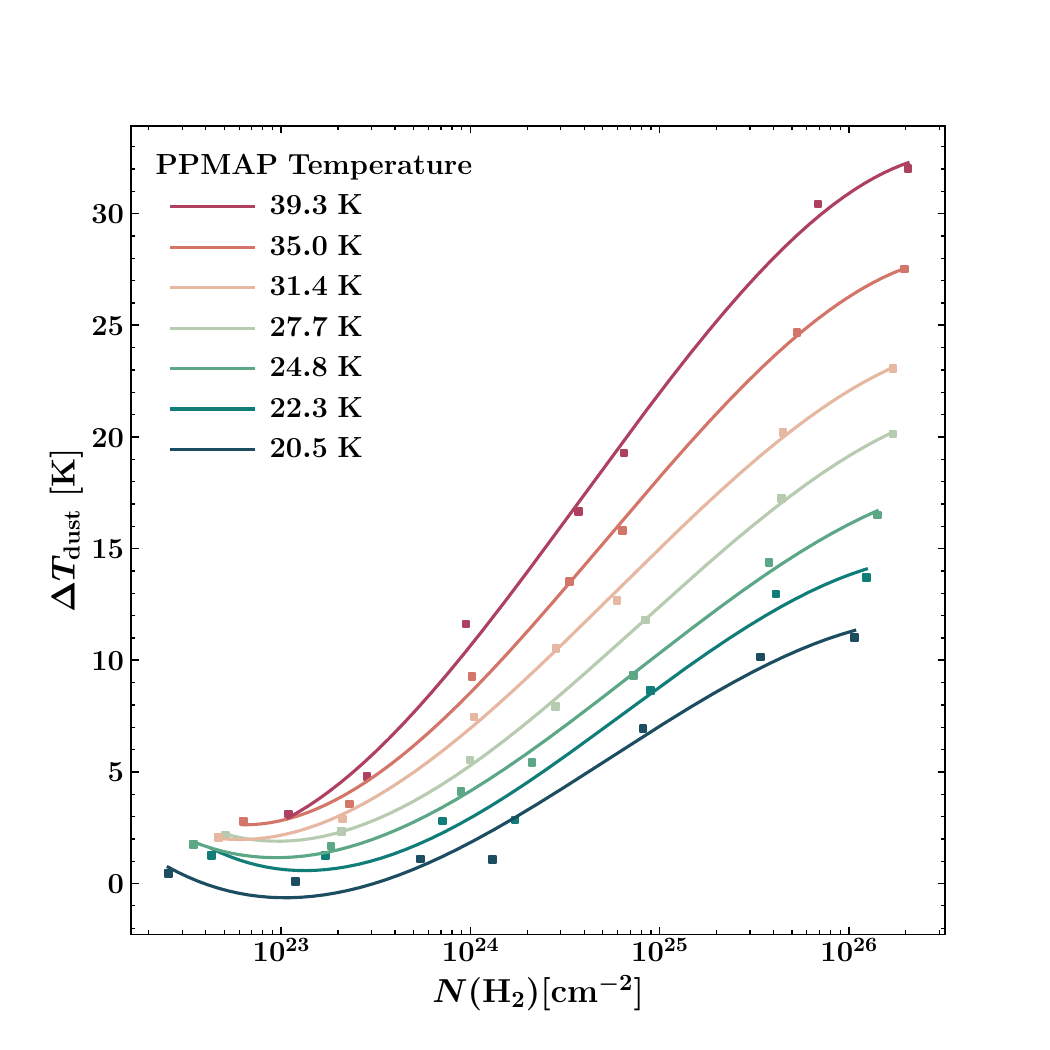}
      \caption{Temperature correction $\Delta T_\mathrm{dust}$ as a function of the PPMAP column density $N(\mathrm{H_2})$. The correction is shown for different input temperatures, spanning 20.5~K to 39.3~K. Data points are interpolated using cubic splines. N.B.: 99\% of the PPMAP-derived column densities are below $8 \times 10^{23}$~cm$^{-2}$.}
         \label{fig:tcorr}
   \end{centering}
\end{figure}  

\subsection{Dust temperature post-correction}\label{appendix:tempcorr}

PPMAP relies on the assumption that the dust emission is optically thin. However, this assumption is not met in regions with high-density spikes, such as cores in high-mass star-forming regions. To account for the effect of larger optical depths, we used a procedure established in previous work (first applied by \citealt{Motte2018b} in the analysis of W43-MM1) to correct the dust temperature maps generated by PPMAP.
The temperature correction table used in this procedure was constructed based on a grid of model sources with varying temperatures ($T_\mathrm{dust}^\mathrm{model}$), optical depths, sizes, and measurement errors. The synthetic sources were represented by the simplest model possible: a uniform dust slab with specified temperature and optical depth. Using this simple dust model, we produced synthetic flux density maps between 70~$\upmu$m and 3~mm. Then, we applied PPMAP to the synthetic flux density maps to provide an estimate for the column density $N(\mathrm{H_2})$ and temperature $T_\mathrm{dust}^\mathrm{PPMAP}$. Using different dust model parameters, we sampled the relationship $\Delta T_\mathrm{dust} = f(N(\mathrm{H_2}), T_\mathrm{dust}^\mathrm{PPMAP})$ between the temperature bias $\Delta T_\mathrm{dust} = T_\mathrm{dust}^\mathrm{PPMAP} - T_\mathrm{dust}^\mathrm{model}$, the PPMAP-derived column density $N(\mathrm{H_2})$ and the PPMAP-derived dust temperature $T_\mathrm{dust}^\mathrm{PPMAP}$. This investigation enabled the implementation of an a posteriori correction. The established relationship is depicted in Fig.~\ref{fig:tcorr}, covering the initial range of explored temperature and column densities. In fact, the original correction table provided by \citealt{Motte2018b} is constrained within certain limits, insufficient to accommodate our PPMAP products. To extend the correction beyond these limits, we performed an extrapolation using cubic splines while assessing the quality and reliability of the extrapolated results.
Subsequently, we applied the extrapolated correction table to the PPMAP outputs a posteriori to account for the effect of optical depth on the dust temperature estimates. In each pixel of the PPMAP outputs, temperatures are updated using the tabulated value $\Delta T_\mathrm{dust} = f(N(\mathrm{H_2}), T_\mathrm{dust})$. The impact of the correction on the column density is considered negligible, since the added warm dust represents a minor contribution to the total mass.
It is essential to note that this correction is based on a simplified model and may not be fully suitable for the diverse environments observed in the ALMA-IMF fields, including \HII regions, internally heated protostellar envelopes and prestellar cores (Motte et al., in prep.). Despite its limitations, this correction helps compensate for the assumption of optical thinness of PPMAP and should be considered when aiming to constrain the temperature in regions with potential optical depth effects such as ALMA-IMF protoclusters.

\subsection{PPMAP ring-like artifacts and post-processing}\label{appendix:post-processing}

\begin{figure}[htb]
   \begin{centering}
      \includegraphics[width=\hsize, trim={1.75cm 1.75cm 1.75cm 8.5cm},clip]{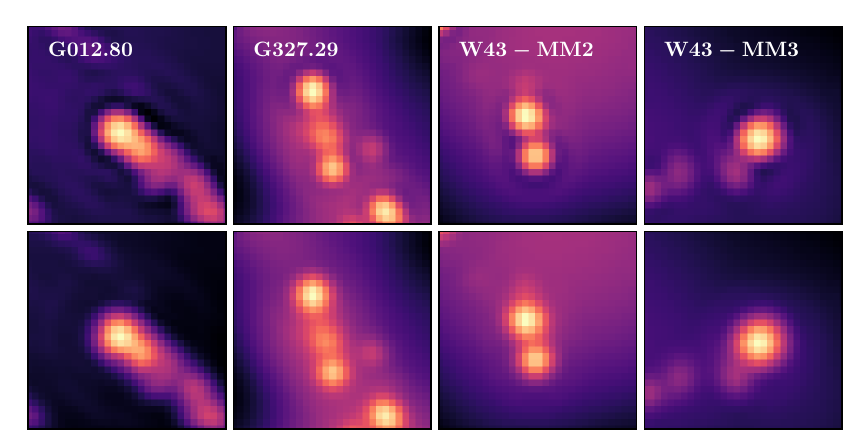}
      \caption{Zooming in on characteristic ring-like artifacts, we present four column density maps (G012.80, G327.29, W43-MM2, and W43-MM3) in a left-to-right sequence. The top panel illustrates the initial PPMAP outputs, while the bottom panel displays the maps after undergoing post-processing as outlined in Appendix~\ref{appendix:post-processing}. The color scale is logarithmic, to enhance visibility.}
         \label{fig:artifacts}
   \end{centering}
\end{figure}

Temperature and column density maps generated with PPMAP often exhibit ring-like artifacts, as shown in Fig.~\ref{fig:artifacts}. These artifacts consistently appear around sources of significant brightness, and their intensity grows with the brightness of the source they are associated with. They generally occur either as complete or partial spurious rings around a source, and in the most extreme case are accompanied by secondary peaks (cf. Fig.~\ref{fig:artifacts}, first panel).
The underlying cause of these ring-like artifacts may be attributed to several inter-twinned factors. Firstly, PSF errors and missing spatial frequencies between the ALMA images and the rest of the observations could produce these artifacts. Secondly, due to the strong dependence of emission on temperature, PPMAP may fail to reproduce steep temperature gradients around bright sources, thus causing ring-like artifacts. Lastly, the violation of PPMAP's assumption of optically thin emission toward spikes in column density may also contribute to errors \citep{Howard2020}. As reported in Appendix~\ref{appendix:convergence}, elevating the number of PPMAP iterations beyond a certain threshold amplifies these spurious patterns. This emphasizes the importance of accurately assessing measurement errors and controlling the number of iterations to prevent over-fitting.
Most importantly, the presence of ring-like artifacts may introduce a bias in determining the temperatures of cores. To gauge the extent of this effect, we quantified the discrepancy between our actual maps and PPMAP outputs resulting from an  increased number of iterations ($N_\mathrm{iter} = 25~000$). This yielded an estimated upper limit of $\pm 1$~Kelvin for potential temperature biases stemming from these artifacts (cf. Table~\ref{table:systematic-errors}). This estimate of $\pm 1$~Kelvin effectively corresponds to the maximum value of the temperature variation measured between the crest and anti-crest of the artifacts in the temperature maps.

We implemented a conservative post-processing treatment to improve the quality of the PPMAP column density and luminosity products, specifically to reduce spurious features around high-density peaks. The post-processing method is defined as follows:
\begin{enumerate}
    \item We first convolved the original map with a Gaussian kernel of size 1.25$^{\prime \prime}$ (equivalent to 1 pixel).
    \item Next, we flagged pixels $(i, j)$ that met two specific conditions, identifying them as potentially spurious and in need of correction:
    \begin{equation}
    \begin{split}
    &(a)~\frac{\lvert P(i, j) - P'(i, j) \lvert}{P(i, j)} \geq t , \\
    &(b)~\mathrm{B6}(i, j) \leq \sigma_\mathrm{MAD} ,
    \end{split}
    \end{equation}
    where $P$ and $P'$ are respectively the PPMAP output (column density, temperature and luminosity maps) and its convolved counterpart, $\mathrm{B6}$ is the ALMA 1.3~mm map smoothed at the PPMAP's angular resolution of 2.5$^{\prime \prime}$, $\sigma_\mathrm{MAD}$ is the median absolute deviation measured in $\mathrm{B6}$, and $t = 0.05$ is a suitable threshold.
    \item The flagged pixels were then corrected by replacing their values with the corresponding values from the convolved product ($P'(i, j)$).
\end{enumerate}
The post-processing approach smoothed out artifacts selectively, ensuring that only spurious pixels were modified. Pixels associated with statistically significant signal, as measured in the ALMA B6 continuum map, remained unchanged. The result of post-processing is shown in Fig.~\ref{fig:artifacts} for a small sample of hand-selected artifacts.

\subsection{Resampling}\label{appendix:resampling}
Finally, we resampled the PPMAP column density and luminosity products onto new grids using the \citealt{Deforest2004} adaptive, anti-aliased resampling algorithm implemented through the \texttt{reproject\_adaptive} routine from the \texttt{Python} package \texttt{reproject}. In this step, the sampling of the beam is increased from 2 pixels per beam to 3 pixels per beam (that is, the pixel size is lowered from 1.25$^{\prime\prime}$ to 0.83$^{\prime\prime}$). The main motivation for this increased sampling was to achieve better source detection by the \textsl{getsf} algorithm (see Sect.~\ref{sect:GETSF}). The adaptive resampling process leads to minimal map variations, typically around 1-2\% on average, and potentially a few Kelvin toward warm sources in the temperature maps. To address this limitation, we employed the \texttt{reproject\_exact} routine on the PPMAP temperature maps. This approach enabled us to conserve temperatures at the positions of point-like sources, such as cores and protostars, albeit at the expense of introducing (negligible) aliasing artifacts.

\section{Supplementary tables and figures}\label{appendix:supplementary}

\begin{figure*}[htb]
   \begin{centering}
      \includegraphics[width=\hsize, trim={0.25cm 4.25cm 1cm 4.75cm},clip]{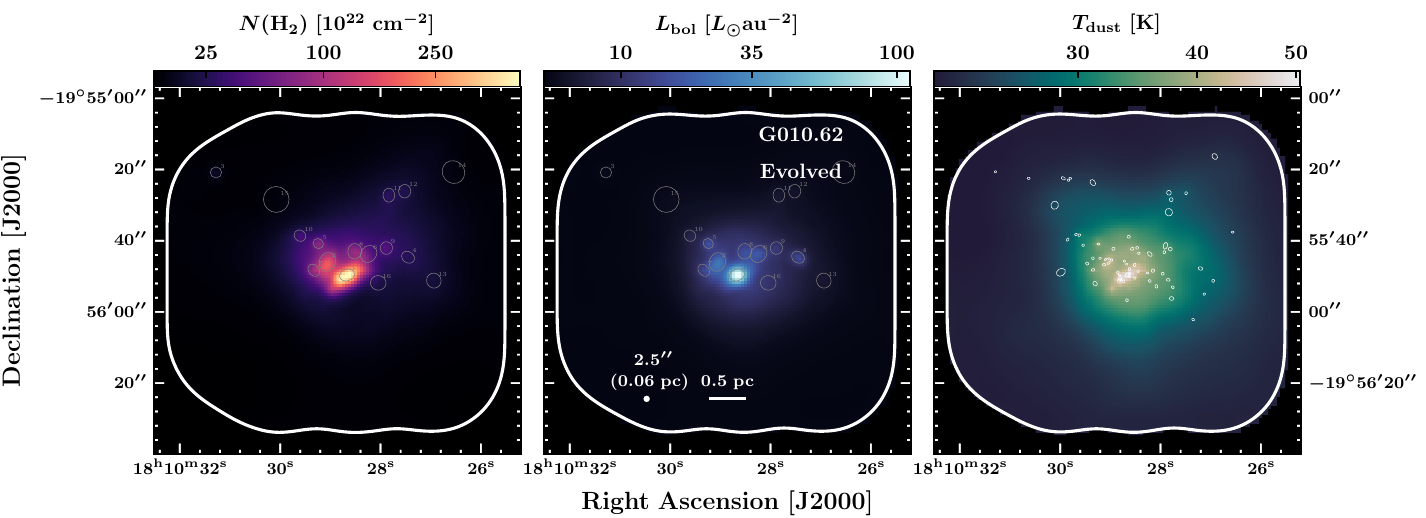}
      \includegraphics[width=\hsize, trim={0.25cm 4.25cm 1cm 4.75cm},clip]{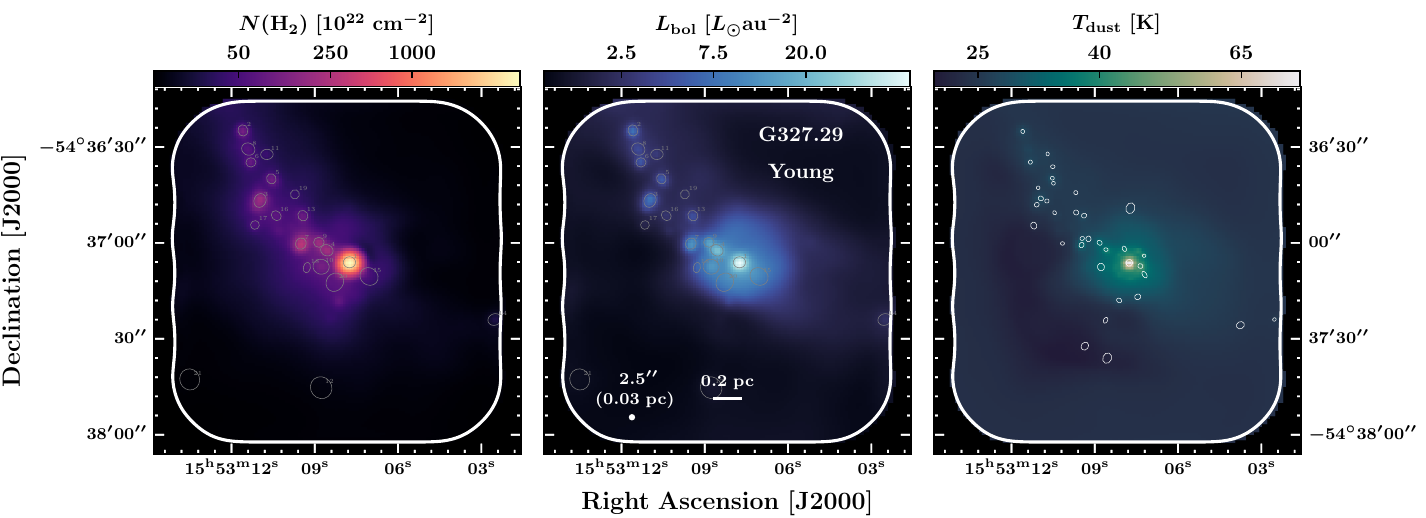}
      \includegraphics[width=\hsize, trim={0.25cm 4.25cm 1cm 4.75cm},clip]{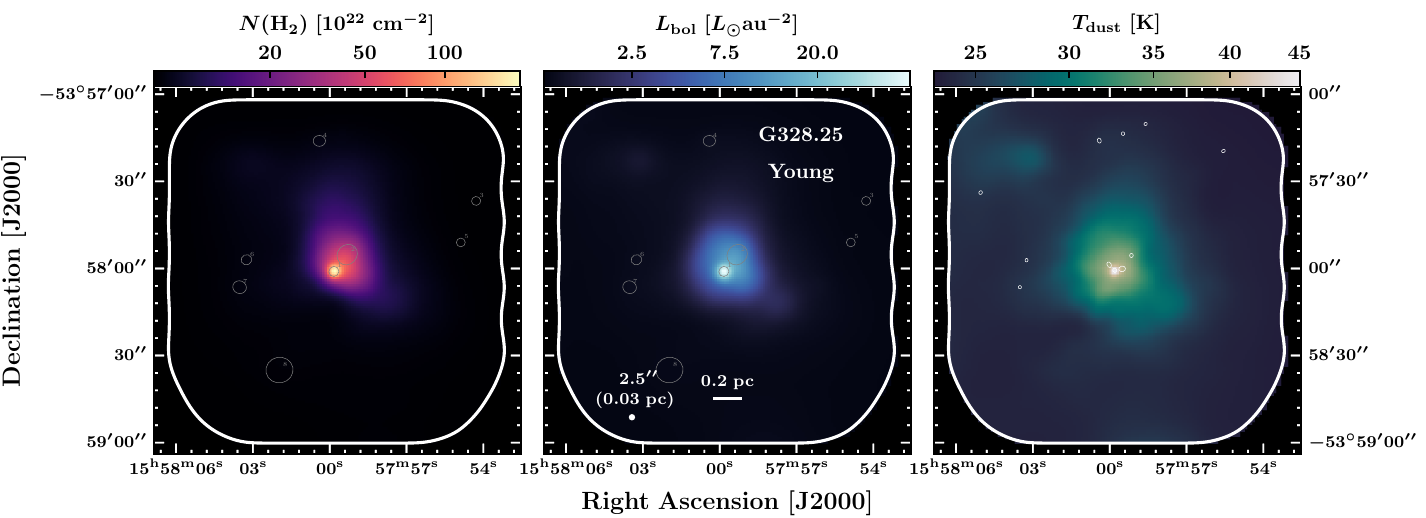}
      \caption{PPMAP products illustrated for the remaining protoclusters. From left to right: column density map ($N(\mathrm{H_2}$)), bolometric luminosity ($L_\mathrm{bol}$), corrected dust temperature ($T_\mathrm{dust}$). White continuous contours outline the ALMA 1.3~mm mosaic areas. The luminosity peaks extracted from the PPMAP luminosity maps (see Sect.~\ref{sect:GETSF}) are overlaid in gray. The continuum cores identified by \citet{Louvet2024} in the ALMA 1.3~mm images are overlaid in white. The size of the ellipses reflects the FWHM of the sources.}\label{fig:PPMAPs-supplementary}
   \end{centering}
\end{figure*}

\begin{figure*}[htb]\ContinuedFloat
   \begin{centering}
      \includegraphics[width=\hsize, trim={0.25cm 4.25cm 1cm 4.75cm},clip]{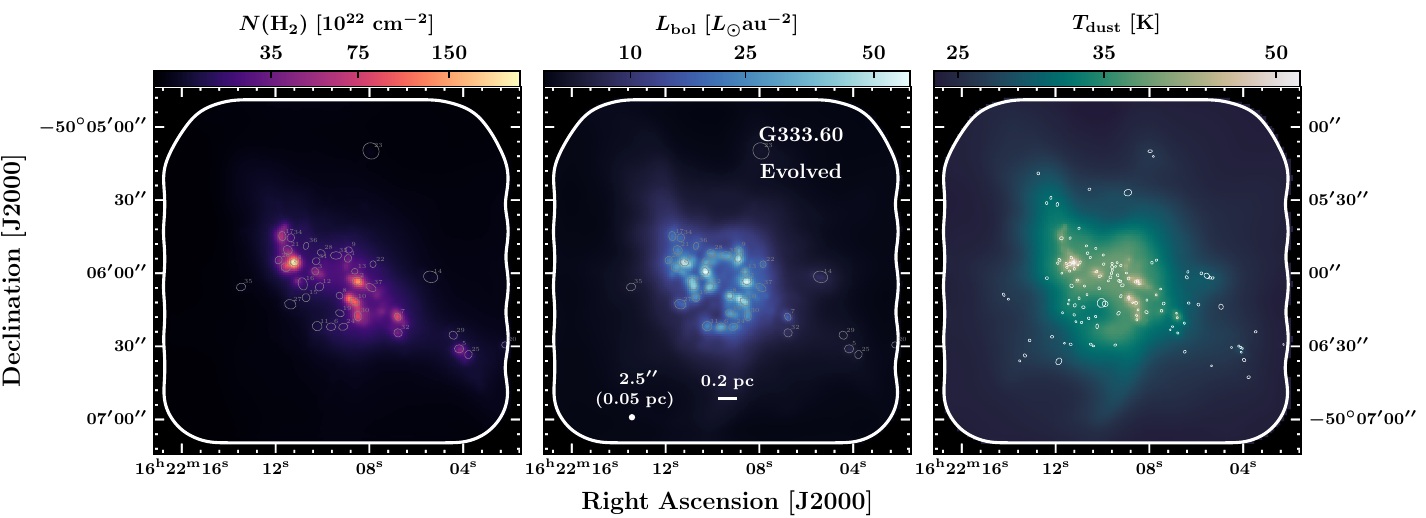}
      \includegraphics[width=\hsize, trim={0.25cm 4.25cm 1cm 4.75cm},clip]{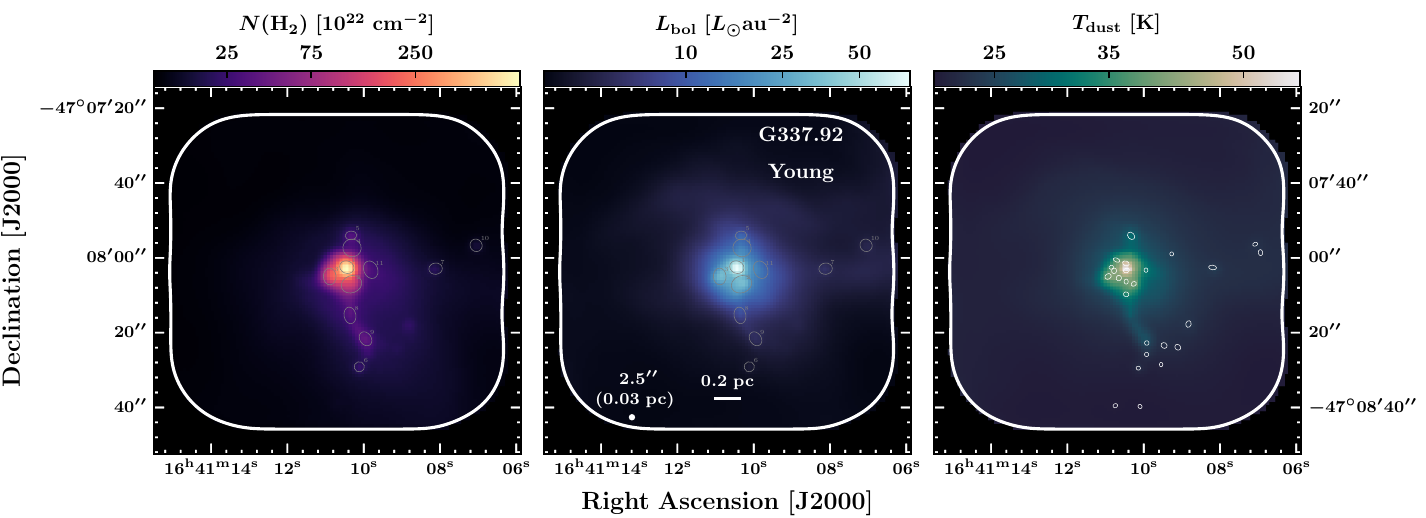}
      \includegraphics[width=\hsize, trim={0.25cm 4.25cm 1cm 4.75cm},clip]{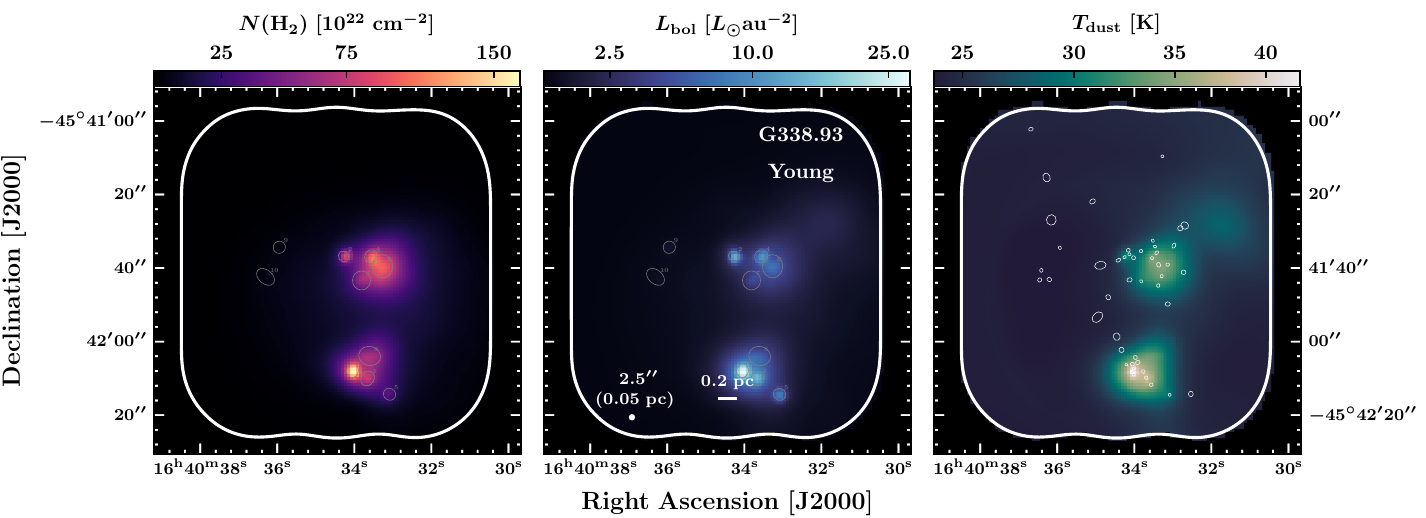}
      \caption{(Continued)}
   \end{centering}
\end{figure*}  

\begin{figure*}[htb]\ContinuedFloat
   \begin{centering}
      \includegraphics[width=\hsize, trim={0.25cm 4.25cm 1cm 4.75cm},clip]{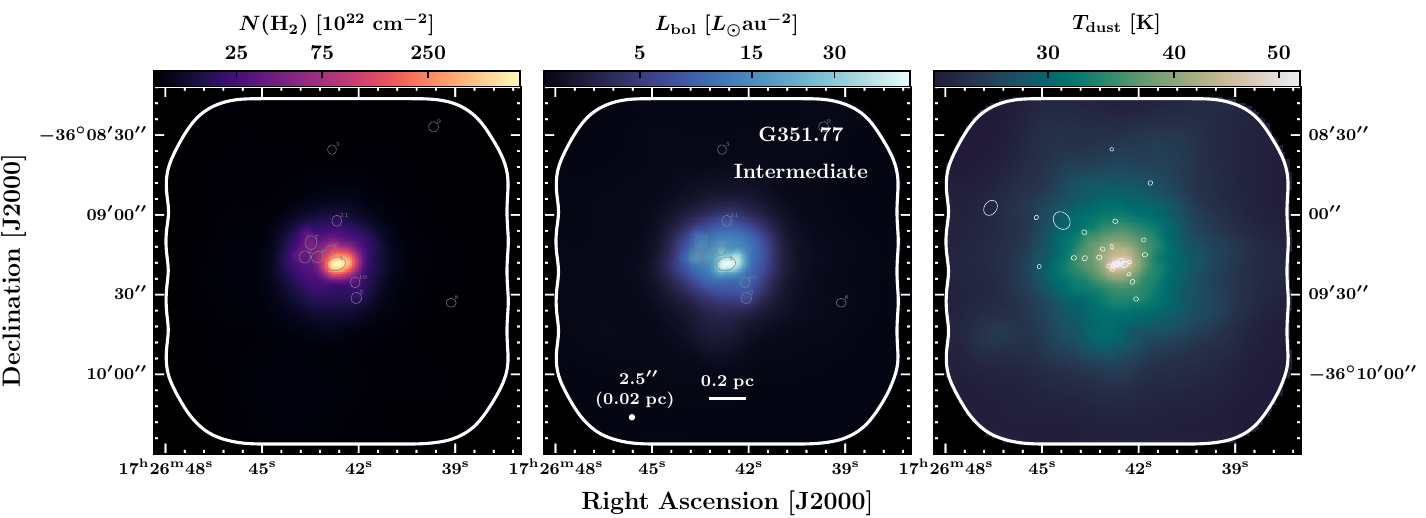}
      \includegraphics[width=\hsize, trim={0.25cm 4.25cm 1cm 4.75cm},clip]{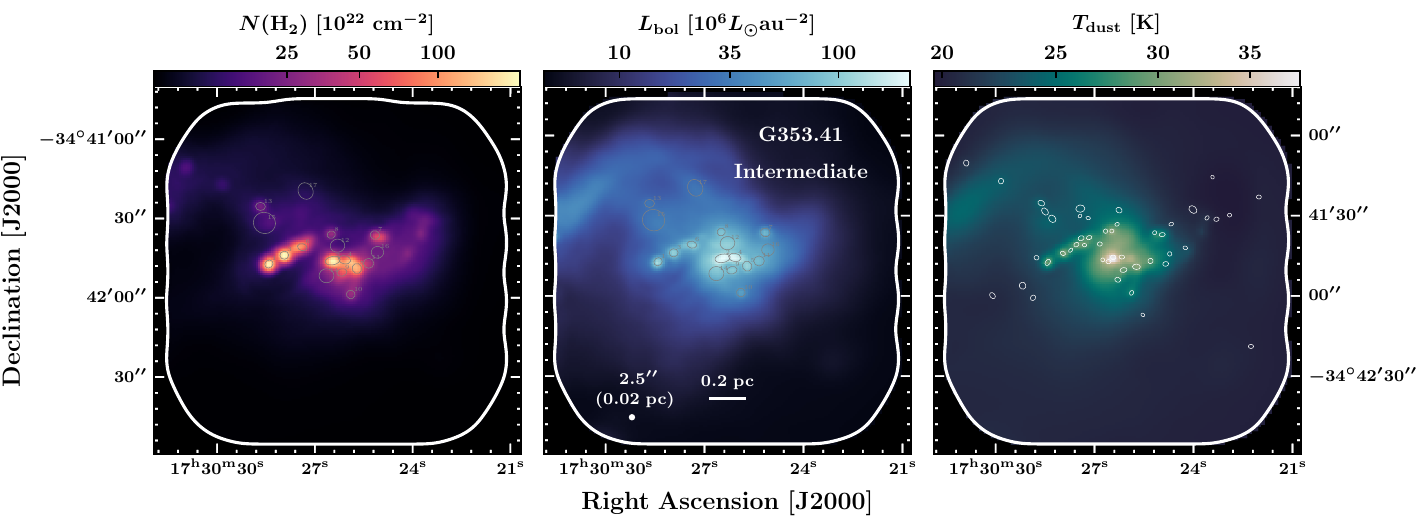}
      \includegraphics[width=\hsize, trim={0.25cm 4.25cm 1cm 4.75cm},clip]{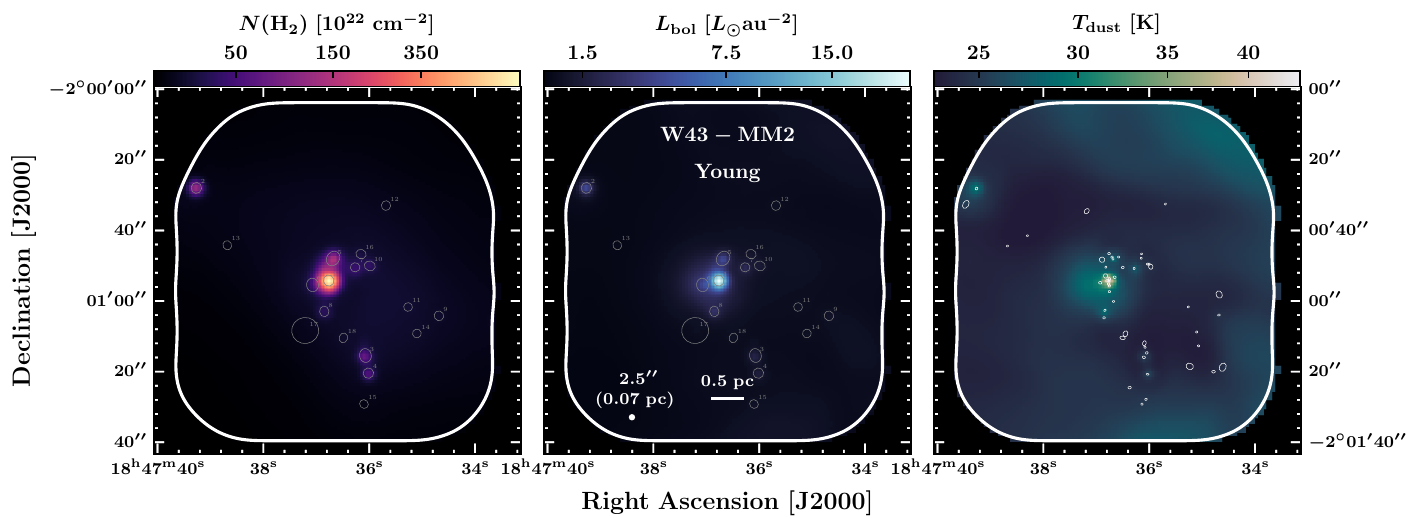}
      \caption{(Continued)}
   \end{centering}
\end{figure*}  

\begin{figure*}[htb]\ContinuedFloat
   \begin{centering}
      \includegraphics[width=\hsize, trim={0.25cm 4.25cm 1cm 4.75cm},clip]{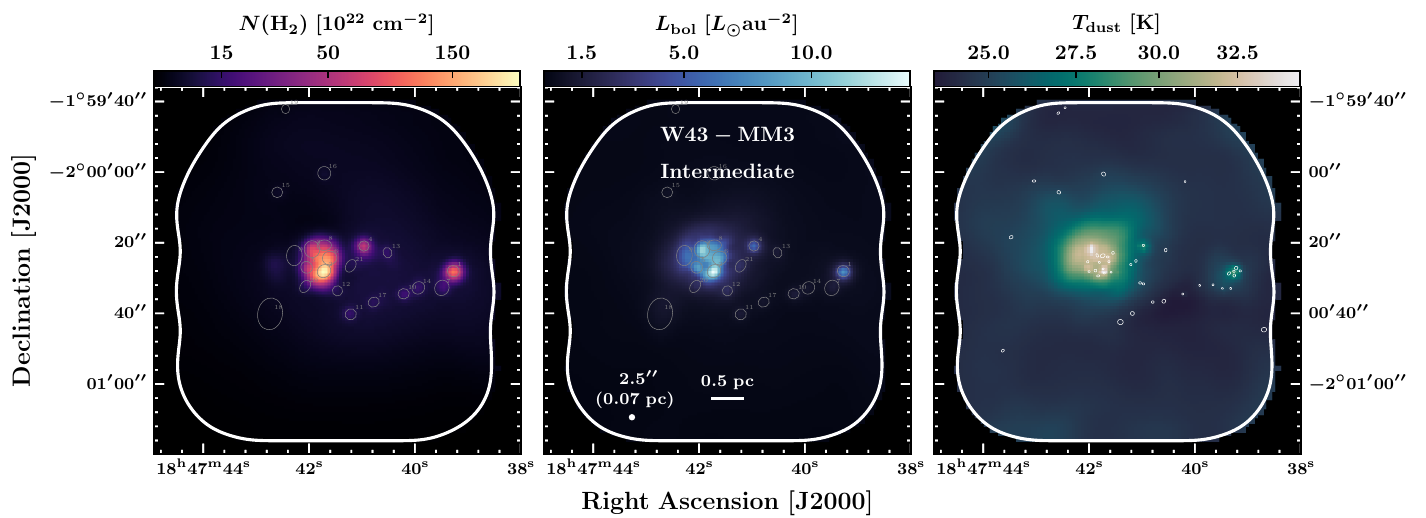}
      \includegraphics[width=\hsize, trim={0.25cm 4.25cm 1cm 4.75cm},clip]{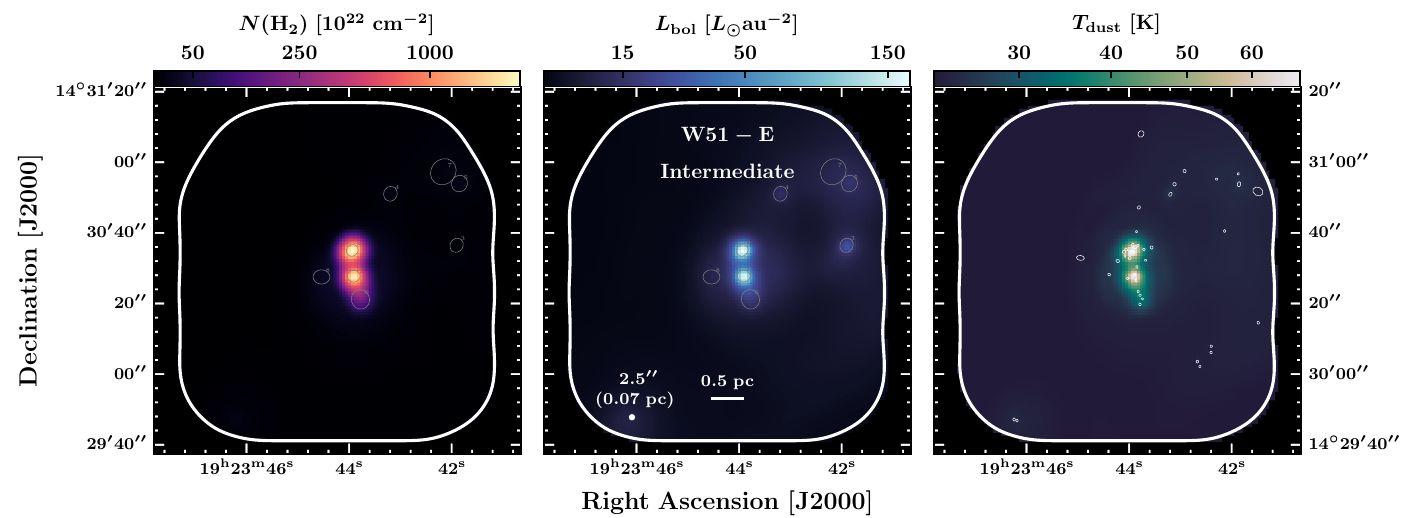}
      \includegraphics[width=\hsize, trim={0.25cm 4.25cm 1cm 4.75cm},clip]{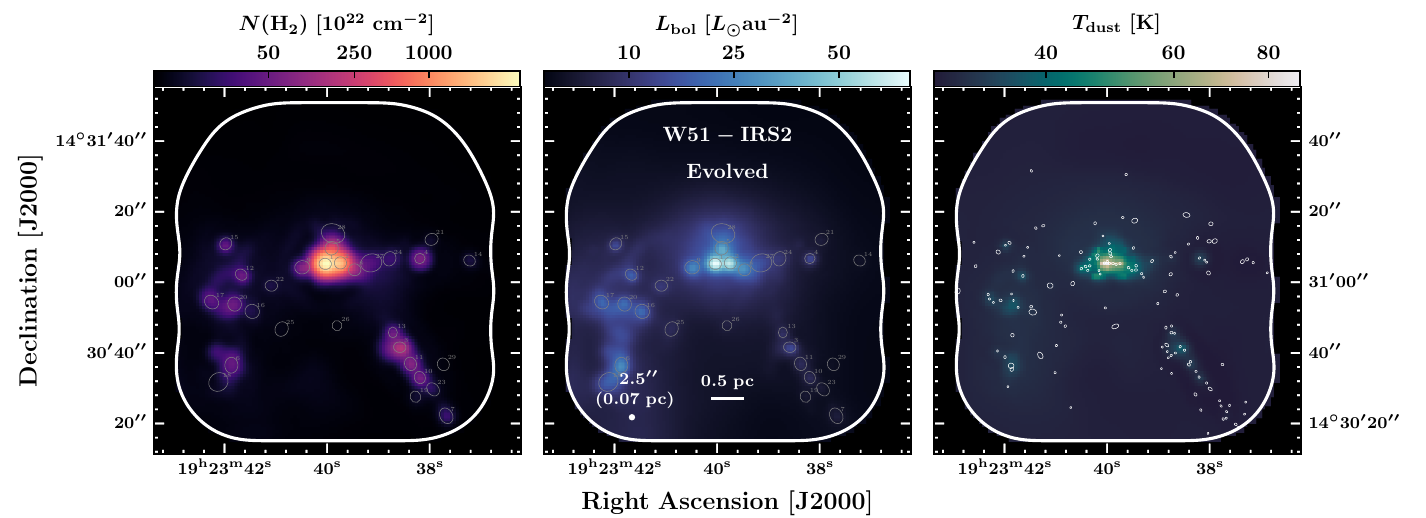}
      \caption{(Continued)}
   \end{centering}
\end{figure*}  

\begin{figure*}[htb]
   \begin{centering}
      \includegraphics[width=\hsize, trim={2cm 7.5cm 0cm 1cm},clip]{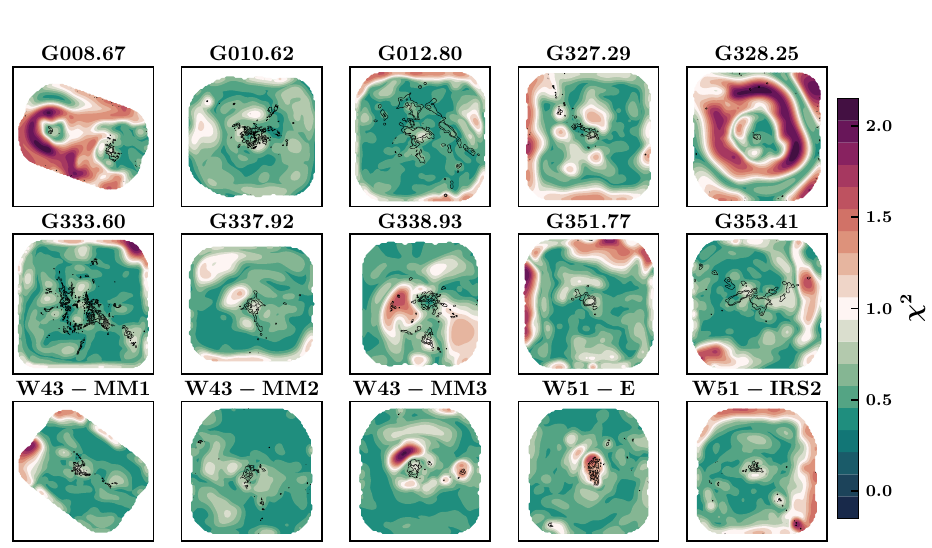}
      \caption{Performance of PPMAP modeling. Reduced $\chi^2$ maps are shown for each region studied. Black contours represent the continuum emission at 1~mm, using logarithmically spaced levels between 1\% and 10\% of the maximum flux.}\label{fig:PPMAPs-chi2}
   \end{centering}
\end{figure*}

\onecolumn
\begin{longtable}{c c c c c c c c c c c c c}
\hline
Region & ID & R.A. (J2000) & Dec. (J2000) & a & b & PA & FWHM & $L_{\rm bol}$ & $M$ & $L / M$ \\ 
       &    & [deg]        & [deg]        & ["] & ["] & [$\deg$] & [au] & [$\times 10^2 L_\odot$] & [$M_\odot$] & [$L_\odot / M_\odot$] \\ \hline \\[-1.0em]
\endfirsthead
\caption{continued.} \\ 
\hline
Region & ID & R.A. (J2000) & Dec. (J2000) & a & b & PA & FWHM & $L_{\rm bol}$ & $M$ & $L / M$ \\ 
       &    & [deg]        & [deg]        & ["] & ["] & [$\deg$] & [au] & [$\times 10^2 L_\odot$] & [$M_\odot$] & [$L_\odot / M_\odot$] \\ \hline \\[-1.0em]
\endhead 
\hline
\endfoot
G008.67 & \footnotesize{1} & \footnotesize{271.57936} & \footnotesize{-21.62552} & \footnotesize{3.6} & \footnotesize{3.4} & \footnotesize{122} & \footnotesize{11800} & \footnotesize{$57\pm18$} & \footnotesize{$290\pm80$} & \footnotesize{19} \\
            & \footnotesize{   2} & \footnotesize{ 271.59800} &  \footnotesize{-21.61969} &  \footnotesize{ 3.6} &  \footnotesize{ 3.3} & \footnotesize{  91} & \footnotesize{ 11700}  
& \footnotesize{ $19\pm6$ } & \footnotesize{ $93\pm26$ } & \footnotesize{ $20$ }  & & \\ 
            & \footnotesize{   3} & \footnotesize{ 271.57747} &  \footnotesize{-21.62720} &  \footnotesize{ 3.4} &  \footnotesize{ 3.0} & \footnotesize{ 268} & \footnotesize{ 10900}  
& \footnotesize{ $7.0\pm3.2$ } & \footnotesize{ $16\pm6$ } & \footnotesize{ $43$ }  & & \\ 
            & \footnotesize{   4} & \footnotesize{ 271.57764} &  \footnotesize{-21.62601} &  \footnotesize{ 3.8} &  \footnotesize{ 3.5} & \footnotesize{ 161} & \footnotesize{ 12400}  
& \footnotesize{ $6.3\pm2.5$ } & \footnotesize{ $18\pm7$ } & \footnotesize{ $35$ }  & & \\ 
            & \footnotesize{   5} & \footnotesize{ 271.57863} &  \footnotesize{-21.62740} &  \footnotesize{ 3.7} &  \footnotesize{ 3.4} & \footnotesize{ 191} & \footnotesize{ 12100}  
& \footnotesize{ $7.4\pm2.9$ } & \footnotesize{ $17\pm6$ } & \footnotesize{ $44$ }  & & \\ 
            & \footnotesize{   6} & \footnotesize{ 271.58083} &  \footnotesize{-21.62373} &  \footnotesize{ 3.2} &  \footnotesize{ 2.8} & \footnotesize{ 148} & \footnotesize{ 10000}  
& \footnotesize{ $2.4\pm1.2$ } & \footnotesize{ $4.3\pm2.5$ } & \footnotesize{ $57$ }  & & \\ 
            & \footnotesize{   7} & \footnotesize{ 271.57859} &  \footnotesize{-21.62235} &  \footnotesize{ 3.4} &  \footnotesize{ 3.1} & \footnotesize{ 103} & \footnotesize{ 11000}  
& \footnotesize{ $1.8\pm0.9$ } & \footnotesize{ $26\pm9$ } & \footnotesize{ $7$ }  & & \\
\end{longtable}
~\\~\\
\raggedright
\footnotesize{
The quantities a, b, PA, FWHM, $L_\mathrm{bol}$, $M$ and $L / M$ are respectively the major and minor sizes at half-maximum, angle of the major axis (east of vertical), physical full width at half-maximum in astronomical units, bolometric luminosity, mass, and (bolometric) luminosity-to-mass ratio. Some sources overlap between W43-MM2 and W43-MM3.}
\twocolumn


\end{document}